\documentclass[prb,aps,amssymb,amsmath,twocolumn,showpacs,floatfix]{revtex4-1}

\usepackage{graphicx}
\usepackage{bm}
\usepackage{amssymb,amsmath, mathrsfs}
\usepackage{color}
\usepackage{dsfont}
\usepackage{amsfonts}
\usepackage{enumerate}

\newcommand{\Tr}{{\rm Tr}\,} 
\newcommand{\Ln}{{\rm Ln}\,} 
\newcommand{\Det}{{\rm Det}\,} 
\newcommand{\xUnit}{{\mathds 1}} 

\newcommand{\bra}[1]{\left\langle #1 \right\rvert}
\newcommand{\ket}[1]{\left\lvert #1 \right\rangle}

\renewcommand{\Re}{\mathrm{Re}\,} 
\renewcommand{\Im}{\mathrm{Im}\,} 

\begin{document}

\title{Analytically solvable model of an electronic Mach-Zehnder interferometer}

\author{St\'ephane~Ngo~Dinh,$^{1}$ Dmitry A. Bagrets,$^{2}$ and Alexander
D.~Mirlin$^{1,3,4}$}

\affiliation{$^{1}$Institut f\"ur Theorie der \!Kondensierten \!Materie and DFG Center for Functional Nanostructures, Karlsruhe Institute of Technology, 76128 Karlsruhe, Germany\\
$^{2}$Institut f\"ur Theoretische Physik, Universit\"at zu K\"oln, Z\"ulpicher Str.~77, 50937 K\"oln, Germany\\
$^{3}$Institut f\"ur Nanotechnologie, Karlsruhe Institute of Technology, 76021 Karlsruhe, Germany\\
$^{4}$Petersburg Nuclear Physics Institute, 188300 St.~Petersburg, Russia}

\date{\today}

\begin{abstract}

We consider a class of models of non-equilibrium electronic  Mach-Zehnder
interferometers built on integer quantum Hall edges states. The models are
characterized by the electron-electron interaction being restricted to the inner part of
the interferometer and transmission coefficients of the quantum quantum point contacts, defining the interferometer, which may take arbitrary values from zero to one.
We establish an exact solution of these models in terms of
single-particle quantities---determinants and resolvents of Fredholm integral
operators. In the general situation, the results can be obtained numerically. In
the case of strong charging interaction, the operators acquire the block
Toeplitz form. Analyzing the corresponding Riemann-Hilbert problem, we reduce
the result to certain singular single-channel determinants (which are a
generalization of Toeplitz determinants with Fisher-Hartwig singularities), and
obtain an analytic result for the interference current (and, in particular, for
the visibility of Aharonov-Bohm oscillations). Our results, which are in good
agreement with experimental observations, show an intimate connection between
the observed ``lobe'' structure in the visibility of Aharonov-Bohm oscillations
and multiple branches in the asymptotics of singular integral determinants. 

\end{abstract}

\maketitle

\section{Introduction}

\begin{figure}[t]
 \includegraphics[width=3.3in]{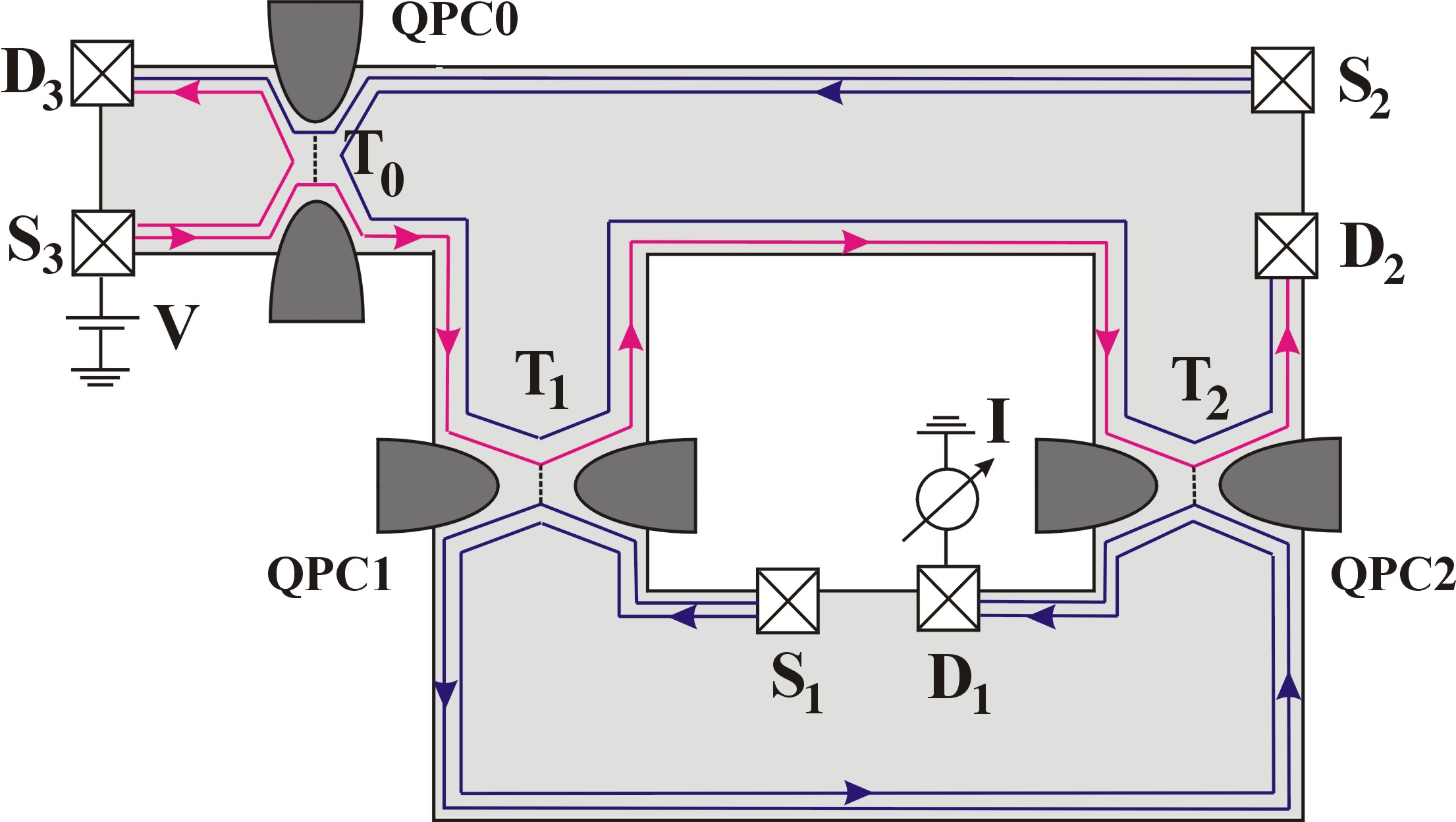}
 \caption{Layout of an electronic Mach-Zehnder interferometer
built on quantum-Hall edge states at filling factor $\nu=2$.
Quantum point contact (QPC1 and QPC2) characterized by transparencies
$T_{1(2)}$ are used to partially mix the outer edge channels. All
Ohmic contacts are grounded, except for the source terminal $S_3$
which is kept at voltage $V$. The current is 
measured in the drain terminal $D_1$. The QPC0 can be used to 
dilute the incoming current in the outer channel by changing the transparency $T_0$.}
\label{fig:MZI_layout}
\end{figure}

Electronic Mach-Zehnder interferometers (MZIs) realized with edge states in the
integer quantum Hall (QH) 
regime have attracted a lot of attention recently because of a striking
interplay between the quantum coherence 
and effects of electron-electron interaction observed in these mesoscopic  
devices~\cite{Ji:2003, Neder:2006, Neder:2007, Neder:2007a, Roulleau:2007,Roulleau:2008, Roulleau:2008a,
Roulleau:2009,Litvin:2007,Litvin:2008,Litvin:2010,Bieri:2009,Huynh:2012,Helzel:2012}. 
By analogy to the optical interferometer, 
the chiral edge states in the electronic MZI, playing the role of light beams, 
are coupled by quantum 
point contacts (QPCs), which act as electron beam-splitters (see
Fig.~\ref{fig:MZI_layout}). The 
differential conductance measured in the above experiments shows strong
Aharonov-Bohm (AB) oscillations. 
They are a manifestation of quantum coherence of electrons, propagating through
different arms of interferometer, 
and are quantified in terms of visibility.  The most remarkable experimental
observation is that the out-of-equilibrium  
visibility does not decrease monotonically with voltage but rather
demonstrates a sequence of decaying  
oscillations (``lobes''). Such a dependence cannot be explained within an
assumption of non-interacting electrons.

Investigation of quantum interference and decoherence in Aharonov-Bohm
rings and interferometers has a long history~\cite{Aronov:1987}. In
particular, much attention has been paid to sources of dephasing that may
arise from the external 
noise~\cite{Seelig:2001,Seelig:2003,Marquardt:2004,Marquardt:2004a,Foerster:2005,Neder:2007b} 
or are the result of the intrinsic electron-electron 
interaction~\cite {Ludwig:2004,LeHur:2005,LeHur:2006,Texier:2005,Dmitriev:2010}. 
The advent of QH interferometers has renewed the interest in this problem, with
a considerable number of recent theoretical 
works~\cite{Chalker:2007,Sukhorukov:2007,Neder:2008,Youn:2008,Levkivskyi:2008,Levkivskyi:2009,Kovrizhin:2009,Kovrizhin:2010,
Schneider:2011,Rufino:2012} aiming at a  resolution of  the ``visibility
puzzle'' in MZIs. 
These recent theories can be subdivided into the
approaches assuming contact~\cite{Levkivskyi:2008,Levkivskyi:2009,Rufino:2012} and 
long-range~\cite{Chalker:2007,Sukhorukov:2007,Kovrizhin:2009,Kovrizhin:2010,
Schneider:2011} Coulomb interaction. Despite the fact that the model
of contact {\it e-e} interaction may successfully describe the related experiments on the
energy relaxation in the QH edge states at filling factor 
$\nu$=2~\cite{Altimiras:2010,Altimiras:2010a,leSueur:2010,Kovrizhin:2012,Levkivskyi:2012}, 
results of Refs.~\onlinecite{Schneider:2011,Rufino:2012} indicate 
that the account of the long-range character of Coulomb interaction is
of central importance for a full understanding of non-equilibrium phenomena in
MZIs.

The natural choice of a theoretical approach to one-dimensional (1D) interacting
electrons
in the QH edge states is that of bosonization~\cite{Wen:2004}. However, in the
case under interest, one faces serious complications when trying to apply this
approach. First, already in the single-channel problem, the bosonized action of
the theory ceases to be Gaussian under non-equilibrium conditions
\cite{Gutman:2008,Gutman:2010a,Gutman:2010} (see also a related earlier work
\cite{Abanin:2005} on the non-equilibrium Fermi-edge singularity). This difficulty
has been solved by development of the non-equilibrium bosonization formalism
yielding results for physical observables in terms of single-particle Fredholm
determinants which are of Toeplitz type for a short-range interaction
model\cite{Gutman:2010a,Gutman:2010}. Second, an even more severe obstacle
arises when one describes electron scattering at QPCs. Specifically, electron
tunneling between two edge channels yields the $\cos$-like term in the
bosonized Hamiltonian (an the action), impeding a solution to the
problem. For this reason, almost all recent theories of MZIs consider the limit
of weakly coupled edge states 
where the perturbative treatment of electron tunneling at QPCs is justified.
This is rather unfortunate  
since in the experiment transmission coefficients of both QPCs are usually close
to one half. While in a very restricted set of models exact solutions via the
Bethe ansatz are available
\cite{Fendley:1995,Ponomarenko:2009,Ponomarenko:2010}, the systems we are
interested in do not belong to this class. On the other hand, it would be
clearly highly advantageous to have an analytically treatable mode for
integer QH MZI for an arbitrary number of edge channels, arbitrary interaction
range and strength, and arbitrary transmissions at QPC.

In this paper we consider the model of the MZI operating at integer filling
factor $\nu$ where electrons interact only when they are {\it inside} the interferometer.
The model is specified by two single-particle scattering matrices of the QPCs defining the
interferometer and by the model of Coulomb interaction inside the
interferometer. 
We focus on the model of ``maximally long-range'' interaction
when the interaction energy depends only on total
charges collected within each of the arms and is characterized by an
electrostatic charging energy $E_c$. Let us emphasize that this restriction is
not crucial: within this approach one can, in principle,  consider any
interaction within the interior region of MZI.  

For the case $\nu=1$, such a model was introduced in Refs.~\onlinecite{Neder:2008,Youn:2008}. In Ref.~\onlinecite{Kovrizhin:2010} an exact solution to it at $\nu=1$
was obtained by using a combination of bosonization and refermionization techniques and
was expressed in terms of single-particle determinant and resolvent that were evaluated numerically.
Our way to treat the problem is different in many aspects. We consider a MZI
with an arbitrary number of edge states and use the non-equilibrium functional
bosonization approach developed by us previously~\cite{NgoDinh:2012}. Within
this framework we demonstrate that 
an interfering current can be expressed in terms of a Fredholm functional determinant 
of the single-particle ``counting'' operator which bears 
resemblance to the problem of electron full counting statistics (FCS) of
mesoscopic transport~\cite{Levitov:1996}. In general, this determinant should
be evaluated numerically. In the limit of strong interaction
$E_c \gg 1/\tau$, where $\tau$ is the electron flight time through the MZI, the
``counting'' operator takes the block  Toeplitz form.
Under this condition, a fully analytical treatment turns out to be possible.
By solving the Riemann-Hilbert problem, we get rid of the matrix structure
and express the result in terms of a determinant of a single-channel singular
integral operator generalizing Toeplitz determinants with Fisher-Hartwig
singularities. Determinants of this  type have been studied in
Ref.~\onlinecite{Protopopov:2012} where a conjecture on their asymptotic
behavior was formulated and supported by a large body of analytical and
numerical arguments. This allows us to obtain the result in a closed analytic
form. 

Our analytical result demonstrates that the ``lobe'' pattern in visibility is a
many-body interference effect  
resulting from the quantum superposition of (at least) two many-particle scattering amplitudes 
with the mutual phase difference which is linear in voltage.
In the limit of strong interaction we find the scaling exponents 
which describe power-law dependences of these amplitudes on voltage and obtain the non-equilibrium 
dephasing rate governing the exponential suppression of visibility with bias
(or, equivalently, with the length of the arms of the MZI).
The power-law exponents as well as the dephasing rate depend on the
transmission coefficient of the first QPC
and on the filling factor $\nu$. 

Our analytical findings are corroborated and complemented by numerical
evaluations of the Fredholm determinants determining the exact solution for
arbitrary $E_c \sim 1/\tau$. At $E_c \gg 1/\tau$ our numerical results
provide further support to the aforementioned conjecture of
Ref.~\onlinecite{Protopopov:2012}. 
At moderate charging energy $E_c \gtrsim 1/\tau$ and $\nu$=2 the obtained
results match very well experimental observations. 

The remainder of the paper is organized as follows. Section~\ref{s2} is devoted to the exposition
of our main results and their physical interpretation. In Sec.~\ref{s3} we
present the non-equilibrium
functional bosonization approach. We demonstrate that the MZI problem defined above is exactly
solvable by means of the instanton method. In the limit of strong interaction
(Sec.~\ref{s4}),
the full analytical treatment becomes possible. It is based upon the 
asymptotic results for the generalized Toeplitz determinants. We show the relation of the MZI
problem to the latter theory and evaluate analytically the Aharonov-Bohm
conductance. In Sec.~\ref{s5} we consider the influence of an additional quantum
point contact (QPC0) diluting the incoming current in one of the channels and 
develop numerically exact approach to solve the problem in the case of a moderate
charging energy. Finally, in
Sec.~\ref{concl} we summarize the findings of this work and discuss prospects
for future research.

\section{Results and qualitative discussion}
\label{s2}

In this section we set the stage by defining the theoretical model of the Mach-Zehnder interferometer
and then present our results and give their physical interpretation. This part
of the paper is self-contained and can be read independently of other sections, where
we provide technical details of the calculations.

\subsection{Model of the MZI}
\label{s2.1}

\begin{figure}[b]
 \includegraphics[width=1.5in]{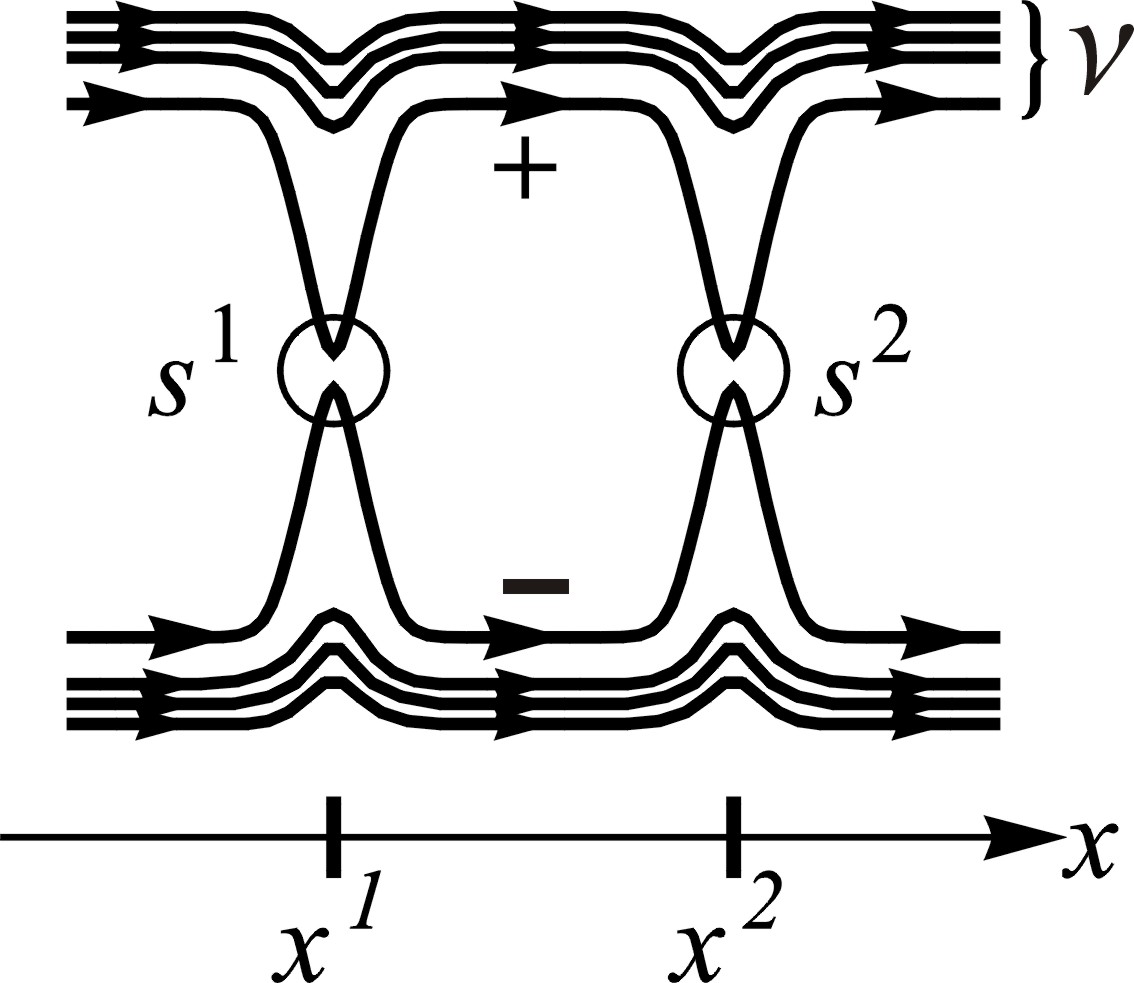}
 \caption{Scheme of an MZI at filling factor $\nu$. Two quantum point contacts are characterized
by the scattering matrices $s^1$ and $s^2$, which connect the outer channels.
Inner channels are fully reflected.}
\label{fig:MZI_scheme}
\end{figure}

We consider the Mach-Zehnder interferometer (MZI)
realized with edge states in the QH regime at integer filling factor  
$\nu$. The experimental layout (in case of $\nu=2$) and the scheme of the MZI are shown 
in Figs.~\ref{fig:MZI_layout} and \ref{fig:MZI_scheme}. In this setup the outer chiral channels, 
propagating along different arms of the MZI (we denote them by the index $\pm$) 
are coupled by means of two quantum point contacts (QPCs), located at points $x_{\pm}^{1(2)}$.
An additional quantum point contact, QPC0, is used to bias the incoming outer
channel at the upper edge by voltage $V$ (and in general also to
dilute the incoming current if the transmission coefficient of the QPC0 is tuned
to a value $T_0>0$).
We further assume that all inner chiral channels are fully reflected from each
QPC and, in particular,
the incoming inner channels are grounded. This layout of the MZI and the bias scheme are
realized in most of the experiments (an exception is Ref.~\onlinecite{Bieri:2009}, where the MZI
setup at $\nu=2$ did not contain additional QPC0). 

\begin{figure*}[t]
\centering{
 \includegraphics[width=7in]{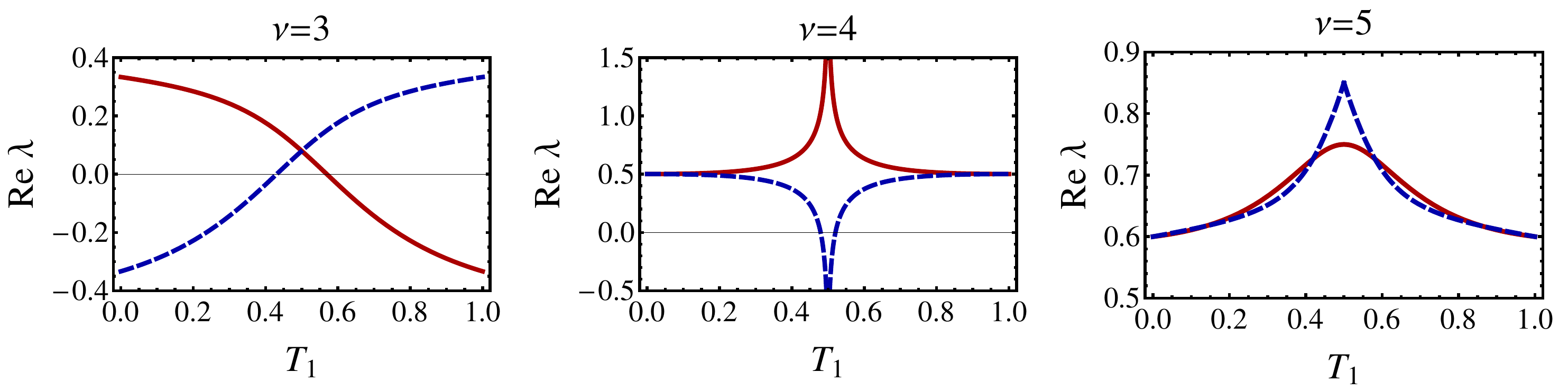}
}
 \caption{Power-law exponents shown as the function of transmission coefficient 
$T_1$. Solid line shows ${\rm Re}\,\lambda_1$, dashed line shows ${\rm Re}\,\lambda_2$ in the case of $\nu=3$,
and $(-{\rm Re}\,\lambda_2)$ in the case of $\nu = 4,5$.  }
\label{fig:Powers}
\end{figure*}

A theoretical model  considered throughout the paper is specified by
the action
${\cal S} = {\cal S}_0 + {\cal S}_{\rm int}$ of interacting 1D fermions,
\begin{eqnarray}
 {\cal S}_0&=&\sum_{\varrho = \pm}\int dt\,dx \,
 \bar\psi_{\varrho}(x) \left(i \partial_t + iv\partial_x \right) \psi_{\varrho}(x), \label{eq:S_0}\\ 
 {\cal S}_{\rm int}&=& \frac{1}{2}E_c \sum_{\varrho=\pm }\int dt\, {\cal N}_\varrho^{\,2}, \label{eq:S_int}
\end{eqnarray}
which are described by the Grassmann fields $\psi_\varrho$ in the arm $\varrho=\pm$. In the case
$\nu\geq 2$ the fermionic fields have the vector structure $\psi_\varrho =
(\psi_{1\varrho}, \dots, \psi_{\nu\varrho})^T$ due to
multiple edge channels. The Coulomb interaction is taken into account by the
electrostatic model with a charging energy $E_c = e^2/C$, such that electrons
interact only when they are inside the interferometer. Thus in
Eq.~(\ref{eq:S_int})
\begin{equation}
{\cal N}_\varrho = \sum_{k=1}^\nu \int_{x_\varrho^1}^{x_\varrho^2} \bar\psi_{k\varrho}(x+0)\psi_{k\varrho}(x)
\end{equation}
is the total number of electrons in the upper/lower arm of the MZI. The action
${\cal A}_{\rm int}$ describes intra- and interchannel Coulomb interaction which is maximally
non-local (or long-range) in space.  At the same time the inter-edge interaction is disregarded in 
our model, which is motivated by the fact that different edges are spatially well separated.     
At the QPCs outgoing fermion fields (with channel index $k=1$) are related to the incoming ones
by the scattering matrices
\begin{align}
\psi_{1\varrho}(x_\varrho^{j}+0) = s_{\varrho\mu}^j \psi_{1\mu}(x_\mu^{j}-0),  \label{eq:S_matrix_relation}\\
 \hat s^j = 
   \left(
   \begin{array}{cc}
     i R_j^{1/2} & T_j^{1/2}\\
     T_j^{1/2} & i R_j^{1/2}
   \end{array} 
  \right), 
  \label{eq:s_matrix}
\end{align}
where $T_j$ and $R_j$ are reflection and transmission coefficients at the $j$-th
QPC. 

The model of the MZI with the above action ${\cal A}$ is exactly solvable for any value of the
charging energy $E_c$ and transmission coefficients $T_j$, as we show in 
Sections~\ref{s3}, \ref{s4} and \ref{s5}. For simplicity, we consider an
interferometer
with equal arms, $x_+^{(2)}-x_+^{(1)} = x_-^{(2)}-x_-^{(1)} =L$, which is predominantly
the experimental situation. In the limit $E_c\tau \gg 1$, where $\tau = L/v$ is the electron
dwell time in the MZI, fully analytical treatment is possible. In the more general case
$E_c\tau \sim 1$ we have developed a numerically exact scheme to evaluate the visibility in 
the MZI as a function of voltage (Sec. \ref{s5}).  Before going into
details of the
calculations (Sec.~\ref{s3}), we summarize our main results.

\begin{figure*}[t]
\centering{
 \includegraphics[width=2.5in]{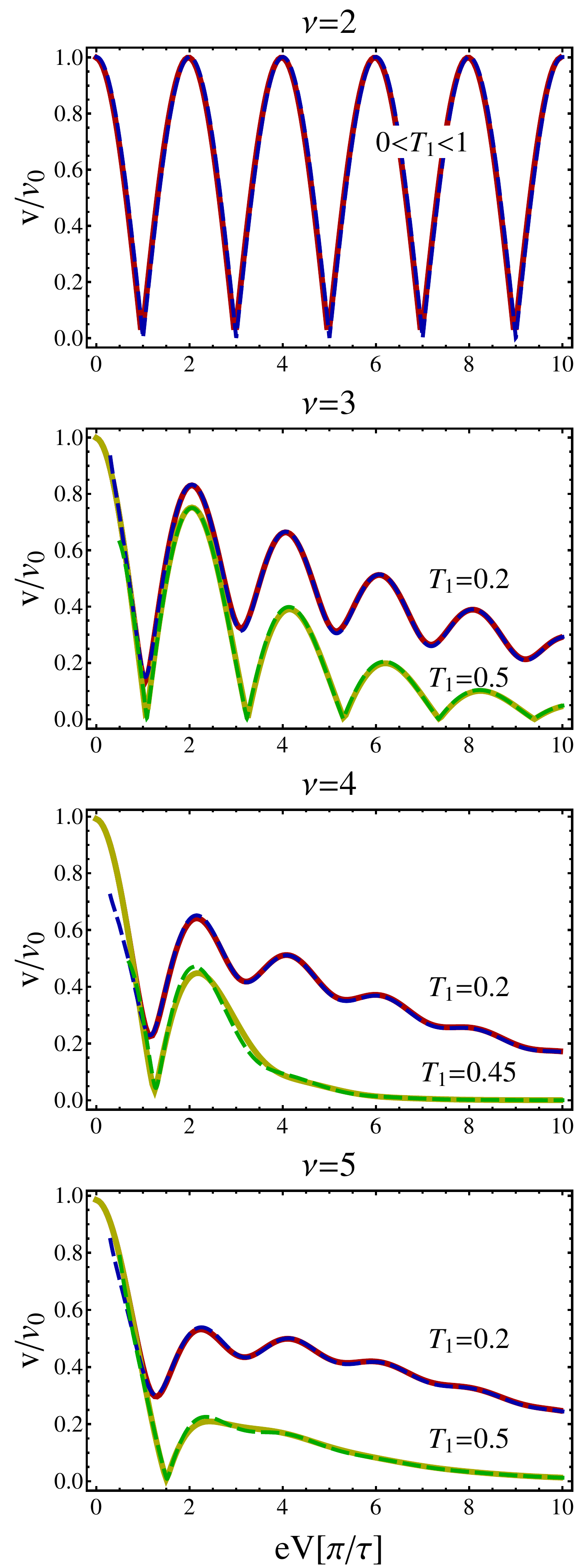}
 \hspace{2cm}
 \includegraphics[width=2.5in]{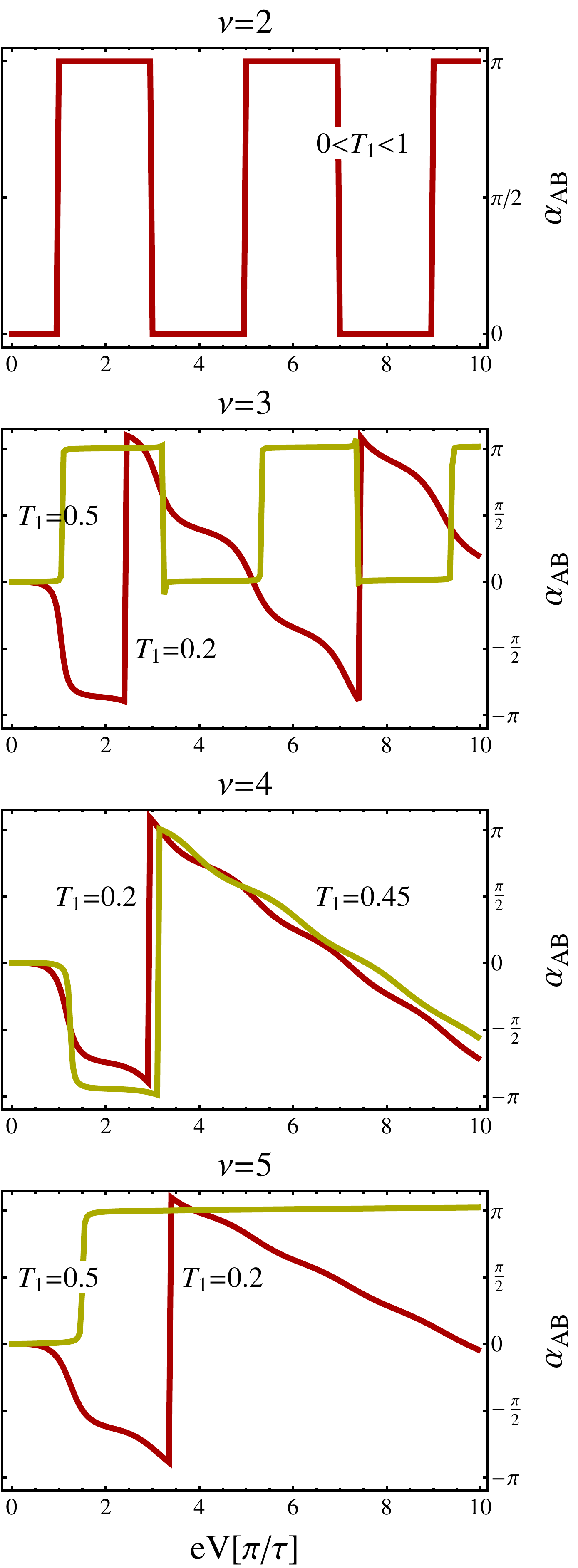}
}
 \caption{The voltage dependence of the visibility (normalized to its
non-interacting value; left) and of the phase (right) of AB oscillations for
$\nu=2,3,4,5$ in the strong-interaction limit, $E_c \tau \gg 1$. Solid lines
show numerically exact results, dashed lines represent analytical results. The
latter are strictly speaking valid in the asymptotic high-voltage limit $eV \tau
\gg 1$ but turn out to work almost perfectly already at very modest values of
voltages, $eV\tau/\pi \gtrsim 0.5$. }
\label{fig:MZI_grid}
\end{figure*}

\subsection{Limit of strong interaction}
\label{s2.2}

First, we discuss the results in the limit $E_c\tau \gg 1$. In the case of not too low
voltages, namely at $eV\tau \gtrsim~1$, our model predicts the asymptotic expansion for 
the differential conductance $dI/dV$ of the MZI in the form 
\begin{eqnarray}
{\cal G}(V) &=& \frac{e^2}{2\pi\hbar} \Bigl( T_1 R_2  + T_2 R_1 \nonumber\\
&+& 2(T_1 R_1 T_2 R_2)^{1/2}
{\rm Re}\left[e^{i\Phi} \frac{\partial{\cal I}_0 }{\partial (eV\tau) }\right]\Bigr),
\end{eqnarray}
where $\Phi$ is the magnetic flux and ${\cal I}_0$ is the amplitude of the interference
Aharonov-Bohm (AB) contribution to the current, 
\begin{equation}
{\cal I}_0 = e^{i eV\tau(\beta_1 +1/\nu)}\Bigl(C_1 (eV\tau)^{\lambda_1} + 
C_2 (eV\tau)^{\lambda_2} e^{\pm i eV\tau }\Bigr). \label{eq:I_0}
\end{equation}
The choice of the sign $\pm$ in the exponent of the last term will be explained
below Eq.~(\ref{eqn:MZIExpHighNu}). 
Equation (\ref{eq:I_0}) contains the two leading terms of a series (in general,
infinite). The dependence of each term of this series on $eV\tau$ is characterized
by a certain power-law exponent $\lambda_i$ and by a certain oscillating factor.

Interpretation of the different ingredients in this expression is as follows.
The coefficient 
\begin{eqnarray}
\beta_1= \frac{1}{2 \pi i}\ln \left( R_1 e^{-4\pi i/\nu} + T_1\right)\\
\nonumber
\end{eqnarray}
describes the non-equilibrium dephasing of the AB oscillations induced by a combined
effect of inelastic {\it e-e} scattering and the quantum shot noise generated at the 1st QPC. 
If $\nu\geq 3$ then ${\rm Im}\beta_1 > 0$ and, by defining the out-of-equilibrium dephasing
rate as 
\begin{equation}
\tau_\phi^{-1} = eV \,{\rm Im}\,\beta_1,
\end{equation}
we see that AB oscillations are suppressed 
by the factor $e^{-\tau/\tau_\phi}$ in the high-bias limit $eV \gg 1/\tau$. 

It is worth stressing that the exponential suppression of the interference current is
directly related to the full counting statistics (FCS) of electrons passing through the QPC1
at the time interval $\tau$.
Indeed, defining the FCS cumulant generating function (CGF) of the backscattering current as
\begin{equation}
\chi_\tau(\lambda) = \left[ 1 + R_1 (e^{i\lambda}-1)\right]^{eV\tau/2\pi},
\end{equation}
where $\lambda$ is the so-called ``counting field''~\cite{Levitov:1996}, we see
that the damping factor is equal to 
\begin{equation}
e^{i\beta_1 eV\tau} = \chi_\tau(-4\pi/\nu).
\end{equation}

The exponents $\lambda_{1,2}$, which set the power-law dependence of the
interference current on bias, belong to the class of non-equilibrium quantum
critical exponents. Physically, they can be understood as being due to
the Anderson orthogonality catastrophe which happens each time when an electron
enters or leaves the interior part of the MZI where it strongly interacts with
all other electrons. It is worth mentioning that in the considered simplified
model, where the {\it e-e} interaction is present only inside MZI, the
orthogonality catastrophe is absent for the incoherent contribution to the
current, which stays linear in voltage as in the case of non-interacting
fermions.  

The exponents $\lambda_{1,2}$ are functions of both, the filling factor $\nu$
and the transparency $T_1$ of the first QPC, and are shown in
Fig.~\ref{fig:Powers}. The explicit expressions read
\begin{eqnarray}
\label{eqn:MZIExpLowNu}
\lambda_{1,2}  =  &-& 2\Bigl( \frac{1}{\nu}-\frac{1}{2} +\beta_1 \pm \frac{1}{2}\Bigr)^2  \\
 &+& 1 - \frac{2}{\nu} + \frac{2}{\nu^2}  ,  \qquad  2 \leq \nu <4, \nonumber
\end{eqnarray}
for low filling factors and
\begin{eqnarray}
\label{eqn:MZIExpHighNu}
\lambda_1 &=& -2\Bigl( \frac{1}{\nu} + \beta_1\Bigr)^2 +  1 - \frac{2}{\nu} + \frac{2}{\nu^2}, \\
\lambda_2 &=& -2\Bigl( \frac{1}{\nu} + \beta_1 \pm \frac{1}{2}\Bigr)^2 - \frac{1}{2} + \frac{2}{\nu^2},
\quad \nu \geq  4 \nonumber
\end{eqnarray}
in the case of higher $\nu$. In the case $2\leq \nu < 4$ the voltage dependent
phase factor in Eq.~(\ref{eq:I_0})
has to be taken with the sign (+). For $\nu \geq 4$ the $\pm$ sign corresponds to the case
$T_1 >1/2$ and $T_1 < 1/2$, respectively. The coefficients $C_{1,2}$ in Eq.~(\ref{eq:I_0}) are
some bias-independent complex numbers which depend solely on $\nu$ and $T_1$ and
can be found
from the fit of this asymptotic expansion to its numerically exact counterpart.
In the limit of strong interaction, $E_c \tau \gg 1$, the case $\nu=1$ is very
special. Specifically, one has then ${\cal I}_0 = (eV\tau)$ and the MZI behavior
is the same as in the absence of {\it e-e} interaction.

Experimentally, one usually quantifies the coherence of the interferometer in
terms of the
visibility ${\cal V}$ and the phase $\alpha_{AB}$ of the AB oscillations of the
conductance. The visibility is defined as the ratio of the amplitude of the AB
oscillations to the mean value of the conductance. In our model 
\begin{equation}
{\cal V} = {\cal V}_0 \frac{\partial\, {\cal I}_0}{\partial (eV\tau)}, \quad 
{\cal V}_0 = \frac{2(T_1 R_1 T_2 R_2)^{1/2}}{T_1 R_2  + T_2 R_1},
\end{equation}
where ${\cal V}_0$ is the non-interacting value of ${\cal V}$, and 
\begin{equation}
\alpha_{AB} = {\rm arg}\left[{\partial\, {\cal I}_0}/{\partial (eV\tau)} \right].
\end{equation}
In terms of the above quantities the conductance ${\cal G}(V)$ takes the form
\begin{equation}
{\cal G}(V) = \frac{e^2}{2\pi\hbar}\Bigl(R_1 T_2 + R_2 T_1\Bigr)
\Bigl(1 + {\cal V}(V) \cos\bigl[\Phi + \alpha_{AB}(V)\bigr]\Bigr).
\end{equation}
In Fig.~\ref{fig:MZI_grid} we show the visibility ${\cal V}$ (normalized to its
non-interacting value ${\cal V}_0$) and the phase $\alpha_{\rm AB}$ 
of the AB oscillations as functions of bias for different filling factors
$\nu=2,3,4,5$ and for different transmissions $T_1$. 

In each plot we have fitted the exact visibility (obtained numerically) by its
analytic form based on Eq.~(\ref{eq:I_0}) with two free parameters
$C_1$ and $C_2$.
Although Eq.~(\ref{eq:I_0}) is strictly speaking an asymptotic formula valid in
the high voltage limit $eV \tau \gg 1$,
we see in Fig.~\ref{fig:MZI_grid}  that the analytical result is an excellent
approximation already starting from very modest values of voltages,
$eV\tau/\pi \gtrsim 0.5$. For still smaller voltages, the visibility
saturates at its non-interacting value ${\cal V}_0$.

The most spectacular feature of  Fig.~\ref{fig:MZI_grid} are oscillations
of visibility which become particularly strong yielding a ``lobe structure
with the visibility reaching zero at minima for $\nu=2$ (for any value of
$T_1$) and for $T_1=0.5$ (for any $\nu$). In this cases, the cusps in the
visibility at its minima are accompanied by $\pi$-jumps in the phase $\alpha_{\rm
AB}$.
As we explain below, the special role of $\nu=2$ is a characteristic
feature of the strong-interaction limit. On the other hand, the point $T_1=0.5$
remains special for a generic interaction.  

At $\nu=2$ we have $\lambda_1 = \lambda_2 = 0$ and $C_1 = -C_2$. This gives the
oscillatory visibility ${\cal V} = {\cal V}_0 \cos(eV\tau/2)$ which is independent of the
transparency $T_1$ and does not decay with bias. The behavior of the MZI in this
case is analogous to that in a model with short-range electron
interaction which is also exactly solvable at $\nu=2$ by means of the
method of refermionization, as has been recently shown in
Ref.~\onlinecite{Rufino:2012}. 
On a technical level, the absence of dephasing and independence of visibility
on $T_1$ at $\nu=2$ in the limit $E_c\tau \gg 1$ comes from the fact that the
counting phase becomes $-4\pi/\nu = -2\pi$ in this case.  At a moderate charging
energy $E_c\tau \sim 1$ both the dephasing and the dependence on $T_1$ in the
visibility of $\nu=2$ MZI are restored, see Sec.~\ref{s2.3}.

In the case $\nu=3$ an infinite number of lobes is observed. As discussed
above, the  visibility reaches zero at minima when (and only when) the
transmission is $T_1=1/2$. The reason for this special role of the point
$T=1/2$ is as follows: in this case the real parts of the two exponents are equal, 
$\Re\lambda_1=\Re\lambda_2$.

For $\nu \geq 4$ our model predicts only one central and one side lobe. Note, that 
at $\nu=4$ the exponents $\lambda_{1,2}$ logarithmically diverge at $T_1 \to 1/2$. 
This is the reason why at $\nu=4$ we have chosen to plot ${\cal V}(V)$ 
and $\alpha_{AB}(V)$ for a slightly different value $T_1=0.45$ (see
Fig.~\ref{fig:MZI_grid}).

\begin{figure}[b]
 \includegraphics[width=3.2in]{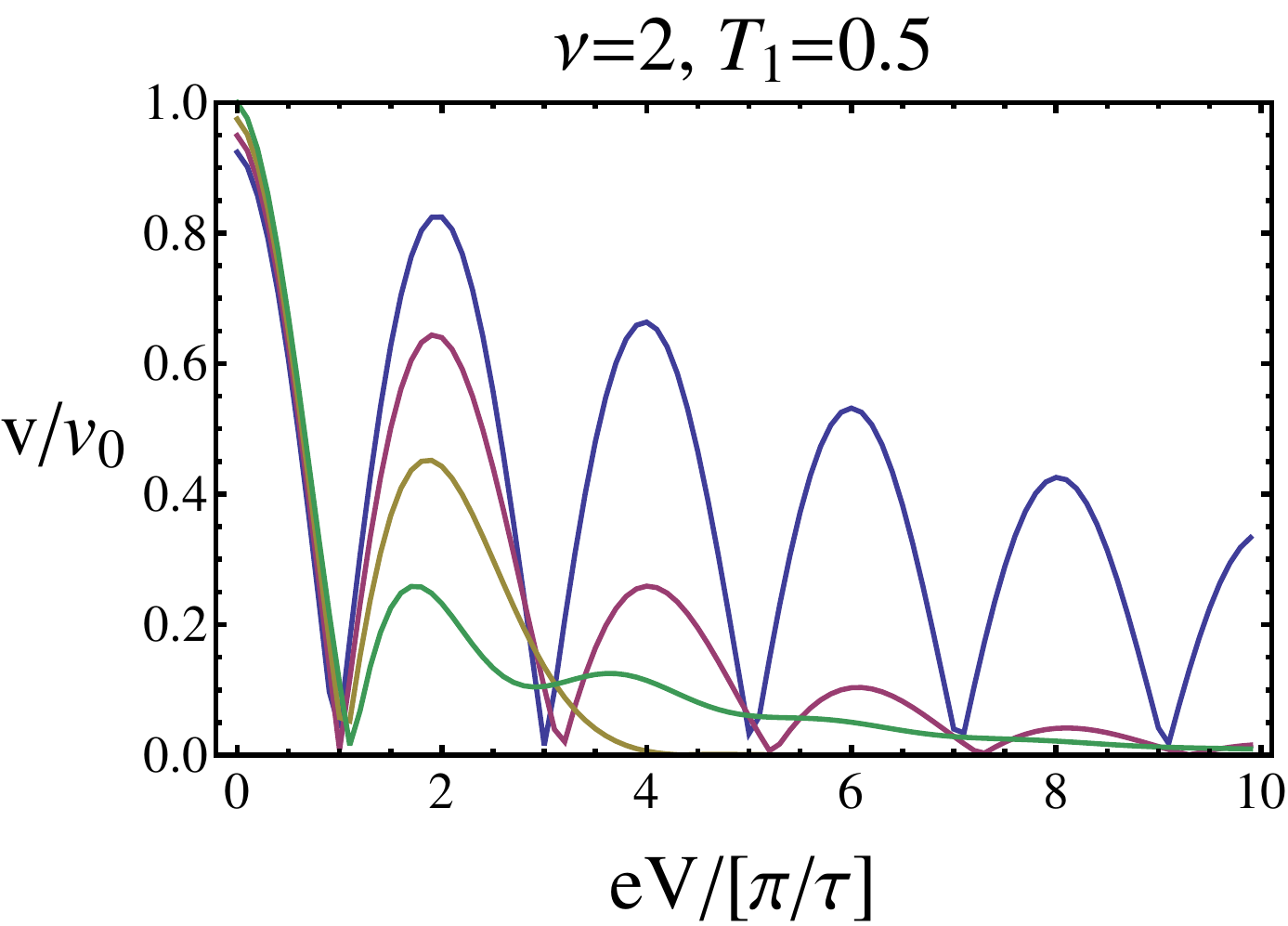}
 \caption{Visibility of AB oscillation at $\nu=2$,
$T_1=0.5$, and strong interaction, $E_c \tau \gg 1$,
in the presence of an additional quantum point contact, QPC0.
The curves from up to down were evaluated numerically for the reflection
coefficient of QPC0 equal to $R_0=0.9, 0.7, 0.5$ and $0.3$.}
\label{fig:Inj_noneq}
\end{figure}

\begin{figure*}[t]
\centering{
\includegraphics[width=3.0in]{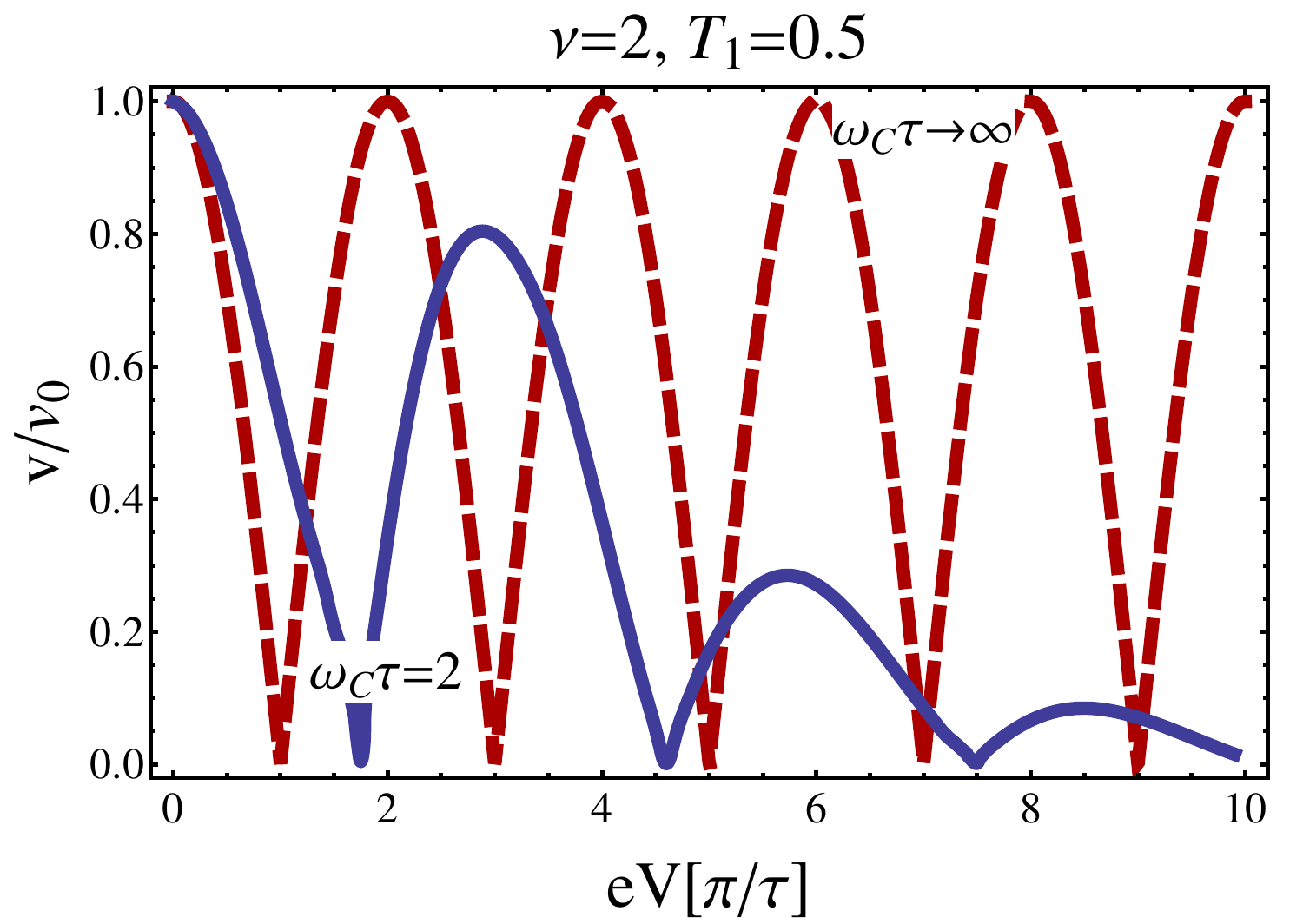}
\hspace{0.5cm}
\includegraphics[width=3.0in]{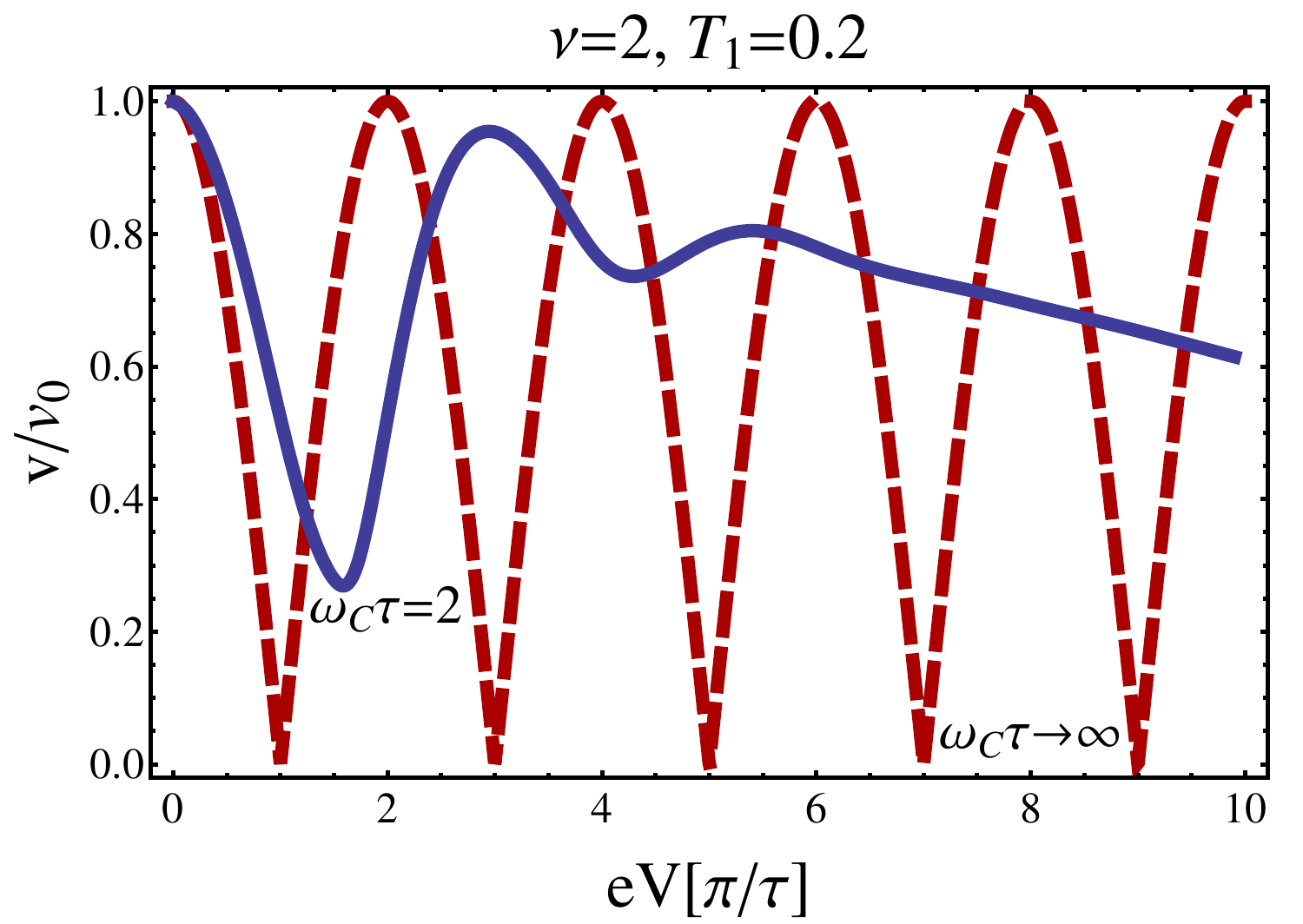}
}
\caption{Bias dependence of visibility of MZI with $\nu=2$ for the moderate
interaction strength ($\omega_c\tau=2$, solid blue curve)
in comparison to the limit of strong interaction ($\omega_c\tau\to
\infty$, dashed red line), where $\omega_c
= \nu E_c/\pi$ is the charge relaxation frequency. Note that at finite
$\omega_c$ the nodes in visibility are present only at $T_1=1/2$.}
\label{fig:Vis_Intermediate}
\end{figure*}

We turn now to the effect of a dilution of the impinging
current due to the electron scattering at an additional quantum point contact,
QPC0, which is put outside of the interferometer.
At $R_0<1$ the QPC0 generates the double-step distribution function
$f_+(\epsilon) = T_0 \theta(-\epsilon) + R_0 \theta(eV-\epsilon)$ for incoming
electrons, which affects the power-law exponent and serves as an additional source
of dephasing. The effect of QPC0 is particularly noticeable in the
strong-interaction model with $\nu=2$, since in this situation no dephasing and
no power-law decay of oscillations is found in the absence of QPC0 (see above).
In Fig.~\ref{fig:Inj_noneq} we show the visibility in this situation,
with half-transmitting QPC1, $T_1=1/2$,  and for different values of the
reflection coefficient $R_0$ of QPC0. In the case $R_0 > 1/2$ the suppression
of visibility with voltage can be roughly characterized by the dephasing rate
$1/\tau_\phi = (eV/2\pi) \ln(2R_0-1)$, which diverges logarithmically at $R_0\to 1/2$.
At this value of $R_0$ the behavior of the MZI visibility changes from the
regime with multiple
side lobes, characterized by periodic oscillations in ${\cal V}(V)$ with a typical
period $\sim 2\pi/\tau$, to the regime with only one node. Such a transition in
the behavior of visibility under variation of $R_0$ has been first found in 
Ref.~\onlinecite{Levkivskyi:2009} in the weak-tunneling regime $T_1\ll 1$ for the 
short-range {\it e-e} interaction model. 
Recently, such an effect of QPC0 on the visibility was observed experimentally\cite{Helzel:2012}. We believe that the experimental conditions of long-range interaction and $T_1\approx 1/2$ are closer to the ones studied within our model.

As discussed in more detail Sec.~\ref{s4.3.3} the appearance of the visibility fringes in our model stems 
from the superposition of two {\it multi-particle} amplitudes having the relative phase shift $eV\tau$.
These amplitudes correspond to processes with different number of particle-hole excitations between two Fermi edges. 
In other words, the ``lobe'' pattern in the visibility is a many-body effect linked to {\it e-e} interaction.
   
\begin{figure*}[t]
\centering{
\includegraphics[width=3.25in]{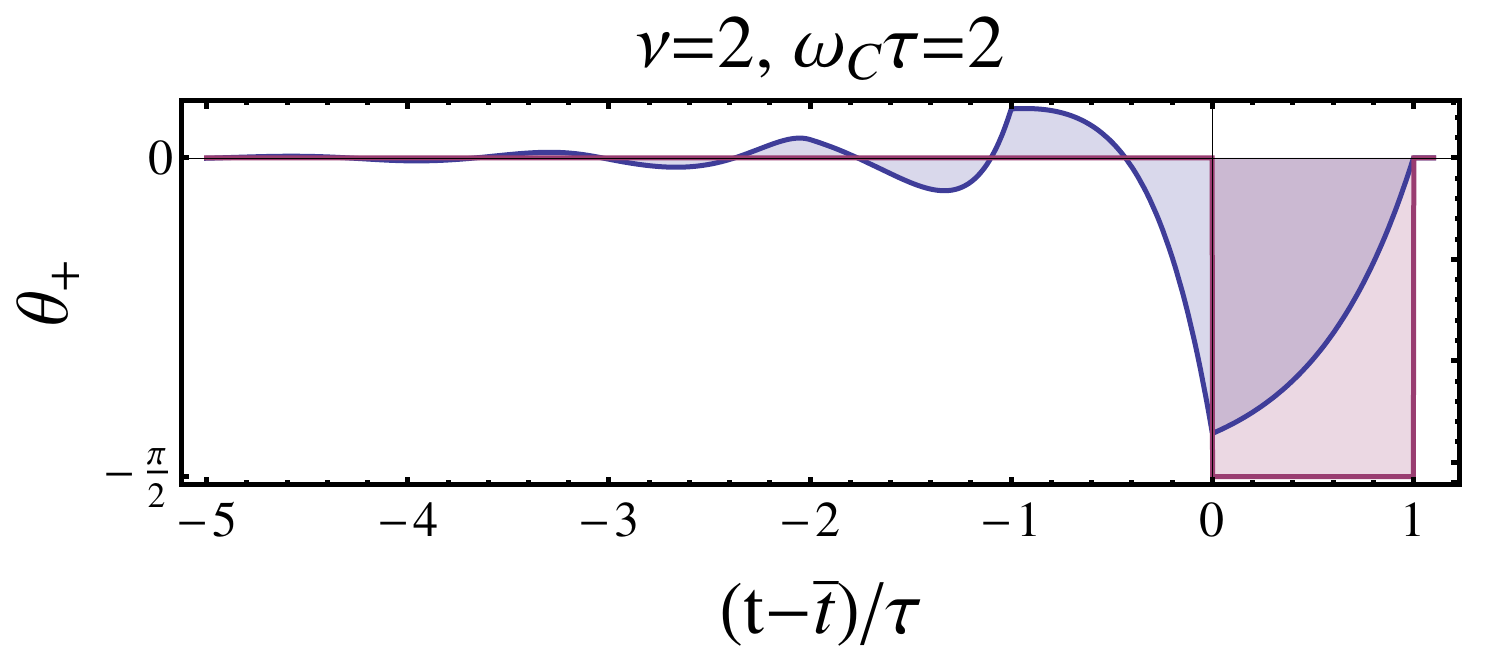}
\hspace{0.25cm}
\includegraphics[width=3.25in]{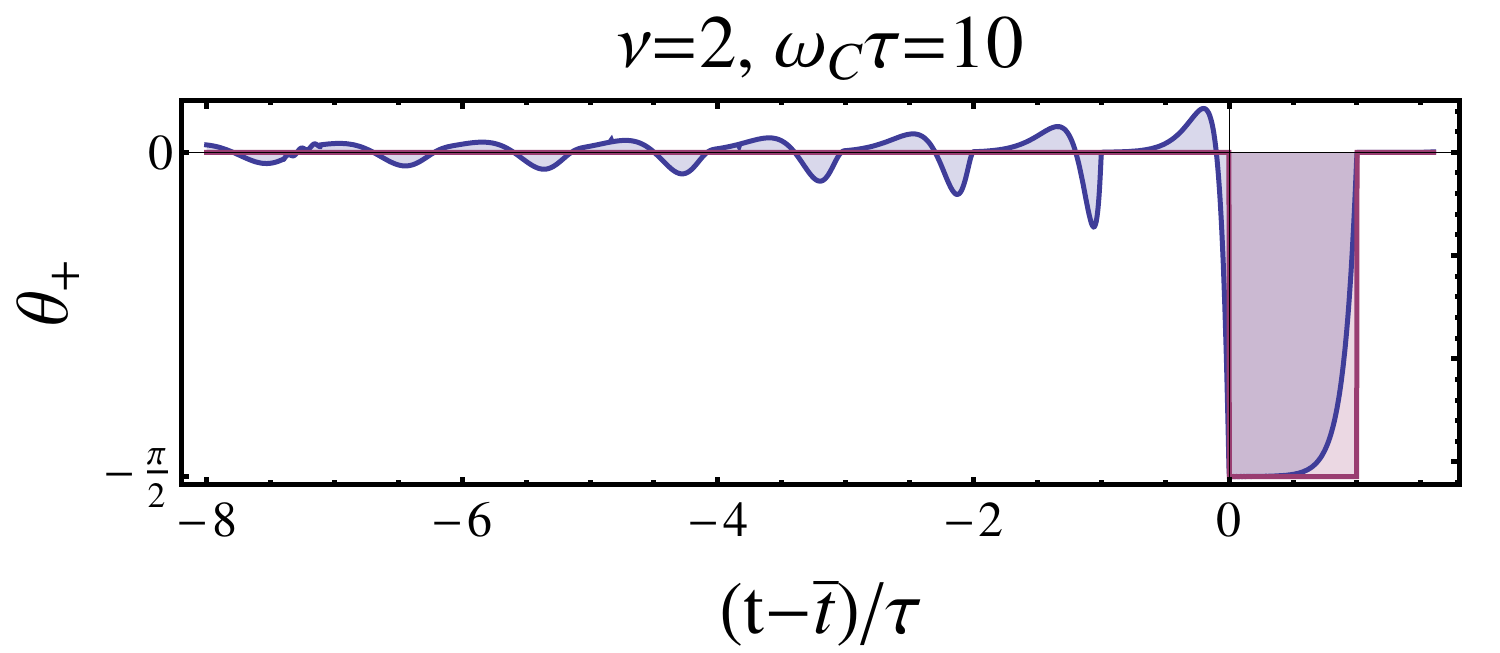}
}
\caption{Time-dependent ``counting'' phase for  $\nu=2$ and two strengths of
Coulomb interaction, $\omega_c\tau = 2$ and $\omega_c\tau=10$, where $\omega_c =
\nu E_c/\pi$ is the charge relaxation frequency. 
The phase approaches the ``window'' function, Eq.~(\ref{eq:window_function}), in
the strong-interaction limit $E_c \to \infty$.}
\label{fig:Phase_plus}
\end{figure*}

\subsection{The case of moderate strength of interaction}
\label{s2.3}

In this subsection we discuss the results for visibility in the case of not
too strong {\it e-e} interaction, $E_c\tau \sim 1$. We consider here only the
case $\nu=2$ for the following two reasons. First, the majority of
experimental data for MZIs has been collected for this filling factor.
Second, contrary to higher filling factors, at $\nu=2$ a finite (rather
than infinite) value of $E_c\tau$ changes the result qualitatively, since there
is no dephasing and no $T_1$ dependence at $E_c\tau \to\infty$.

The numerically calculated plots of visibility for
transparencies $T_1=0.5$ and $T_1=0.2$ are shown in Fig.~\ref{fig:Vis_Intermediate}.
It is seen that the finite charging energy $E_c$ gives rise to the decay of
visibility ${\cal V}(V)$ with
bias contrary to its behavior at $E_c \tau \to \infty$ discussed in the previous subsection~\ref{s2.2}. 
Note also that nodes (zero-visibility points) in ${\cal V}(V)$ are
generally present only in the case $T_1=1/2$.
At the transmission coefficient close (but not equal) to one half the nodes are
superseded by deep minima. Further, the period of oscillations increases
with decreasing $E_c$. However, the estimate $e(\Delta V) \sim 2\pi/\tau$ for
the scale of oscillations remains valid up to the moderate charging energy $E_c
\sim 1/\tau$. As an example, Fig.~\ref{fig:Vis_Intermediate} shows that at $E_c
\tau = \pi$ the period is larger than its strong-interaction limiting value 
$e(\Delta V) = 2\pi/\tau$ by a factor $\simeq 1.5$. 

The dephasing rate $1/\tau_\phi$ describing the exponential suppression
($\propto e^{-\tau/\tau_\phi}$) of the visibility with bias is found to be
\begin{equation}
\tau_\phi^{-1} = -\frac{eV}{2\pi\tau}\int_{-\infty}^{\tau+\bar t}\,{\rm Re}\Bigl[\ln
\left( 1 + R_1 (e^{4i\theta_+(t)}-1)\right)\Bigr]\,dt ,
\label{eq:Deph_rate_Charging}
\end{equation}
where $\bar t$ is the time when the electron enters the interferometer
and $\theta_+(t)$ is the time-dependent ``counting'' phase 
given by
\begin{equation}
 \label{theta-plus}
\theta_+(t) = -{1\over\nu} {\rm Im}\left[J^>(\bar t-t)-J^>(\tau-t+\bar
t)\right],
\end{equation}
with the function $J^>(t)$ defined below in Eq.~(\ref{eq:J-greater-t}). 
The phase $\theta_+(t)$ is shown in Fig.~\ref{fig:Phase_plus} for $\omega_c\tau
= 2$ and $\omega_c\tau=10$. In the limit $E_c\tau \gg 1$ the time dependence of
$\theta_+(t)$ approaches the ``window'' function
\begin{eqnarray}
\theta_+ (t)= \left\{
\begin{array}{ccl}
-\pi/\nu, &\quad & t\in [\bar t, \bar t + \tau] \\
0, &\quad & t \notin [\bar t, \bar t + \tau], 
\end{array}\right. \label{eq:window_function}\\ \nonumber
\end{eqnarray}
causing the dephasing rate to vanish at $\nu=1,2$. If one introduces the effective
charge $e^*(t)/e = \theta_+(t)/\pi$, then it can be physically interpreted as
the optimal charge fluctuation on the upper arm of the MZI which promotes scattering of the
transport (``trial'') electron from one arm of the interferometer into the
other.
\footnote{We note that in view of the specific chiral geometry of the MZI the
charge from the internal interacting region of the interferometer can always
freely leak into the source or drain, and thus the issue of Coulomb blockade
phenomenon is not relevant here.}
Loosely speaking, if such scattering event starts at a time instant $\bar t$,
then it finishes no later than $\bar t + \tau$
(cf. the upper bound of the time integral in Eq.~(\ref{eq:Deph_rate_Charging})).
It means that an electron entering the MZI at time $\bar t$ cannot be influenced by those
electrons which enter at times larger than $\bar t + \tau$, since by the
latter time the trial electron leaves the interior interacting region of the
system through the second QPC.
On the other hand, and perhaps somewhat counterintuitively, a typical arm-to-arm
electron scattering is generally preceded by a rearrangement of 
the charge $e^*(t)$ on the MZI at all times $t < \bar t$. We thus see that the single electron
transfer through the MZI in the presence of {\it e-e} interaction is a
collective many-body process involving many electrons.

Our results match well experimental observations in many designs of
Mach-Zehnder
interferometers at filling factor $\nu$=2, which happen to be rather universal. Namely,
at $T_1$ close to $1/2$ the experimentally observed dependence of the
visibility on voltage shows a number of ``lobes'' whose 
amplitude gets suppressed with the increase of bias. At the same time, the
voltage dependence of the AB phase is close to a piecewise constant
function with jumps of a magnitude $\pi$ at minima of
the visibility. Further, we estimate the period of oscillations. As follows from
Fig.~\ref{fig:Vis_Intermediate}, the characteristic energy
scale corresponding to the first minimum in the visibility is $e(\Delta V) \sim
2\pi/\tau = 2\pi v/L$.
An estimate for the drift velocity in our phenomenological model,
$v \sim {\nu e^2}/{\epsilon \pi \hbar}$,
can be obtained following Ref.~\onlinecite{Aleiner:1994}, where the excitation
spectrum of the compressible Hall liquid has been studied 
(see Sec. IV.C of our previous work, Ref.~\onlinecite{Schneider:2011}, for a
more detailed discussion).
Taking $\epsilon=12.5$ for the dielectric constant of the GaAs heterostructure, one obtains
$v \sim 1.1\cdot 10^5$~m/s. Note, that this estimate agrees well with an
effective velocity 
$v_{\rm eff}=6.5\cdot 10^4$~m/s found in Ref.~\onlinecite{Kovrizhin:2012} from the analysis
of data on energy relaxation in QH edge states at $\nu$=2. For a typical size of
the interferometer
$L \sim 10~\mu$m we then get $\Delta V \sim 40\mu$V, which is of the same order
as the experimentally observed energy scale of the visibility oscillations.

Having completed a presentation and discussion of  our key
results, we now turn
to the exposition of the method and of technical aspects of the derivation.

\section{Non-equilibrium functional bosonization for Mach-Zehnder
interferometer}
\label{s3}

In this Section we show how the method of the non-equilibrium functional
bosonization can be used
to solve the model of the MZI defined in Sec.~\ref{s2.1}. First, we present
the Keldysh action of the problem and derive the expression for the direct and
interference current (Sec.~\ref{s3.1}). Then we give details of the real-time
non-equilibrium instanton approach
(Sec.~\ref{s3.2}). Using the special structure of the Keldysh action, we show
that this method 
becomes exact in the case of the simplified model of the Coulomb interaction
(considered in the present paper) in which electrons interact only 
in the interior region of the MZI. Finally, we specify the form of the instanton
for the case of the constant interaction model.

\subsection{Keldysh action and current}
\label{s3.1}

The theoretical model of the MZI, which we consider throughout the paper, is defined by 
Eqs.~(\ref{eq:S_0}) and~(\ref{eq:S_int}). To make our discussion more
general, we will first assume 
the arbitrary interaction potential $U_\pm(x-x')$ between two electrons in the same edge, which however 
is non-zero only if $x,x'\in [x_1^\pm, x_2^\pm]$. Because of the non-equilibrium character of the problem,
we proceed within the Keldysh-type framework~\cite{Kamenev:2009,Kamenev:2011}.  We decouple
the interaction term ${\cal S}_{\rm int}$ using the Hubbard-Stratonovich transformation with
fields $\varphi_\varrho(x,t)$, where the index $\varrho=\pm$ labels two arms of
the interferometer.  Following the logic of the
Keldysh formalism, we then double the number of Grassmann fields,
$\psi_\varrho=(\psi_\varrho^f, \psi_\varrho^b)$,
as well as of the bosonic fields $\varphi_\varrho =
(\varphi_\varrho^f,\varphi_\varrho^b)$, 
where indices $f$ and $b$ denote the fields residing 
on the forward and backward branches of the Keldysh contour ${\cal C}$,
respectively. 
These steps lead us to the MZI action in the form
\begin{widetext}
\begin{eqnarray}
{\cal S} = \sum_{\varrho=\pm} \int_{\cal C}\,dt dx\, \bar\psi_\varrho (i\partial_t + iv\partial_x - \varphi_\varrho)\psi_\varrho
+ \frac{1}{2} \int_{\cal C} dt dx dx' \,\varphi_\varrho(x) {U_\varrho}^{-1}(x-x')\varphi_\varrho(x'). \nonumber
\end{eqnarray}
\end{widetext}
Integration along ${\cal C}$ is to be understood as $\int_{\cal C} dt' A(t')=\int dt'\,(A^f(t')-A^b(t'))$.  

In terms of fermion fields $\psi_\varrho$ the action ${\cal S}$ is quadratic, thus they can be integrated out.
In this way we obtain the Keldysh action ${\cal A}[\varphi]$ of the MZI which depends on the electrostatic
potentials $\varphi_\pm(x,t)$ on two arms. The outlined method is known as the functional bosonization.
The integration over the Grassmann fields $\psi_\varrho$ should to be performed
with taking into account the relation~(\ref{eq:S_matrix_relation}) at QPCs;
this relation has to be satisfied independently on each
branch of the Keldysh contour.  The action ${\cal A}[\phi]$ in the case of
a generic non-equilibrium setup formed by 1D electronic channels coupled by a
number of local scatters and by electron-electron interaction (``quantum wire
network'') has been found in our previous work~\cite{NgoDinh:2012}. In
particular, 
the Keldysh action ${\cal A}[\phi]$ describing the MZI can be expressed in terms of the time-dependent
single-particle scattering matrix of the interferometer in the given configuration of the fields
$\varphi_\varrho^{f/b}$, which we denote as $S_{f/b}=S[\varphi^{f/b}](t,t')$. This $S$-matrix describes
electron scattering at both QPCs and the propagation of electrons along the arms of the MZI. 
The bosonized action ${\cal A}[\phi]$ has the form 
${\cal A} = {\cal A}_{\rm int} + {\cal A}_{\rm ferm} + \delta{\cal A}$.
Explicitly,
\begin{widetext}
\begin{equation}
{\cal A}_{\rm int}  = 2\sum_{\varrho} \int d\xi d\xi'\, \varphi^c_\varrho(\xi)
\left( U_\varrho^{-1}(x,x')\delta(t-t')-\Pi^A_{\varrho}(\xi,\xi') \right)\varphi^q_\varrho(\xi'),
\label{eqn:ActInt}
\end{equation}
where we have denoted $\xi=(x,t)$, and
\begin{equation}
 {\cal A}_{\rm ferm} + \delta{\cal A} = -i\,{\rm Tr}\ln\left[\xUnit-\hat f +S_{b}^\dagger e^{i\chi} S_{f} \hat f\right]
-2\sum_{\varrho}{\rm Tr} \,\vartheta^q_{0\varrho} f_0. \label{eqn:ActFerm}
\end{equation}
\end{widetext}
Let us explain notations and comment on different terms in the action,
Eqs.~(\ref{eqn:ActInt}) and (\ref{eqn:ActFerm}).  
In the above expression for ${\cal A}_{\rm int}$ we have used the Keldysh basis $\varphi_\varrho=(\varphi_\varrho^c,\varphi_\varrho^q)$,
with ``classical'' and ``quantum'' components being defined as
$\varphi_\varrho^{c,q}=(\varphi_\varrho^f\pm\varphi_\varrho^b)/2$.
The advanced component $\Pi^A$ of the 1D polarization operator, when written in
the frequency-coordinate representation, has the explicit form
\begin{eqnarray}
\Pi^A(\omega,x,x') = \frac{\nu}{2\pi v}\Bigl[ &-&\delta(x-x') \label{eq:PI_wxx}\\
&+& \frac{i\omega}{v}\,e^{-i\omega(x-x')/v}
\times\theta(x-x')\Bigr]. \nonumber
\end{eqnarray}
The combination of the polarization operator and the bare interaction potential
entering Eq.~(\ref{eqn:ActInt}) determines the RPA screened interaction
potential,
\begin{equation}
V^{-1,A}_\varrho(\xi,\xi') = U_\varrho^{-1}(x,x')\delta(t-t')-\Pi^A_{\varrho}(\xi,\xi').
\end{equation}

The action ${\cal A}_{\rm ferm}$ has the form of the fermion determinant and
bears a close
connection with the problem of electron full counting
statistics~\cite{Levitov:1996}.
We have introduced the distribution functions of the source reservoirs,
$\hat f={\rm diag}(\hat f_+, \hat f_-)$, where $\hat f_\pm$ are diagonal matrices in the
channel space. Without QPC0, $T_0=0$ their components are $\hat f_\pm^n(t,t') = e^{-ieV_\pm^n(t-t')}f_0(t-t')$.
Here $V_\pm^n$ is the voltage applied to the $n$-th channel in the upper/lower arm and
\begin{equation}
	f_0(t-t')=\frac i{2\pi} \frac 1{t-t'+i 0}.	
\end{equation}
is the time representation of the equilibrium Fermi distribution function.
For instance, for the MZI presented in Fig.~\ref{fig:MZI_layout} in the case of vanishing transparency
$T_0=0$, the only non-zero voltage is $V_+^1=V$. At $T_0>0$, when QPC0 is used to dilute
the impinging current, the function $f_+^1$ is the double-step distribution
in the energy domain and is given by
\begin{equation}
f_+^1(t-t') = \left[ R_0\,e^{-ieV(t-t')} + T_0\right]f_0(t-t')
\end{equation}
in the time representation.
We have also introduced the auxiliary ``counting fields''
$\hat \chi = {\rm diag}(\chi_+,\chi_-)$ in the drains which enable us to find the number of electrons
transferred through the MZI. The determinant in Eq.~(\ref{eqn:ActFerm}) is evaluated with respect to
time, channel and arm indices. 

Next, we  specify the $S$-matrices $S_{f/b}$ of the MZI which enter
Eq.~(\ref{eqn:ActInt}) and encode all information about electron
scattering. We introduce the phases $\vartheta_\varrho^{f/b}(t)$ accumulated
between the QPCs $1$ and $2$ due 
to interaction,
\begin{equation}
	\vartheta_\varrho^{f/b}(t)=-v^{-1} \int dx'\, \varphi^{f/b}_\varrho(x',t+(x'-x^1_\varrho)/v).
\end{equation}
Let now $x_\varrho^S$ and $x_\varrho^D$ be the coordinates of the source and drain reservoirs. We also define 
the ``time-delay'' operator $\Delta^{lk} = {\rm
diag}(\Delta^{lk}_+,\Delta^{lk}_-)$, where
\begin{equation}
\Delta^{lk}_\varrho(t,t')=\delta(t-t'-(x^l_\varrho-x^k_\varrho)/v),
\label{eq:Delta}
\end{equation}
with indices $k,l \in \{S,1,2,D\}$. It coincides with a transfer matrix from $x_\varrho^k$ to
$x_\varrho^l$ along the arm $\varrho$ of the MZI in the non-interacting limit. Assuming the 
absence of interaction outside of the interferometer, we obtain the total
$S$-matrices
\begin{equation}
S_{f/b}=\Delta^{D2} \hat S_2 \Delta^{21} e^{i\hat\Phi} e^{i\hat\vartheta^{f/b}} \hat S_1\Delta^{1S}.
\end{equation}
Here $\hat S_1$ and $\hat S_2$ are the local scattering matrices of the 1st and
2nd QPC. Further, $\hat\Phi$ is the diagonal flux matrix $\hat\Phi = {\rm
diag}(\Phi/2, -\Phi/2)\oplus \xUnit_{2\nu-2}$, where $\Phi$ denotes the 
Aharonov-Bohm phase and the sign $\pm$ distinguishes between the upper and
lower arms. The direct sum ($\oplus$) refers to the channel space and $\xUnit_n
$ is the $n\times n$-unity matrix. 
 The matrix $\hat\vartheta^{f/b} = {\rm diag}(\hat \vartheta^{f/b}_+,
\hat \vartheta^{f/b}_-)$ has an analogous structure. 
For the MZI scheme shown in Fig.~\ref{fig:MZI_scheme}, only the outer channels
are mixed by scattering. In this case, $\hat S_j = \hat s^j \oplus {\xUnit}_{2\nu-2}$, with $\hat s^j$ given by Eq.~(\ref{eq:s_matrix}).

Finally, the counter-term $\delta{\cal A}$ in the action~(\ref{eqn:ActFerm}) is
included to cancel the equilibrium
Fermi-sea contribution which does not affect the non-equilibrium electron
transport. The ``quantum'' phase
$\vartheta_{0\varrho}^q$ entering $\delta{\cal A}$ is defined as  
$\vartheta_{0\varrho}^q(t)\equiv \vartheta_{\varrho}^f(t+(x_\varrho^D -
x_\varrho^1)/v)-\vartheta_{\varrho}^b(t+(x_\varrho^D - x_\varrho^1)/v)$; 
the trace (${\rm Tr}$) is taken over channel and arm indices and also includes
integration over time.

The bosonized action ${\cal A}[\varphi,\chi]$ enables us to find the generating function of the 
interferometer's FCS  as a functional integral over $\varphi$,
\begin{equation}
{\cal Z}(\vec\chi) = \int {\cal D}\varphi^{f/b}_{\pm}(x,t)\,
\exp\left\{ i{\cal A}(\,\varphi,\vec\chi) \right\}.
\label{eq:Z_PathI}
\end{equation}
Then the number of electrons transferred to, say, the lower drain during a long observation time
$t_0 \gg {\rm max}\{\hbar/eV, \hbar v/L\}$ is obtained as
\begin{equation}
N_- = -i\partial_{\chi_-} \ln \mathcal Z\Big\rvert_{\chi=0} = 
\left\langle \partial_{\chi_-}{\cal A}_{\rm ferm}\Big\rvert_{\chi=0}\right\rangle_\varphi.
\label{eq:Nmean}
\end{equation}
The quantum mechanical average $\langle \dots \rangle_\varphi$ here is
understood as the path integral over $\varphi$ with the weight $e^{i\cal A}$,
see Eq.~(\ref{eq:Z_PathI}), 
but with ``counting fields'' $\chi$ put to zero. 
Since the Coulomb interaction is assumed to be absent outside the interferometer cell,
${\cal A}_{\rm ferm}$ simplifies considerably (in the rest we will not explicitly state $\chi=0$ any longer):
\begin{eqnarray}
{\cal A}_{\rm ferm} = -i\ln \Det\mathcal D, \quad
 \mathcal D \equiv \xUnit-\hat f+\hat S_1^\dagger e^{2i\hat \vartheta^q} \hat S_1 \hat f.
 \label{eq:A_zero_chi}
\end{eqnarray}
The same action can be represented in the equivalent form as ${\cal A}_{\rm ferm} = -i\ln \Det\tilde{\mathcal D}$,
with
\begin{eqnarray}
\tilde{\cal D} \equiv \hat S_1{\cal D} \hat S_1^\dagger = \xUnit-\tilde f +e^{2i\hat \vartheta^q} \tilde f,
\end{eqnarray}
where $\tilde f$ plays a role of the non-equilibrium density matrix of the interferometer.
If the voltage is applied to the outer channels only, then 
$\tilde f = \tilde f^1 \oplus (\xUnit_{2\nu-2}\cdot f_0)$,
with
\begin{equation}
\tilde f^1 \equiv \hat s_1 \hat f^1 \hat s_1^\dagger = 
			\begin{pmatrix}
				R_1 f_+^1 + T_1 f_-^1& i\sqrt{R_1T_1} (f_+^1-f_-^1)\\
				-i\sqrt{R_1T_1}(f_+^1-f_-^1) & T_1 f_+^1+R_1 f_-^1
			\end{pmatrix}.
\end{equation}
Note, that ${\cal A}_{\rm ferm}$ at $\chi=0$ depends on the scattering matrix $s_1$ only.
It also depends solely on the ``quantum'' field $\varphi^q$, but not on the
``classical'' one. These special features stem from the chiral nature of the
MZI. The independence of ${\cal A}_{\rm ferm}$ on the classical component of
the field will play a crucial role in the sequel, as it will allow us
to find an exact solution of the problem. 

Using the definition~(\ref{eq:Nmean}) and the full $\chi$-dependent fermion
action~(\ref{eqn:ActFerm}), one obtains the following intermediate expression
for $N_-$:
\begin{eqnarray}
N_-= \Bigl\langle &\Tr& \mathcal D^{-1} s_1^\dagger e^{-i\hat\vartheta^b}e^{-i\hat\Phi} {\Delta^{21}}^\dagger  
s_2^\dagger|-\rangle\nonumber\\
&\times& \langle -|s_2 \Delta^{21} e^{i\hat\vartheta^f} e^{i\hat\Phi} s_1 \hat f \Bigr\rangle_\varphi. 
\end{eqnarray}
To derive this result we have taken into account that only outer channels with a
non-equilibrium distribution matrix $\hat f^1$ may contribute to the transport of electrons and
have introduced  the basis $|\pm\rangle$ in this linear subspace.
Taking the trace over the channels and using 
the explicit form of $\Delta^{12}$, given by Eq.~(\ref{eq:Delta}),  
we find $N_-=\sum_{\mu\kappa} N_{\mu\kappa}$, where
\begin{widetext}
\begin{equation}
N_{\mu\kappa} = \bra-s_2\ket \mu \bra\kappa s_2^{\dagger}\ket- \left\langle \int d \bar t\ e^{i\vartheta_\mu^f(\bar t)-i\vartheta_\kappa^b(\bar t+\tau_\mu-\tau_\kappa)}\ e^{i\Phi_\mu-i\Phi_\kappa}\ \bra\mu s_1 
\left(\hat f \circ \mathcal D^{-1}\right)(\bar t,\bar t+\tau_\mu-\tau_\kappa )\, s_1^\dagger \ket\kappa\right\rangle_\varphi.
\label{eqn:NMuKappa}
\end{equation}
\end{widetext}
In this expression we have introduced $\Phi_\pm = \pm \Phi/2$, which are just the AB phases accumulated at each arm of the MZI, and defined $\tau_\pm = (x_\pm^2 - x_\pm^1)/v$ --- the flight time of electrons along upper/lower arm.  
The sign ``$\circ$'' denotes the convolution in the time and channel space.
Clearly, the diagonal ($N_{\mu\mu}$) and off-diagonal ($N_{\mu,-\mu}$) elements give, respectively, the
direct and interference contributions to the total current.

\subsection{Exact solution: from many-body to single-particle problem}
\label{s3.2}

In general, the functional integral for interacting electrons in a quantum wire
network cannot be evaluated exactly. In
Refs.~\onlinecite{NgoDinh:2010,Schneider:2011,NgoDinh:2012} an instanton
approach has been developed which yields a controllable approximation to the
problem for the case of weak tunneling between the channels. It turns out that
for the problem considered here this method becomes exact (for any tunneling
strength). Specifically, we show below that the functional integral can
be exactly evaluated in a fashion similar to the exact solution of
the problem of non-equilibrium Luttinger liquids in
Refs.~\onlinecite{Gutman:2010a,Gutman:2010,Gutman:2010b}. In fact, we will see
later that there is a deep connection between the two problems. 

We have shown above that the number of electrons transferred through the MZI is given by Eq.~(\ref{eqn:NMuKappa}).
This expression implies the path integral over all realization of the fields $\varphi^{f/b}_\pm(x,t)$.
As we reveal below this functional integral can be performed exactly. 
What crucially simplifies the calculation of the quantum mechanical average
is the fact that $\mathcal D$ and hence the $\mu\kappa$-matrix  elements in
Eq.~(\ref{eqn:NMuKappa}) together with ${\cal A}_{\rm ferm}$ do not contain
$\varphi^c$ or, 
equivalently, $\vartheta^c$. The classical fields enter the RPA action
${\cal A}_{\rm int}$ and the phases
$\vartheta^{f/b}_\varrho=\vartheta^c_\varrho\pm \vartheta^q_\varrho$, and appear there only linearly. 
Therefore $\varphi^c$ can be exactly integrated over. To this end let us
introduce sources $J$ so that
\begin{eqnarray} 
{\cal A}_{J;\mu\kappa}&\equiv &\vartheta_\mu^c(\bar t)-\vartheta^c_\kappa(\bar t+\tau_\mu-\tau_\kappa)\nonumber\\
&=&\sum_\varrho \int\!\! d \xi\,J^q_{\varrho;\mu\kappa}(\bar t;\xi)
\varphi^c_\varrho(\xi).
\label{eqn:ActJ}
\end{eqnarray}
We see that for $\mu=\kappa$ the source vanishes. On the contrary, at
$\mu=-\kappa$ the
explicit expression for $J$ reads
\begin{eqnarray}
J^q_{\mu;\mu\kappa}(\bar t,\xi)&=& -v^{-1} \delta(\bar t+(x-x_\mu^1)/v-t), \\ 
J^q_{\kappa;\mu\kappa}(\bar t,\xi) &=& v^{-1} \delta(\bar t+\tau_\mu-\tau_\kappa+(x-x_\mu^1)/v-t). \nonumber
\end{eqnarray}
The physical meaning of the source terms in the action is rather obvious. They
describe an electron transfer
between two Hall edges of the MZI, thereby creating a hole in the arm $\kappa$
and adding an extra electron into the arm~$\mu$. 

Decomposing $\vartheta^{f/b}$ in Eq.~(\ref{eqn:NMuKappa}) 
into the ``classical'' and ``quantum'' parts, we rewrite the formula for the
particle numbers $N_{\mu\kappa}$ in the form
\begin{widetext}
\begin{equation}
N_{\mu\kappa}= \bra-s_2\ket \mu \bra\kappa s_2^{\dagger}\ket- \int {\cal D} \varphi^c{\cal D}\varphi^q\, 
\int  d \bar t\ e^{i{\cal A}_{\rm int}+i{\cal A}_{J;\mu\kappa}} \times
\left\{ e^{i{\cal A}_{\rm ferm} + i\delta{\cal A}} 
\mathcal A_{\mu\kappa}(\bar t,\bar t+\tau_\mu-\tau_\kappa)\right\}, 
\end{equation}
where the prefactor $\mathcal A_{\mu\kappa}$ is defined as
\begin{equation}
\mathcal A_{\mu\kappa}(t_1,t_2) = e^{i\vartheta_\mu^q(t_1)+i\vartheta_\kappa^q(t_2)}\ e^{i\Phi_\mu-i\Phi_\kappa}\ \bra\mu s_1 \left(\hat f \circ \mathcal D^{-1}\right)(t_1,t_2) \,s_1^{\dagger} \ket\kappa. \label{eqn:preExp}
\end{equation}
As has been emphasized previously, the fermion action ${\cal A}_{\rm ferm} +
\delta {\cal A}$ and the matrix ${\cal D}$
are functionals of $\varphi^q$ only, see Eq.~(\ref{eq:A_zero_chi}). Hence, one can first perform
the integration over the ``classical'' field $\varphi^c$. Taking into account
that ${\cal A}_{\rm int}+{\cal A}_{J;\mu\kappa}$ is linear in $\varphi^c$,
we obtain
\begin{equation}
\int {\cal D} \varphi^c\, \exp\left[i{\cal A}_{\rm int}+i{\cal A}_{J;\mu\kappa}\right]=\int {\cal D} 
\varphi^c\, \exp\left[2i\varphi^c (V^A)^{-1}\varphi^q+i\varphi^c J^q\right] \propto \delta\left(\varphi^q-\varphi^q_\ast\right),
\end{equation}
where the $\delta$-function fixes the quantum component $\varphi^q$ to be equal
to the saddle-point trajectory 
\begin{equation}
\varphi^q_{\ast\varrho}(\xi)=-\frac12 \sum_\sigma \int 
d\xi'\, V^A_{\varrho\sigma}(\xi,\xi')  J^q_{\sigma;\mu\kappa}(\bar t,\xi'). \label{eqn:instPot}
\end{equation} 
The $\delta$-function constraint renders trivial the subsequent integration over
$\varphi^q$.  
Taking quantum-mechanical average $\langle\ldots\rangle_\varphi$ is therefore
reduced to the evaluation 
of the integrand $e^{i{\cal A}_{\rm ferm} + i\delta{\cal A}} \mathcal A_{\mu\kappa}$ on the 
optimal trajectory $\varphi^q=\varphi^q_\ast$. The particle numbers are then simplified to
\begin{equation}  
\label{eqn:CurrentExpPreExp}
N_{\mu\kappa}= \bra-s_2\ket \mu \bra\kappa s_2^\dagger\ket-  \int d \bar t\ 
e^{i{\cal A}_{\rm ferm} + i\delta{\cal A}}\ \mathcal A_{\mu\kappa}(\bar t,\bar t+\tau_\mu-\tau_\kappa)
\Big\rvert_{\vartheta^q=\vartheta^q_\ast},
\end{equation}
\end{widetext}
with the ``quantum'' saddle-point phase (or ``instanton'')
\begin{equation} 
\vartheta^q_{\ast\varrho}(t) = -v^{-1} \int d x'\, \varphi_{\ast\varrho}^q(x',t+(x'-x^1_\varrho)/v_\varrho).
\label{eq:Inst_phase}
\end{equation}
In what follows we will frequently refer to $\vartheta^q_*$ as to ``counting
phase'' in view of an analogy between the action ${\cal A}_{\rm ferm}$ and the
theory of the FCS.

To reiterate the logic, we have reduced the path integration over $\varphi$ to the evaluation
of the integrand for the numbers $N_{\mu,-\mu}$ on the ``quantum'' saddle-point 
trajectory $\varphi_*^q$. This is the main result of the present subsection.  
The optimal ``quantum'' electrostatic field is related via the RPA interaction 
potential to the source terms $J$ which describe the electron transfer between two edges of the MZI,
see Eq.~(\ref{eqn:instPot}).
The bare interaction potential enters the result through the RPA kernel $V^A(\xi,\xi')$,
thus the outlined method is very general, provided the {\it e-e} interaction can
be disregarded outside of the MZI cell. The result is expressed in terms of
determinants and resolvents that are of single-particle complexity; thus, we have
achieved a dramatic simplification as compared to the original many-body
problem. Needless to say, the obtained apparently single-particle quantities
carry all the physical information about the many-body physics of the problem,
including, in particular, non-equilibrium orthogonality-catastrophe exponents
and non-equilibrium dephasing.  

Let us now discuss the direct (incoherent) contribution to the current which arises from the
diagonal numbers ($N_{++}$ and $N_{--}$). Within our model, they are not affected by {\it e-e} interaction.
Indeed, in this case the sources vanish, $J^q_{\sigma;\mu\mu} = 0$. Hence the instanton trajectory is trivial,
$\vartheta^q_*~=~0$. One thus get ${\cal D}={\mathds 1}$, ${\cal A}_{\rm ferm} + \delta{\cal A} =0$ and
$\mathcal A_{\mu\mu}=R_1 f_\mu^1+T_1 f^1_{-\mu}$, that yields
\begin{align}
N_{++}= T_2 \Tr\left[R_1f_++T_1 f_-\right],\nonumber \\
N_{--}= R_2 \Tr\left[R_1 f_-+T_1 f_+\right].
\end{align}
Taking the ${\rm Tr}$ in the time space, we have
\begin{equation}
N_{++} + N_{--} = \frac{eV\, t_0}{2\pi\hbar}\,\left( T_2 R_1 + R_2 T_1\right).
\end{equation} 
Thus, the direct current is linear in bias; it is the same as in the
non-interacting limit.

\subsection{Constant interaction model}
\label{s3.3}

\begin{figure}[t]
\includegraphics[scale=0.38]{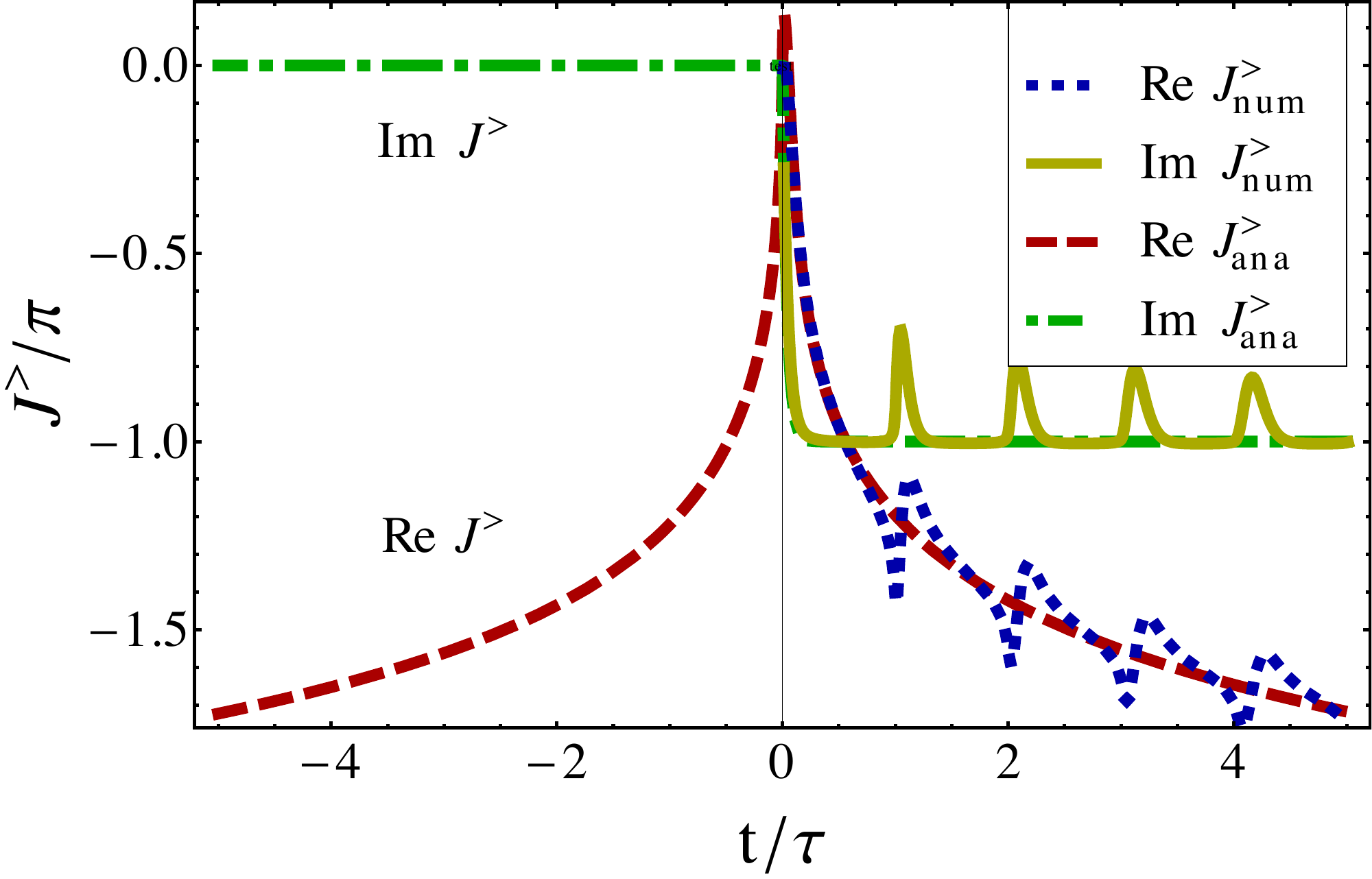}
\caption{Correlation function $J^>$: comparison of analytic approximation $J^>_{\rm ana}$ 
given by Eq.~(\ref{eq:J_log}) and numerical results $J^>_{\rm num}$ for $\omega_c\tau=25$.}
\label{fig:JNumAna}
\end{figure}

In this subsection we specify the ``counting phase'' for the constant
interaction model with
the charging energy $E_c$, as it is defined by Eq.~(\ref{eq:S_int}). We assume that both arms of the MZI
have the same length $L$, thus $\tau = \tau_+=\tau_-$ and $x_+^{1,2}=x_-^{1,2}$. Then the bare interaction
potential $U_\pm(x,x') = E_c$ if both $x,x'\in[x^1,x^2]$ and $U_\pm(x,x')=0$ otherwise.
This form leads to the RPA potential $V^A(\omega)$ which is non-zero
only if $x,x'\in[x^1,x^2]$ and constant inside this region,
\begin{equation}
		V^A_\pm(\omega) = E_c\left(1-E_c\int_{x^1}^{x^2} dx dx'\, \Pi^A(\omega;x,x') \right)^{-1}.
\end{equation}
Using Eq.~(\ref{eq:PI_wxx}) one has
\begin{equation}
\int_{x^1}^{x^2} dx dx'\,\Pi^A(\omega;x,x') = \frac {i\nu}{2\pi}\frac{1-e^{-i \omega\tau}}\omega.
\end{equation}
Defining the charge relaxation frequency as $\omega_c = \nu E_c/(2\pi)$ we obtain
\begin{equation}
	V^A(\omega) = \frac {2\pi\omega_c}\nu \frac{\omega}{\omega-i\omega_c(1-e^{-i\omega\tau})}.
\end{equation}
We now use this RPA result to find the instanton potential $\varphi_*^q$ and the phase $\vartheta_*^q$.
By virtue of Eq.~(\ref{eqn:instPot}) we have
\begin{equation}
\varphi^q_{*\eta}(\xi) = -\xUnit_{[x^1,x^2]}(x) \frac{\eta \kappa }{2v} \int_{x^1}^{x^2} \!\!dx'\,
V^A(t-\bar t-(x'-x^1)/v),
\end{equation}
where $\xUnit_{[x^1,x^2]}(x)$ is the projector on the interval where {\it e-e} interaction 
and thus the potential are present. Taking into account the relation~(\ref{eq:Inst_phase}) between 
the phase and the potential, one finds 
\begin{align}
\vartheta^q_{*\eta}(t) &= \frac{\eta\kappa}{v^2} \int_{x^1}^{x^2}\!\! dx' dx''\int\frac{d\omega}{2\pi}\, 
e^{-i\omega(t-\bar t+(x'-x'')/v)} V^A(\omega)\nonumber\\
				&= \frac{\kappa\eta}\nu {\rm Im}\left[J^>(\bar t-t)-J^>(\tau-t+\bar t)\right],	
\end{align}
where we have introduced the phase-phase correlation function
\begin{equation}
\label{eq:J-greater-t}
J^>(t) =\int_0^{+\infty}\!\!d\omega \frac{i\omega_c (1-e^{i\omega\tau})}
{\omega\bigl(\omega+i\omega_c (1-e^{i\omega \tau})\bigr)} \left(e^{-i\omega
t}-1\right).
\end{equation}
The function $J^>(t)$ has appeared and was analyzed in our previous work
(see Sec.~4.3.4 of Ref.~\onlinecite{NgoDinh:2012}) in the context of theory of
Fabry-Perot QH interferometer.  In the limit of strong coupling
$\omega_c \tau \gg 1$ and long time $\omega_c t \gg 1$ it can be well approximated by the logarithmic asymptotic
\begin{equation}
J^>(t) \simeq -\gamma - \ln\left[-\frac{t + i a}a\right],\quad a\sim \omega_c^{-1},
\label{eq:J_log}
\end{equation}
with $a$ being the short-time cut-off, see Fig.~\ref{fig:JNumAna}.
Therefore, except for times $t$ in a close vicinity of either $\bar t$ or $\bar
t + \tau$, the ``counting phase'' simplifies to 
\begin{equation}
\vartheta_{*\eta}^q(t) = \bar \vartheta_\eta w_{\bar t,\tau}(t), \quad \bar \vartheta_\eta=\kappa\eta \pi/\nu
\end{equation}
with the $\kappa$-dependent constant $\bar \vartheta_\eta$ and the unit
``window'' function
\begin{equation}
w_{\bar t,\tau}(t)=\theta(t-\bar t)-\theta(\bar t+\tau-t).
\label{eq:w_func}
\end{equation}
In the case of a moderate charging energy, $\omega_c\tau \sim 1$, one has to
resort to a numerical evaluation of
the (imaginary part of) correlation function $J^>(t)$. We have $\vartheta_{*\eta}^q(t)=-\kappa\eta\,\theta_+(t)$,
where the function $\theta_+(t)$ is independent of $\kappa$ and $\eta$. 
Typical plots of $\theta_+(t)$ (at $\nu=2$) are shown in
Fig.~\ref{fig:Phase_plus} of Sec.~\ref{s2.3}. For brevity we will omit the $\ast$ index when denoting the counting phase $\vartheta_{\eta}^q(t)=\vartheta_{*\eta}^q(t)$ in the following.

In passing we note, that the time-dependent counting phase $\vartheta_+(t)$ in our theory is
the analog of the kernel $Q(x)$, introduced in Ref.~\onlinecite{Kovrizhin:2010} (cf. Fig.~7 in this work).
Similar to the phase $\vartheta_+(t)$, this kernel depends on the nature of interaction potential
and is used to describe the phase, which an electron, passing the MZI, accumulates due to
interaction with other electrons.
 
\section{Strong interaction: analytical solution}
\label{s4}

As shown in Sec.~\ref{s3}, the considered model of a MZI with
inside-only interaction can be exactly solved by the non-equilibrium functional
bosonization method. The result is expressed in terms of single-particle
objects: determinants and resolvents of Fredholm operators. While it is not
too difficult to evaluate such quantities numerically, it would be
highly advantageous to have a fully analytical solution of the problem. Such a
solution will be obtained in the present section for the regime of strong
interaction, $E_c\tau \gg 1$. This solution will allow us to understand much
better the physics of the problem, including the formation of the visibility
oscillations (taking a form of ``lobes'' in certain situations as discussed
above) and their decay with voltage. Further, while determining the exact
solution, we will establish a deep connection of the present problem with that
of non-equilibrium Luttinger liquid and, more generally, with a broader class of
non-equilibrium many-body problems. 

As we have demonstrated in Sec.~\ref{s3}, the ``counting phase'' 
in the strong-interaction regime, $E_c\tau \gg 1$, is reduced
to the ``window'' function on the interval $[\bar t, \bar t + \tau]$. 
We will show in Sec.~\ref{s4.1} that under this condition the interference current can be
represented in terms of singular Fredholm determinants generalizing
Toeplitz determinants with Fisher-Hartwig singularities.
Using asymptotic properties of such determinants (Sec.~\ref{s4.2}), we
further derive the high-voltage form of the AB contribution to the current
(Sec.~\ref{s4.3}). The result is Eq.~(\ref{eq:I_0}) which has been already
discussed in detail in Sec.~\ref{s2.2}. 

\subsection{Reduction to a single-channel problem}
\label{s4.1}

In Sec.~\ref{s3.2} we have expressed the interference current in terms of
the operator 
${\cal D}$ and the non-equilibrium density matrix $\tilde f$ which, in addition to being 
the matrices in the time space, have also a nontrivial channel structure: for
given times $t_1, t_2$ they are 
matrices from $\mathbb C^{2\times 2}$. 
(Since all the relevant non-equilibrium physics arising due to scattering at QPCs
concerns only the outer channels, we focus on these two channels. In what follows 
we consider projections of all operators, such as $\hat\vartheta^q$, $\hat f$, 
onto the two outer channels; thus they retain the smaller $2\times 2$-channel structure.) The double
index structure (times and channels) very seriously complicates 
the computation of the determinant and finding the inverse of ${\cal D}$.
In this section, using the Riemann-Hilbert technique, we reduce the  matrix
determinant and resolvent to a product of certain single-channel
determinants. We show that the corresponding operators belong to the class of
singular Fredholm operators that may be considered as a generalization
of Toeplitz matrices with Fisher-Hartwig singularities. The reduction to a
single channel problem will allow us to calculate analytically the current in
MZI, see Sec.~\ref{s4.2} and \ref{s4.3} below. 

\subsubsection{Heuristic argument: Relation to full counting statistics}
\label{s4.1.1}

Before presenting a rigorous derivation of the
reduction formula, we will put forward a more heuristic argument in its favor.
This argument is based on a connection between the fermion action ${\cal A}_{\rm
ferm}$ and the theory of the full counting statistics (FCS). Consider the
cumulant generating function (CGF) for the statistics of 
$N_\pm$, the numbers of non-interacting electrons which tunnel through the QPC1 during the time interval
$[\bar t, \bar t + \tau]$ into the upper/lower arm, resp.
\begin{equation}
\chi_\tau(\lambda_+,\lambda_-)=\langle e^{i \lambda_+ \hat N_+ + i\lambda_- \hat N_-}\rangle
\label{eq:chi_tau}
\end{equation}
with corresponding ``counting fields'' $\lambda_\pm$. The brackets here mean
a quantum-statistical average.
It is known~\cite{Levitov:1996} that this CGF can be represented as a
functional determinant,
\begin{align}
\chi_\tau(\lambda_+,\lambda_-) = 
{\rm Det} \left[ {\mathds 1} - \hat f + s_1^\dagger e^{i\hat\lambda} s_1 \hat f\right],
\label{eq:chi_tau_LLL}
\end{align}
where $\hat f = {\rm diag}(f_+, f_-)$ is the incoming distribution matrix of the 1st (outer) channel of the MZI.
The counting fields here are assumed to have a time-dependence given by the
``window'' function~(\ref{eq:w_func}). In this way
the measuring time is encoded in the above formula. Let $f_-^1$ is set to the equilibrium Fermi distribution,
and $f_+^1$ is the Fermi distribution with the chemical potential $eV$. 
Asymptotically, at $eV\tau \gg 1$, and dropping
the equilibrium contribution $i(\lambda_+ + \lambda_-)(\tau/2\pi)\int_{-\infty}^0 d\epsilon$
(which is infinite because of the chirality), we obtain
\begin{equation}
\ln \left[\chi_\tau(\lambda_+,\lambda_-)\right] \simeq  \frac{eV\tau}{2\pi}\ln\left[ R_1 e^{i\lambda_+} + T_1 e^{i\lambda_-}\right].
\end{equation}
By comparing the Eqs.~(\ref{eq:chi_tau_LLL}) and (\ref{eq:A_zero_chi}) we conclude that in the limit $E_c\tau \gg 1$
\begin{equation}
e^{i{\cal A}_{\rm ferm}} ={\rm Det}\,{\cal
D}=\chi_\tau\left(2\bar\vartheta_+,2\bar\vartheta_-\right).
\label{eq:det_D}
\end{equation}
Because of this relation to the theory of FCS we will frequently refer to the
matrix ${\cal D}$ as the ``counting operator''.
In the presence of scattering at QPC1 this operator
possesses a $2\times 2$-matrix structure in the channel space. Consider further
a single chiral channel with some (in general, non-equilibrium) distribution 
function $f$ and the phase $\delta(t)= w_{\bar t,\tau}(t) \delta$. Then the generating function 
$\Delta_\tau[\delta,f] = \langle e^{i\delta \hat N} \rangle$ of 
the number $N$ of electrons passing by some observation point during the time interval $[\bar t, \bar t + \tau]$
is given by the determinant of a ``scalar'' counting operator of the kind
\begin{equation}
\Delta_\tau[\delta,f] = {\rm Det} \left[ {\mathds 1} +(e^{i\delta}-1) f\right].
\label{eq:Det_chiral}
\end{equation}
We now argue that the determinant of the matrix counting operator ${\cal
D}$, evaluated on the Fermi-like
distribution functions (which is the case of reflectionless QPC0 with $R_0=1$) can be factorized into 
a product of ``scalar'' determinants of the type (\ref{eq:Det_chiral}).  

Let us take a closer look on the random numbers $\hat N_\pm$ in the CGF given by Eq.~(\ref{eq:chi_tau}).
In the strongly non-equilibrium situation which we consider, i.e. when voltage dominates over the temperature,
they should be significantly negatively correlated (due to partition at QPC1), while their sum, 
$\hat N=\hat N_+ + \hat N_-$ should be much less sensitive to scattering and will only weakly fluctuate
around $\langle \hat N \rangle = eV\tau/(2\pi) + N_0$  with $N_0$ being some equilibrium contribution.
(At the strictly zero temperature and long time limit $\hat N$ does not fluctuate at all.)
We thus expect that $\hat N$ and $\hat N_+$ are only weakly
correlated. Taking further into account that 
$\bar \vartheta_+ = - \bar \vartheta_-$, we obtain
\begin{align}
&\chi_\tau\left(2\bar\vartheta_+,2\bar\vartheta_-\right) = 
\langle e^{2i\vartheta_{+}^q \hat N_+ + 2i\vartheta_{-}^q \hat N_-}\rangle = \nonumber\\
&\langle e^{4i\vartheta^q_{+}\hat N_+ - 2i\vartheta^q_{+} \hat N}\rangle \simeq
\langle e^{4i\vartheta^q_{+}\hat N_+}\rangle \langle e^{-2i\vartheta^q_{+}\hat N}\rangle.
\label{eq:Det_fact_heur}
\end{align}
This representation maps our 2-channel problem to three single-channel ones.
The term $\langle e^{4i\bar\vartheta_+\hat N_+}\rangle$ requires us to count the charge
in a single channel of the upper arm.  It is characterized by the distribution
function 
$$
f_{++}(\epsilon) = R_1\theta(eV-\epsilon) + T_1\theta(-\epsilon)
$$ 
established by the QPC1, thus we conclude that
\begin{equation}
\langle e^{4i\bar\vartheta_+\hat N_+}\rangle = \Delta_\tau[4\bar\vartheta_+, f_{++}].
\label{eq:Det_4}
\end{equation}
The second term, 
$\langle e^{-2i\bar\vartheta_+ \hat N}\rangle = \langle e^{-2i\bar\vartheta_+ (\hat N_+ + \hat N_-)}\rangle$,
counts the total charge in both arms. In view of charge conservation this total charge can
be also measured in the incoming channels the contribution of which are uncorrelated. The corresponding
distribution functions are $f_+(\epsilon)=\theta(eV-\epsilon)$ and $f_-(\epsilon)=\theta(-\epsilon)$.
Therefore, we get
\begin{equation}
\langle e^{-2i\bar\vartheta_+ \hat N}\rangle = \Delta_\tau[-2\bar\vartheta_+,f_+]
\, \Delta_\tau[-2\bar\vartheta_+,f_-].
\label{eq:Det_2}
\end{equation}
Combining the relations~(\ref{eq:Det_fact_heur}-\ref{eq:Det_2}) we arrive at
the following result:
\begin{align}
\chi_\tau(2\bar\vartheta_+,-2\bar\vartheta_+)&=\Delta_\tau[4\bar\vartheta_+, f_{++}] \nonumber\\
&\times\Delta_\tau[-2\bar\vartheta_+,f_+] \, \Delta_\tau[-2\bar\vartheta_+,f_-].
\label{eq:chi_D_Split}
\end{align}
Finally, we complete the argument by using Eq.~(\ref{eq:det_D}), which yields 
the desired decomposition of the matrix determinant ${\rm Det}\,{\cal D}$ into
a product of scalar determinants: 
\begin{align}
{\rm Det}\,{\cal D} &=\Delta_\tau[4\bar\vartheta_+,
f_{++}] \nonumber\\
&\times\Delta_\tau[-2\bar\vartheta_+,f_+] \, \Delta_\tau[-2\bar\vartheta_+,f_-].
\label{eq:D_Split}
\end{align}

The following remark is of order here. The equalities
between the determinants arising in our context and in the context of the
counting statistics are, strictly speaking, valid for not too large values of
the counting phase. For larger phases the counting statistics determinants show
singularities and switch to another Riemannian sheet, while our determinants
behave analytically, see Refs.~\onlinecite{Gutman:2010,Gutman:2010b}  for an extended
discussion. Physically, this is due to the fact that the counting statistics
``knows'' about the charge quantization, whereas for our problem the charge
quantization is of no relevance. This remark does not affect the validity of the
final result (\ref{eq:D_Split}), since both sides of this equation are analytic
functions of the phases.

\subsubsection{Rigorous proof: Riemann-Hilbert method}
\label{s4.1.2}

We are now going to prove Eq.~(\ref{eq:D_Split}) rigorously by analyzing
the associated Riemann-Hilbert (RH) problem.
We consider the function
\begin{equation}
Y(t)=\exp\left[\frac{\bar\vartheta_+}\pi \ln\frac{t-\bar t}{t-\bar t-\tau}\right],
\end{equation}
which is analytic and non-zero in the complex plane $t\in{\mathbb C}$, except
for the
interval of real times $[\bar t, \bar t + \tau]$. It has the property $Y(t)\to 1$ at $|t|\to \infty$.
Next we define the functions $Y_\pm(t) = Y(t\pm i 0)$ on the real axis, $t\in{\mathbb R}$.
They solve the (scalar) RH problem $Y^{-1}_-(t) Y_+(t) =
e^{-2i\vartheta_+^q(t)}$, where
$\vartheta_+^q(t) = \bar\vartheta_+ w_{\bar t,\tau}(t)$. The functions $Y_\pm(t)$ obey important identities,
\begin{eqnarray}
f_0^< \,Y_- \,f_0^< &=& f_0^< \,Y_-, \quad  f_0^> \,Y_-\, f_0^> = Y_-\,f_0^>,  \nonumber \\
f_0^< \,Y_+ \,f_0^< &=& Y_+ \,f_0^< , \quad f_0^> \,Y_+\, f_0^> = f_0^>\, Y_+,
\label{eq:Y_id}
\end{eqnarray}
where the convolution in time on the left-hand side of these relations is
implied and 
we set $f_0^<(t)\equiv f_0(t)$ and  $f_0^>(t) = \delta(t)-f_0(t) = f_0(-t)$.
As an example, the first relation in Eq.~(\ref{eq:Y_id}) reads in explicit
notations as follows:
\begin{equation}
\int dt \frac{i/2\pi}{t_1-t+i0}Y_-(t)\frac{i/2\pi}{t-t_2+i0} = \frac{i/2\pi}{t_1-t_2+i0}Y_-(t_2).
\end{equation}
Due to analytical properties of $Y_-(t)$ this integral is defined by the residue at $t=t_2-i0$
in the lower half-plane.
We also note that in the energy domain at zero temperature, $f^<_0$ is the projector on occupied states, whereas
$f_0^>$ projects on 
unoccupied states. Therefore, we have $(f_0^>)^2=f_0^>$,  $(f_0^<)^2=f_0^<$, 
$f_0^> f_0^<=0=f_0^<f_0^>$, and $f_0^<+f_0^>=1$. The same relations hold in the 
time domain as well, where the product of two operators is understood in the sense of
convolution. Using the basic identities (\ref{eq:Y_id}) one derives another two useful
relations,
\begin{eqnarray}
f_0^< \,Y_- \,f_0^> =0, \quad  f_0^> \,Y_+\, f_0^< =0. \label{eq:Y_id0}
\end{eqnarray}

Let us now turn to the analysis of the counting operator ${\cal D}$ defined by
Eq.~(\ref{eq:A_zero_chi}).
We introduce the gauge matrix $\Lambda = {\rm diag}(e^{-ie V_+ t},e^{-ie V_- t})$ comprising 
voltages $V_\pm$ applied to the outer channel in the upper/lower arms. By using these gauge
factors one rewrites ${\cal D}$ as
\begin{equation}
{\cal D} = \Lambda \left(f_0^> + \Lambda^{-1} s_1^\dagger e^{2i\hat\vartheta^q}s_1 \Lambda \,f_0^<\right)\Lambda^{-1}.
\end{equation}
With the use of solution to the RH problem we have the identity ($\vartheta_+^q = -\vartheta_-^q$)
\begin{eqnarray}
&& \Lambda^{-1} s_1^\dagger \begin{pmatrix} e^{2i\vartheta^q_+} & 0\\ 0 &
e^{2i\vartheta^q_-} \end{pmatrix} s_1 \Lambda  \nonumber \\
&& \hspace*{1cm} =\Lambda^{-1}
s_1^\dagger \begin{pmatrix} e^{4i\vartheta^q_+} & 0\\ 0 & \xUnit \end{pmatrix}
s_1\Lambda\ Y_-^{-1} Y_+.
\label{eq:Ls_theta_sL}
\end{eqnarray}
Bearing in mind that $Y_-^{-1}$ is a local in time operator without matrix structure in the channel space,
one can commute it to the left of Eq.~(\ref{eq:Ls_theta_sL}). In this way we find
\begin{align}
{\mathcal D} = \Lambda Y_-^{-1}\Bigl[Y_- f_0^>
+\Lambda^{-1} s_1^\dagger \begin{pmatrix} e^{4i\vartheta^q_+} & 0\\ 0 & \xUnit \end{pmatrix} s_1\Lambda Y_+ f_0^< \Bigr] \Lambda^{-1}.
\end{align}
To proceed further we apply the unitary transformation, $\tilde{\cal D} = s_1 {\cal D} s_1^\dagger$, and
factorize the operator $\tilde{\cal D}$ into a product of  scalar counting
operators.  This is possible by virtue of the identity
\begin{align}
\left( f_0^> + A f_0^< \right)\,\left(Y_- f_0^> + Y_+ f_0^<\right) = 
Y_- f_0^> + A Y_+ f_0^<,
\end{align}
which is valid for a local in time $2\times 2$ matrix $A(t)$. As one can check, the above relation
follows directly from the projector properties, given by Eqs.~(\ref{eq:Y_id}) and (\ref{eq:Y_id0}).
By setting
\begin{equation}
A = \Lambda^{-1} s_1^\dagger \begin{pmatrix} e^{4i\vartheta^q_+} & 0\\ 0 & \xUnit \end{pmatrix} s_1\Lambda,
\end{equation}
we obtain 
\begin{align}
\tilde{\cal D} &= Y_-^{-1} s_1 \Lambda \left[f_0^>+\Lambda^{-1} s_1^\dagger \begin{pmatrix} e^{4i\vartheta^q_+} & 0\\ 0 & \xUnit \end{pmatrix} s_1\Lambda f_0^< \right] \Lambda^{-1} s_1^\dagger\nonumber \\
		&\times s_1\Lambda \left[Y_- f_0^>+ Y_+ f_0^< \right] \Lambda^{-1} s_1^\dagger.
\end{align}
If one further introduces operators
\begin{align}
	{\cal D}_\ast &\equiv \xUnit +\begin{pmatrix} e^{4i\vartheta^q_+}-\xUnit & 0\\ 0 & 0 \end{pmatrix} \tilde f, \\
	\mathcal D_0 &\equiv\xUnit+\left(e^{-2i\vartheta^q_+}-\xUnit\right) \hat f,
	\label{eq:D_0_def}
\end{align}
where 
\begin{align*}
		\tilde f = s_1 \hat f s_1^\dagger = \begin{pmatrix} f_{++} & f_{+-} \\ f_{-+} & f_{--} \end{pmatrix}
\end{align*}
is the non-equilibrium density matrix in the MZI cell.
Explicitly, we have $f_{++} = R_1 f_+^1 + T_1 f_-^1$ and $f_{+-}=i(R_1 T_1)^{1/2}(f_+^1 - f_-^1)$.
Then $\tilde{\cal D}$ is equivalently rewritten as
\begin{equation}
\tilde{\cal D} = Y_-^{-1} \mathcal D_\ast s_1 Y_- \mathcal D_0 s_1^\dagger \label{eqn:DDecomposition}.
\end{equation}
To obtain the operator $\mathcal D_0$ we have used here once again the solution of the RH problem.
We hence conclude that 
\begin{equation}
{\rm Det}\,\tilde{\cal D} = {\rm Det}\,{\cal D}_* \, {\rm Det}\,{\cal D}_0.
\label{eq:Det_D}
\end{equation}
It is now straightforward to evaluate two determinants appearing on the
right-hand side of this relation.
In the case of matrix ${\cal D}_*$ we obtain
\begin{equation}
{\cal D}_* = \begin{pmatrix}
			\mathcal D_{\ast\ast} & \left(e^{4i\vartheta^q_+}-\xUnit\right) f_{+-}\\
			0 & \xUnit
		\end{pmatrix},
		\label{eq:D_star3}
\end{equation}
where the scalar (in the channel space) counting operator 
\begin{equation}
\mathcal D_{\ast\ast} \equiv  \xUnit+\left(e^{4i\vartheta^q_+}-\xUnit\right) f_{++}
\label{eq:D_star2}
\end{equation}
is expressed solely in terms of the upper diagonal block of the density matrix $f_{++}$,
which, obviously, has the meaning of the non-equilibrium distribution function in the upper
arm of the MZI. As the result,
\begin{equation}
{\rm Det}\,{\cal D}_* = \Delta_\tau[4\bar\vartheta_+, f_{++}].
\label{eq:Det_D_star}
\end{equation}
In the case of matrix ${\cal D}_0$ the incoming density matrix $\hat f$ is diagonal in the
channel basis, that yields
\begin{equation}
{\rm Det}\,{\cal D}_0 = \Delta_\tau[-2\bar\vartheta_+, f_{+}]\,\Delta_\tau[-2\bar\vartheta_+, f_{-}].
\label{eq:Det_D0}
\end{equation}
Combining together Eqs.~(\ref{eq:Det_D}),
(\ref{eq:Det_D_star}), and (\ref{eq:Det_D0}), we obtain the relation
(\ref{eq:D_Split}). The proof of this formula is thus completed.

\subsubsection{Inversion of the matrix counting operator}
\label{s4.1.3}

In the preceding subsection we have proven that the determinant
${\rm Det}\,\tilde{\cal D}$ (or, equivalently, the fermion
action ${\cal A}_{\rm ferm}$) can be expressed in terms
of determinants of single-channel (scalar) operators. For the
evaluation of the
interference current one also needs to consider off-diagonal matrix elements ($\mu=-\kappa$)
$\langle\mu| \tilde f\, \tilde{\cal D}^{-1}(\bar t,\bar t)|\kappa\rangle$, see Eq.~(\ref{eqn:preExp}). The
goal of this subsection is to show that, similar to the action, 
the above matrix elements can be also expressed via the scalar counting operator ${\cal D}_{**}$,
given by Eq.~(\ref{eq:D_star2}).

According to Eq.~(\ref{eqn:DDecomposition}), the inverse of
$\tilde{\cal D}$ can be written as 
\begin{align}
\tilde{\cal D}^{-1} = s_1 \mathcal D_0^{-1} Y_-^{-1} s_1^\dagger \mathcal D_\ast^{-1} Y_-. 
\end{align}
Making use of the solution to RH problem we then represent the
counting operator~(\ref{eq:D_0_def}) in the form
\begin{equation}
\mathcal D_0 = \Lambda Y_-^{-1} \left( Y_- f_0^> + Y_+ f_0^<\right)\Lambda^{-1}.
\end{equation}
The basic relation of the RH method,
\begin{equation}
\label{eq:inverse_formula}
\left( Y_- f_0^> + Y_+ f_0^<\right)^{-1} =  Y_-^{-1} f_0^> + Y_+^{-1} f_0^<,
\end{equation}
easily gives the inverse of ${\cal D}_0$ (one can check the former identity
by multiplying two operators to get the unity, employing for that relations~(\ref{eq:Y_id}) 
and (\ref{eq:Y_id0})). The inverse of $\tilde {\cal D}$ then reads
\begin{align}
	\tilde{\cal D}^{-1} = s_1 \hat\Lambda \left[Y_-^{-1} f_0^>+Y_+^{-1} f_0^<\right] \hat \Lambda^{-1} s_1^\dagger \mathcal D_\ast^{-1} Y_-,
\end{align}
and the subsequent convolution with the MZI's density matrix $\tilde f$ yields
\begin{align}
	\tilde f\tilde{\cal D}^{-1} = s_1 \hat \Lambda f_0 \hat\Lambda^{-1} s_1^\dagger \tilde {\cal D}^{-1} = Y_+^{-1} \tilde f \mathcal D_\ast^{-1} Y_-
\end{align}
The required $(\mu\bar t,\kappa \bar t)$ matrix element of this operator then takes the form
\begin{align}
	\bra \mu \tilde f \tilde {\cal D}^{-1} (\bar t,\bar t) \ket \kappa &= \left(Y_+^{-1} Y_-\right)(\bar t)\ \bra \mu \tilde f\mathcal D_\ast^{-1} (\bar t,\bar t) \ket \kappa \nonumber \\
&= e^{2i\bar \vartheta_+} \bra \mu \tilde f\mathcal D_\ast^{-1} (\bar t,\bar t) \ket \kappa.
\label{eq:mu_fD_kappa}
\end{align}
The inversion of the operator $\mathcal D_\ast$ appearing here is not exactly trivial, 
but it is simplified a lot due to its triangular structure~(\ref{eq:D_star3}) in the channel space.
Note that the relation $\mathcal D_{\ast\ast}(t_1,t_2)=\delta(t_1-t_2)$ for $t_1\notin [\bar t,\bar t+\tau]$ implies the same for the inverse, $\mathcal D^{-1}_{\ast\ast}(t_1,t_2)=\delta(t_1-t_2)$ for $t_1\notin [\bar t,\bar t+\tau]$ (this can be seen by employing the block matrix representation or using the reformulation in terms of the 
Riemann-Hilbert problem). One therefore obtains
\begin{align}
		\bra - \tilde f\mathcal D_\ast^{-1} (\bar t,\bar t) \ket + &= f_{-+} \mathcal D_{\ast\ast}^{-1}(\bar t,\bar t) = f_{-+} w_{\bar t,\tau}\ \mathcal D_{\ast\ast}^{-1}(\bar t,\bar t),
		\label{eq:minus_plus}
\end{align}
and the analogous relation for the conjugated matrix element,
\begin{align}
\bra + \tilde f\mathcal D_\ast^{-1} (\bar t,\bar t) \ket - &= \left[f_{+-}-f_{++} \mathcal D^{-1}_{\ast\ast} \left(e^{4i\vartheta_+^q}-\xUnit\right)f_{+-}\right]_{\bar t,\bar t}\nonumber\\
&=\mathcal D_{\ast\ast}^{-1} w_{\bar t,\tau}\ f_{+-}(\bar t,\bar t).
\label{eq:plus_minus}
\end{align}
It is worth pointing out that the instanton phases $\bar \vartheta_\eta=\pm \eta
\pi/\nu$ in Eqs.~(\ref{eq:minus_plus}) and (\ref{eq:plus_minus}) have opposite
signs (and hence $\mathcal D_\ast$ and $\mathcal
D_{\ast\ast}$ differ between these two equations). Since under complex
conjugation $f_\eta(t)^\ast=f_\eta(-t)$, 
these two matrix elements are indeed complex conjugates of each other.
Relations~(\ref{eq:minus_plus}) and (\ref{eq:plus_minus}) are the final result
of this subsection
and will be used below in Secs.~\ref{s4} and \ref{s5} for evaluation of the
interference current.

\subsection{Toeplitz matrices and their generalizations}
\label{s4.2}

In this subsection we relate the current and the action to the theory of Toeplitz matrices. We review key results on the large-$N$ asymptotic behavior of
Toeplitz determinants with ``Fisher-Hartwig singularities'' and of a more general
class of singular Fredholm determinants. 
These results will be then used to calculate the determinant and the inverse of
the operator ${\cal D}_{**}$, which will serve as the basis for the calculation
of the AB conductance made in Sec.~\ref{s4.3}.

\subsubsection{From integral operators to Toeplitz matrices}
\label{s4.1.4}
We relate first the fermion action ${\cal A}_{\rm ferm}$ to the
theory of Toeplitz matrices.
As we have shown in the previous section, the action is
expressed in terms of single-particle counting operators, see
Eq.~(\ref{eq:D_Split}).

The linearized single-particle spectrum used in our 1D model lacks upper and
lower band edges. Thus, a definition of the determinant of such operators 
requires an ultra-violet (UV)
regularization. One possible way to implement the regularization is
a discretization of the time coordinate $t$ in steps $\Delta t=\pi/\Lambda$,
which amounts to the introduction of an UV cutoff $\Lambda$ and restriction of
energies to the range $[-\Lambda,\Lambda]$. In this regularization procedure
operators with kernels such as $\mathcal D_{\ast\ast}(t_1,t_2)$, 
cf.~(\ref{eq:D_star2}), are turned into (in general, infinite) matrices with
discrete time indices.

In the limit of strong interaction the phase $\vartheta^q_\eta = \bar \vartheta_\eta w_{\bar t,\tau}$ is a piecewise constant function which vanishes outside the interval $[\bar t,\bar t+\tau]$. Introducing the projector $P$ which acts on functions $\phi(t)$ by multiplication with a window function in time, $P\phi=\phi w_{\bar t,\tau}$, and thus satisfies $P^2=P$, we can write
	\begin{align}
		\Det \mathcal D_{\ast\ast}& = \Det \left[\xUnit+P \left(e^{4i\bar \vartheta_+}-1\right) f_{++}\right]\nonumber\\
& = \Det \left[\xUnit+P \left(e^{4i\bar \vartheta_+}-1\right) f_{++}P\right].
	\end{align}
The operator $g_{\ast\ast}\equiv \xUnit+P \left(e^{4i\bar \vartheta_+}-1\right) f_{++}P$ has a block structure, 
namely
	\begin{align} \label{eqn:blockToeplitzCont}
		g_{\ast\ast}(t_1,t_2) = \left\lbrace \begin{array}{ll}
		                                           g(t_1-t_2), & t_1,t_2 \in [\bar t,\bar t+\tau],\\
								\delta(t_1-t_2), & {\rm otherwise},
		                                           \end{array}\right.
	\end{align}
where
\begin{equation}	
g(t_1-t_2) \equiv  \delta(t_1-t_2)+\left(e^{4i\bar \vartheta_+}-1\right) f_{++}(t_1-t_2).
\label{eq:g_symbol}
\end{equation}
The kernel of the operator $g_{\ast\ast}$ is nontrivial only if both $t_1$ and
$t_2$ lie in the interval $[\bar t,\bar t+\tau]$, in which case it depends
solely on the difference $t_1-t_2$. The  determinant of $g_{\ast\ast}$ will be
given by the nontrivial block $g(t_1-t_2)$.  The UV regularization of $g$
as described above will give rise to a large $N\times N$-matrix
$\left(g_{jk}\right)_{1\le j,k\le N}$, $N=\tau/\Delta t=\Lambda\tau/\pi$, whose
elements depend on index differences, $g_{jk}=g_{j-k}$. The matrix $(g_{jk})$ of
such type is known as a \emph{Toeplitz} matrix. 

Matrices of Toeplitz form are ubiquitous in mathematics and physics where they
appear in a variety of contexts (see e.g.\ \cite{Fisher:1968,Krasovsky:2011} for summaries of applications). 
It was shown in Refs.~\onlinecite{Gutman:2010,Gutman:2010b,Gutman:2011} that
observables in a vast range
of problems of 1D non-equilibrium interacting fermions (and bosons)  can be
expressed in terms of Toeplitz determinants $\Delta_N=\det
\left(g_{i-j}\right)_{1\le i,j\le N}$.

The behavior of determinants of such matrices becomes
particularly non-trivial when the corresponding symbol (essentially the Fourier
transform of $g_{j-k}$, see Sec.~\ref{s4.2} for more detail) has singular
points known as Fisher-Hartwig singularities. In our case such  singularities
arise in view of discontinuities of the double-step distribution function
$f_{++}$. In Refs.~\onlinecite{Gutman:2011,Protopopov:2011} the large-$N$ behavior of
Toeplitz determinants with Fisher-Hartwig singularities has been established
analytically and verified numerically. These results (``generalized
Fisher-Hartwig conjecture'') go beyond the ``standard''  Fisher-Hartwig
conjecture (proven in Ref.~\onlinecite{Deift:2011}) as they
contain not only the leading term but also subleading power-law contributions
that have different oscillatory factors. We will see below that taking into
account such contributions will be crucial for obtaining the oscillatory
dependence of visibility of MZI on voltage. A further generalization was
achieved recently in Ref.~\onlinecite{Protopopov:2012} where a broader class of
singular Fredholm determinants (determined by two symbol functions that show
multiple singularities in energy and coordinate spaces, respectively)
was explored and corresponding asymptotics were found. Such determinants will
arise below when we will invert the operator ${\cal D}_{**}$.

\subsubsection{Asymptotics of Toeplitz determinants}
\label{s4.2.1}

For the benefit of the reader we summarize here the relevant results on Toeplitz determinants.

A Toeplitz matrix $g_{jk}$ with $j,k=1,2,\ldots,N$ is defined by its symbol
$g(z)$ as follows:
	\begin{align}
		g_{jk} = g_{j-k}=\int_{-\pi}^\pi\!\frac{d\varphi}{2\pi}\,
g(e^{i\varphi}) e^{-i\varphi(j-k)}.
	\end{align}
The determinant of such a matrix is called Toeplitz determinant.
An important class of Toeplitz matrices (which is of relevance for our work and
for various other non-equilibrium many-body problems) is generated by symbols
with Fisher-Hartwig (FH) singularities, 
	\begin{align} \label{eqn:symbFHForm}
		g(z)=e^{V(z)} \prod_{j=0}^m \left\lvert
z-z_j\right\rvert^{2\alpha_j}\ \gamma_j(z) \left(z/z_j\right)^{\beta_j} 
	\end{align}
	where $V(z)$ is a smooth function, $m+1$ is a positive integer (number
of singular points), $z_j\equiv e^{i\varphi_j}$, ${\rm Re}\, \alpha_j>-\frac
12$, $\beta_j\in\mathbb C$, and 
	\begin{align}
		\gamma_j(z) = \left\lbrace
    \begin{array}{ll}
		            e^{i\pi \beta_j}, & -\pi<\arg z<\varphi_j,\\
					e^{-i\pi\beta_j}, & \varphi_j<\arg
z<\pi.
    \end{array}\right.
	\end{align}
In the context of our work, only the case $\alpha_j=0$ (when the singularities
of $f(z)$ are discontinuities) will be relevant, so that we consider it
henceforth. 
The large-$N$ asymptotic behavior of the corresponding Toeplitz
determinant $\Delta_N$ reads\cite{Gutman:2011,Protopopov:2011}:
\begin{widetext}
\begin{align} \label{eqn:GutmanFH}
		\Delta_N= e^{NV_0} \sum_{n_0+\ldots+n_m=0}\, \prod_{j=0}^m
z_j^{n_jN}\	\left[N^{-\sum_{j=0}^m \beta_j^2}\ \prod_{0\le j<k\le m}
\left\lvert z_j-z_k\right\rvert^{2\beta_j\beta_k}\ \prod_{j=0}^m G(1+\beta_j)
G(1-\beta_j)\right]_{\beta_j\to\beta_j+n_j}(1+\ldots).
	\end{align}
\end{widetext}
where $V_0=\int_{-\pi}^\pi\!\frac{d\varphi}{2\pi}\, V(e^{i\varphi})$ and $G$ is
the Barnes G-function. The summation in Eq.~(\ref{eqn:GutmanFH}) goes over
a set of integers $n_0$, $n_1$, \ldots, $n_m$ (whose sum is zero); we will see
below that they can be understood as labeling branches of $\log g(z)$  in the
intervals of continuity of the symbol. Each of these sets (``branches'') is
characterized by a distinct factor $\prod_{j=0}^m z_j^{n_jN}$ that in our
context will give rise to a distinct oscillatory exponent. Equation
(\ref{eqn:GutmanFH}) presents explicitly the leading asymptotic behavior for
each of the branches. There exist also subleading power-law
corrections within each of the branches (i.e., corresponding to the same
oscillatory exponent); they are abbreviated by $+\ldots$ in the last bracket. 
Such corrections will be of no importance for our consideration, and we discard
them below. 

We return now to determinants of the type (\ref{eq:Det_chiral})
that arise in the course of the study of MZI. Here
$f$ is some distribution function and $\delta(t)=\delta w_{[0,\tau]}(t)$
is a constant in the window of the duration $\tau$ and zero otherwise.
We are interested in the large-$\tau$ asymptotic behavior of
$\Delta_\tau[\delta,f]$. As discussed in Sec.~\ref{s4.1.4}, the UV
regularization is implemented by using a high-energy cut-off $\Lambda$, 
so that the energy is restricted to the range $[-\Lambda,\Lambda]$
and the time is discretized, $t_j= j\Delta t=j\pi/\Lambda$. In energy
representation, the operator 
of interest reads [cf. Eq.~(\ref{eq:g_symbol})]
	\begin{align} \label{eqn:symbolOrig}
		\tilde g(\epsilon) = 1+(e^{i\delta}-1)f(\epsilon).
	\end{align}
This can be identified with a symbol $g(z)$ of a Toeplitz matrix, 
provided energy $\epsilon\in [-\Lambda,\Lambda]$ and angle $\varphi\in
[-\pi,\pi]$ are related by rescaling: $\varphi=\epsilon \pi/\Lambda$. The
introduction of a hard cutoff $\pm\Lambda$ and the above compactification of the
energy axis will give rise to unphysical effects at this energy scale (since
it will generate an unphysical discontinuity at $\varphi=\pm\pi$). These
unphysical effect are eliminated by imposing ``periodic boundary conditions'' in
energy domain\cite{Gutman:2011}, which amounts to the following modification of
the symbol: 
$\lim_{\epsilon\to -\Lambda} g(\epsilon)=\lim_{\epsilon\to \Lambda} g(\epsilon)$:
	\begin{align}\label{eqn:symbolCont}
		g(\epsilon)= e^{i\delta
\epsilon/(2\Lambda)}\left[1+(e^{i\delta}-1)f(\epsilon)\right].
	\end{align}
Here we have taken into account that $\lim_{\epsilon\to -\Lambda}
f(\epsilon)=1$ and $\lim_{\epsilon\to \Lambda} f(\epsilon)=0$. 

To be specific, let us consider explicitly two examples corresponding to two
lowest values $m=0,1$ (i.e., one and two FH singularities.) First, we consider 
the equilibrium distribution function $f(\epsilon)=\theta(\mu-\epsilon)$. The symbol is
\begin{align}
		g(e^{i\varphi}) = \left\lbrace
    \begin{array}{ll}
		e^{i\delta\varphi/(2\pi)} e^{i\delta}, & -\pi < \varphi< \pi \mu/\Lambda,\\
		e^{i\delta\varphi/(2\pi)}, & \pi\mu/\Lambda < \varphi < \pi,
	\end{array} \right.
\end{align}
which is of the form (\ref{eqn:symbFHForm}) 
with $m=0$, $\alpha_0=0$, $\beta_0=\delta/(2\pi)$, $z_0=e^{i\pi\mu/\Lambda}$, 
and $V_0=i\delta(1+\mu/\Lambda)/2$. According to Eq.~(\ref{eqn:GutmanFH}) in the large-$N$ limit 
the $\det (g_{j-k})$ asymptotically behaves as
\begin{align} \label{eqn:GutmanAsSingle}
\Delta[\delta,f_{\rm single}]&= \exp\left[{i\frac\delta{2\pi} (\Lambda+\mu)\tau}\right]\ 
\left(\frac{\Lambda\tau}\pi\right)^{-\left(\frac\delta{2\pi}\right)^2}\nonumber\\
&\times G(1-\frac\delta{2\pi})\,G(1+\frac\delta{2\pi}).
\end{align}

Next, we consider a  double-step distribution function 
$f(\epsilon)=(1-a)\, \theta(\mu_0-\epsilon) +a \,\theta(\mu_1-\epsilon)$, where we assumed that 
$\mu_0<\mu_1$. In this case the symbol reads
	\begin{align}
		g(e^{i\varphi}) & =\left\lbrace \begin{array}{ll}
		                 e^{i\delta \varphi/(2\pi)} e^{i\delta}, & -\pi<\varphi<\frac{\pi\mu_0}{\Lambda},\\
						e^{i\delta \varphi/(2\pi)} \left[1+(e^{i\delta}-1)a\right], 
                         & \frac{\pi\mu_0}{\Lambda}<\varphi<\frac{\pi\mu_1}{\Lambda},\\
						e^{i\delta \varphi/(2\pi)}, & \frac{\pi\mu_1}{\Lambda}<\varphi<\pi.
		                              \end{array}\right.
	\end{align}
Hence, the symbol has two FH singularities $z_j=e^{i\pi \mu_j/\Lambda}$, $j=0,1$, with 
	\begin{align}
		e^{-2\pi i \beta_0}=\frac{1+(e^{i\delta}-1)a}{e^{i\delta}},\quad e^{-2\pi i \beta_1}=\frac 1{1+(e^{i\delta}-1)a}.
	\end{align}
We choose
	\begin{align}
		\beta_1=-\frac i{2\pi} \ln\left[1+(e^{i\delta}-1)a\right], \quad \beta_0=\frac\delta{2\pi} -\beta_1.
	\end{align}
It is easy to see that the symbol has the form (\ref{eqn:symbFHForm})
with $m=1$, $\alpha_j=0$, and
	\begin{align}
		V(z)=V_0=i\delta/2+i\delta\frac{\mu_0}{2\Lambda}+ieV\frac\pi\Lambda \beta_1	
	\end{align}
where we introduced $eV=\mu_1-\mu_0$. 
According to (\ref{eqn:GutmanFH}) the asymptotic behavior of the Toeplitz determinant $\det\left(g_{j-k}\right)$ is given by
\begin{widetext}
	\begin{multline} \label{eqn:GutmanAsDouble}
		\Delta[\delta,f_{\rm double}]=\exp\left[i\frac{\delta}{2\pi}(\Lambda+\mu_0) \tau+\frac{eV\tau}{2\pi} \ln \left[1+(e^{i\delta}-1) a\right]\right]\,
		 \sum_{n=-\infty}^\infty e^{-ieV\tau n}\ \left(\frac{\Lambda\tau}\pi\right)^{-(\beta_0+n)^2-(\beta_1-n)^2}\\
\times \left(\frac{\pi eV}\Lambda\right)^{2(\beta_0+n)(\beta_1-n)}\,
		G(1+\beta_0+n) G(1-\beta_0-n) G(1+\beta_1-n) G(1-\beta_1+n).
	\end{multline}
\end{widetext}
In order to identify in the sum over $n$ the leading contributions in the
long-$\tau$ regime, we consider the exponent
	\begin{equation}
		{\rm
Re}\left[-(\beta_0+n)^2-(\beta_1-n)^2\right]=-2\left(n-n^\ast\right)^2 +{\rm
const},
	\end{equation}
where 
\begin{eqnarray}	
n^\ast &=& \frac 12{\rm Re}(\beta_1-\beta_0)
\nonumber \\ &=&
{1\over {2\pi}}\Im\ln\left[(1-a)+a e^{i\delta}\right]
- \frac\delta{4\pi}.
\end{eqnarray}
Note also that the sum of voltage and time exponents,
$\left[-(\beta_0+n)^2 - (\beta_1-n)^2 \right] - \left[2 (\beta_0+n) (\beta_1-n) \right] = -(\beta_0+\beta_1)^2$
is independent of $n$. Thus, terms dominant for $\Lambda\tau\gg 1$ are also leading for large voltages, $eV\tau\gg1$. For the analysis of the optimal value $n^\ast$, we make the decomposition $\delta=2\pi M+\delta'$ with $M\in \mathbb Z$ and $\lvert \delta'\rvert<\pi$. One can show that the phase
\begin{equation}
\delta''\equiv\Im\ln\left[(1-a)+a e^{i\delta}\right]
\end{equation}
has the same sign as $\delta'$ and satisfies $\lvert \delta''\rvert\le\lvert \delta'\rvert$.
Then the optimal $n^*$ becomes
\begin{equation}
 n^\ast=-\frac {M}{2} -\frac{\delta'-2\delta''}{4\pi}, \quad
\lvert n^\ast+M/2\rvert\le 1/4.
\end{equation}
We see that in the case of even $M$ there is a single contribution with $n=-M/2$
giving the most significant contribution to the asymptotic series; other
contributions have substantially smaller (by real part) exponents. On
the other hand, for odd $M$ one has to take into account two contributions with
$n=-(M\pm1)/2$. Indeed, if $a=1/2$ (and thus
$\delta''=\delta'/2$), then these two contributions come with equal exponents.
When $a$ deviates from 1/2, the exponents become different but still may be
very close.

It was shown in Ref.~\onlinecite{Protopopov:2012} that these results can be
genralized to a broader class of singular Fredholm determinants.
Specifically, consider a matrix
\begin{equation}
\label{eq:def_gen_TM}
g_{j,k}=\int_{-\Lambda}^\Lambda\!\frac{d \epsilon}{2\Lambda}\, e^{-i\epsilon \pi/\Lambda [j-k-\delta(t_j)/(2\pi)]}
\tilde g(t_j,\epsilon)\end{equation}
with the symbol
\begin{equation}
\label{eq:symbol_gen_TM}
\tilde g(t,\epsilon)\equiv1+\left(e^{i\delta(t)}-1\right)f(\epsilon).
\end{equation}
Here the notion of symbol has been generalized to include both time 
and energy dependence (through the function $\delta(t)$ and $f(\epsilon)$,
respectively). Let us focus on the case when both
the phase $\delta(t)$ and the distribution function $f(\epsilon)$ are
piecewise constant functions
with jumps at times $\tau_1<\tau_2<\ldots<\tau_{N_\tau}$ and energies
$\mu_1<\mu_2<\ldots <\mu_{N_\mu}$, 
respectively. They satisfy the boundary conditions $\delta(t)=0$ for
$t\notin[\tau_1,\tau_{N_\tau}]$, $f(\epsilon)=1$ for $\epsilon<\mu_1$, and 
$f(\epsilon)=0$ for $\epsilon>\mu_{N_\mu}$. The UV cutoff and the periodic
boundary conditions in energy domain can be implemented as before. 
The discontinuity points define a grid which subdivides the time-energy plane
in domains with different values of the symbol. The domains can be labeled by   
the time indices $j\in \lbrace 0,\ldots, N_\tau\rbrace$, 
and energy indices $k\in\lbrace 0,\ldots, N_\mu\rbrace$.
One associates with this set of domains a set of number $c_{jk}$, 
\begin{gather} \label{eqn:defCjk}
	c_{jk}\equiv \frac 1{2\pi i} \ln \tilde g(\tau_j+0,\mu_k+0)+n_{jk},\\
	c_{j0}\equiv\delta(t_j+0)/(2\pi),\quad c_{0k}=c_{N_\tau,k}=c_{j,N_\mu}=0.
\end{gather}
where $\lbrace n_{jk}\rbrace$ is an arbitrary set of integers. 
Further, a matrix $\beta_{jk}$ with a
time index $j\in \lbrace 1,\ldots, N_\tau\rbrace$ 
and energy index $k\in\lbrace 1,\ldots, N_\mu\rbrace$ is introduced according to
\begin{eqnarray}
	\beta_{jk}\equiv c_{j,k-1}-c_{j,k}+c_{j-1,k}-c_{j-1,k-1}, \\
	\nonumber
\end{eqnarray}
Physically, each entry of this matrix corresponds to a crossing point of one
energy-space and one time-space singularity. In terms of this matrix, a set  
of time ($p_{jl}$) and energy~($q_{km}$) exponents is defined as follows:
\begin{equation} 
	p_{jl}\equiv\sum_{m'=1}^{N_\mu} \beta_{jm'}\beta_{l m'}, \quad
	q_{km}\equiv\sum_{l'=1}^{N_t} \beta_{l'k}\beta_{l'm}.
\end{equation}
One should keep in mind that $c_{jk}$ and thus $\beta_{jk}$, $p_{jl}$, and $q_{km}$ depend on the set 
of integers $n_{jk}$. In Eq.~(\ref{eqn:defCjk}) the logarithm $\ln\tilde g$ is
understood as evaluated at its principal branch, $\Im\ln \tilde g \in
(-\pi,\pi]$. The summation over integers $n_{jk}$ 
hence amounts to summing over different branches of the logarithms.

We are now ready to state the result.
For large time- and energy differences, $\lvert
(\tau_j-\tau_l)(\mu_k-\mu_m)\rvert\gg 1$ ($j\neq l$, $k\neq m$), 
the asymptotic behavior of $\det (g_{j,k})$ is given
by\cite{Protopopov:2012}
\begin{widetext}
\begin{multline}
\label{eq:non_T_series}
	\Delta[\delta(t),f(\epsilon)] =\sum_{\lbrace n_{jk}\rbrace} \Gamma_{\lbrace n_{jk}\rbrace}\ \exp\left[i\sum_{1\le j<N_t}\left(c_{j0}(\Lambda+\mu_1) +\sum_{1\le k<N_\mu} c_{jk} \left(\mu_{k+1}-\mu_k\right)\right)(\tau_{j+1}-\tau_j)\right]\\
		\times \prod_{1\le j<l<N_t}\prod_{1\le k<m<N_\mu}\ \left\lvert \frac{\Lambda\left(\tau_j-\tau_l\right)}{\pi}\right\rvert^{p_{jl}} \left\lvert\frac{\pi\left(\mu_k-\mu_m\right)}\Lambda\right\rvert^{q_{km}}
\end{multline}
\end{widetext}
with coefficients $\Gamma_{\lbrace n_{jk}\rbrace}$ that are independent on
$\tau_j$ and $\mu_k$.
It is not difficult to check that for the phase $\delta(t)$ proportional to a
window function this formula 
agrees with the asymptotics (\ref{eqn:GutmanAsDouble}) of the Toeplitz
determinant.  While a rigorous mathematical proof of the asymptotic
formula~(\ref{eq:non_T_series}) is still lacking,
Ref.~\onlinecite{Protopopov:2012} presented powerful analytical arguments
in favor of its validity supported by strong evidence based on
numerical evaluation of such determinants. We will use
Eq.~(\ref{eq:non_T_series}) below to get analytical results for the current
through the MZI.

\subsubsection{Inversion of the single-channel counting matrix ${\cal D}_{**}$}
As we have shown in Sec.~{\ref{s4.2}}, the interference current can be expressed in terms of
the inverse of the single-channel counting operator ${\cal D}_{**}$. 
We show here that it is related to a generalized Toeplitz determinant, 
whose asymptotic behavior can be estimated on the basis of results presented above. 
To this end we consider the time discretized expression for the operator 
${\cal D}_{**}$, given by Eq.~(\ref{eq:D_star2}), which has the symbol
\begin{align}
	g(\epsilon)=e^{2i\bar\vartheta_+ \epsilon/\Lambda} \left[1+\left(e^{4i\bar\vartheta_+}-1\right) f_{++}(\epsilon)\right]
\end{align}
corresponding to the Toeplitz matrix,
\begin{align}
g_{j-k}&=\frac i{2\pi} \frac 1{j-k-2\bar\vartheta_+/\pi}\left(e^{4i\bar\vartheta_+}-1\right)\nonumber \\
&\times\left[R_1 e^{-i\pi U/\Lambda[j-k-2\bar\vartheta_+/\pi]}+T_1\right].
\label{eq:g_jk}
\end{align}
The inverse of matrix $g$ reads
\begin{align} 
	\left(g^{-1}\right)_{jk} = (-1)^{j+k} \frac{\det g^\sharp(k,j)}{\det (g)},
\end{align}
where $g^\sharp(k,j)$ is the $(N-1)\times (N-1)$  matrix derived from $g$ by removing 
the $k$-th row and $j$-th column. 	
	
Since our primary interest is the matrix element $D_{**}^{-1}(\bar t, t)$ with $\bar t < t < \bar t + \tau$,
see Eq.~(\ref{eq:plus_minus}), we concentrate specifically on the element $(g^{-1})_{1k}$, i.e. we put $j=1$.
In this case one has 
\begin{align}
	\left(g^\sharp(k,1)\right)_{lm} &= \left\lbrace \begin{array}{ll}
	                             	g_{l,m+1}, & 1\le l <k,\\
					g_{l,m}, & k\le l \le N-1
	                             \end{array}\right.\\
			&= \frac i{2\pi} \frac 1{l-m-\frac 2\pi \vartheta_+(t_l;t_k)}\left(e^{4i\vartheta_+(t_l;t_k)}-1\right)
\nonumber\\
&\times \left[R_1 e^{-i\pi U/\Lambda[j-k-\frac 2\pi \vartheta_+(t_l;t_k)]}+T_1\right],
\label{eq:g_sharp_k1}
\end{align}
where we have introduced the time-dependent phase 
\begin{align}
	\vartheta_+(t_l;t_k)= \left\lbrace \begin{array}{ll}
						\bar\vartheta_+ +\pi/2, &\bar t \le t_l < t_k,\\
						\bar \vartheta_+, & t_k\le t_l < \tau + \bar t.
	                               \end{array}\right.
	                               \label{eq:phase_lk}                     
\end{align}	
with $t_l=\bar t+(l-1)\Delta t$. In the continuous representation the phase $\vartheta_+(t,t_k)$ is
the piecewise function of time $t$. Taking into account Eqs.~(\ref{eq:g_jk}), (\ref{eq:g_sharp_k1})
and the definition~(\ref{eq:def_gen_TM}) one observes that the matrix $\left(g^\sharp(k,1)\right)_{lm}$
is the generalized Toeplitz matrix with the symbol~(\ref{eq:symbol_gen_TM}) where the phase 
$\delta(t)=4\vartheta_+(t,t_k)$. Its determinant can be dealt with by the 
use of results presented in the end of the previous subsection. Hence,
\begin{align}
	\left(g^{-1}\right)_{1k} = (-1)^{1+k} \frac{\Delta[4\vartheta_+(\bullet;t_k),f_{++}]}{\Delta[4\bar\vartheta_+,f_{++}]}.
	\label{eq:inv_g_1k}
\end{align}
	
\begin{figure}[b]
	\includegraphics[width=2.6in]{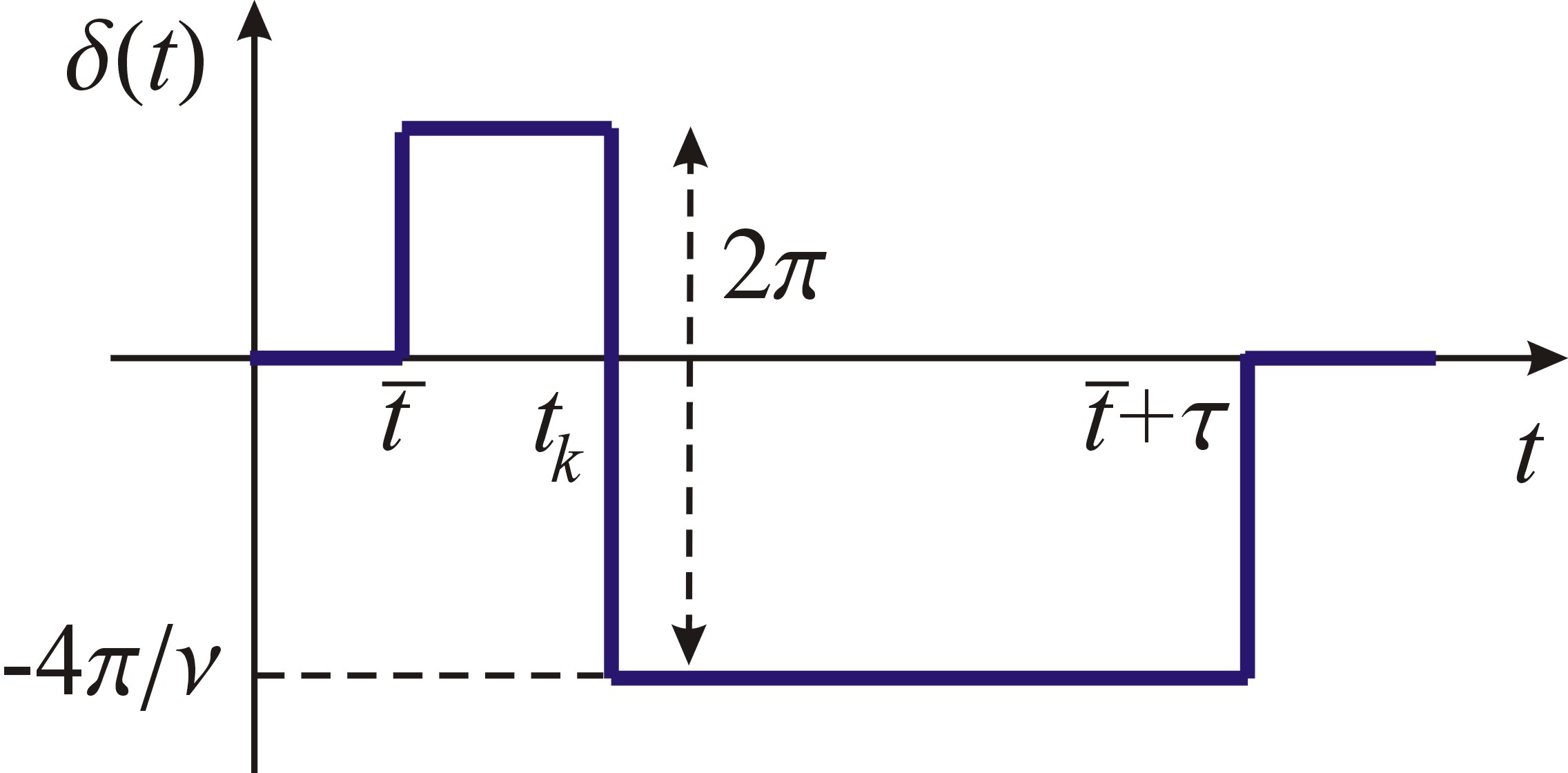}
	\caption{Phase $\delta(t)=4\vartheta_+(t;t_k)$: on top of the ``background phase'' $4\vartheta_+^q(t)$ 
the $2\pi$-pulse accounts for the additional electron during the time interval $0<t<t_k$.} \label{fig:PhaseWPulse}
\end{figure}

It is instructive to apply first the asymptotic relations~(\ref{eqn:GutmanAsDouble}) and (\ref{eq:non_T_series}) 
to invert the matrix $g_{j-k}$ in the limit $T_1\to 0$, when the alternative evaluation can be done via the Riemann-Hilbert method. Under this condition the operator ${\cal D}_{**}$ has the form
\begin{equation}
\left[{\cal D}_{**}\right]_{T_1\to 0} = \Lambda_+\left[ f^>_0 + e^{4i\vartheta_+^q} f_0^<\right]\Lambda_+^{-1}
\end{equation}
with $\Lambda_+(t) = e^{-iV_+t}$. Therefore $\left[{\cal D}_{**}\right]_{T_1\to 0}^{-1}$ can be evaluated in 
the similar fashion as we have found the inverse of the counting operator ${\cal D}_0$ in Sec.~\ref{s4.1.2}.	
By introducing the function
\begin{equation}
\tilde Y(t) = \left(\frac{t-\bar t}{t-\bar t -\tau}\right)^{{2\bar \vartheta_+}/{\pi}},
\label{eq:tilde_Y}
\end{equation}	
which solves the RH problem $\tilde Y_-^{-1} \tilde Y_+ = e^{-4i\vartheta_+^q}$, one further
represents ${\cal D}_{**}$ in the equivalent form
\begin{equation}
\left[{\cal D}_{**}\right]_{T_1\to 0} = \Lambda_+ \tilde Y_-
\left[ \tilde Y_-^{-1} f^>_0 + \tilde Y_+^{-1} f_0^<\right]\Lambda_+^{-1}.
\end{equation}
Using now the relation~(\ref{eq:inverse_formula}), we obtain 	
\begin{equation}
\left[\mathcal D_{\ast\ast}\right]_{T_1\to0}^{-1}(\bar t,t) = \tilde Y_-(\bar t) \left[\xUnit+\left(e^{-4i\vartheta_+^q}-\xUnit\right)f_+\right]_{\bar t,t} \tilde Y_-^{-1}(t).
\end{equation}
Taking into account the explicit form of the solution of the RH problem~(\ref{eq:tilde_Y}),
for times $t\in[\bar t, \bar t +\tau]$  we eventually arrive at	
\begin{align}
	\left[\mathcal D_{\ast\ast}\right]_{T_1\to0}^{-1}(\bar t,t) &= 
		-\frac\Lambda\pi\ e^{-2i\bar\vartheta_+}\sin(2\bar \vartheta_+)\ e^{-ieV(\bar t-t)} \nonumber\\
&\times\left\lvert t-\bar t\right\rvert^{-2\bar\vartheta_+/\pi-1} \left\lvert\bar t+\tau-t\right\rvert^{2\bar \vartheta_+/\pi} \nonumber\\
&\times\tau^{-2\bar \vartheta_+/\pi} \Lambda^{-2\bar \vartheta_+/\pi-1} \label{eqn:DStarStarInverseExact}
\end{align}
where the regularization $\tilde Y_-(\bar t)=\tilde Y_-(\bar t+\Lambda^{-1})$ was chosen.

Let us now demonstrate that the same result can be derived using the properties of the Toeplitz determinants.
For $\mathcal D_{\ast\ast}^{-1}(\bar t,t_k)$ and thus $\left(g^{-1}\right)_{1k}$ one needs to consider the generalized Toeplitz problem with 3 jumps in time domain, $\tau_1=\bar t$, $\tau_2=t_k$, $\tau_3=\bar t+\tau$, and just 1 jump in energy domain $\mu_1=eV$. Further it is $c_{10}=2\bar\vartheta_+/\pi +1$ and $c_{20}=2\bar\vartheta_+/\pi$ and hence $p_{12}=-2\bar\vartheta_+/\pi-1$, $p_{13}=-\left(2\bar\vartheta_+/\pi+1\right)2\bar\vartheta_+/\pi$ and $p_{23}=2\bar\vartheta_+/\pi$. The asymptotic behavior of $\det (g)=\Delta[4\vartheta^q_+,f_+]$ is 
given by Eq.~(\ref{eqn:GutmanAsSingle}) with $\mu=eV$ and $\delta=2\bar\vartheta_+/\pi$.
As the result, one obtains
\begin{align}
	\left[\left(g^{-1}\right)_{1k}\right]_{T_1\to 0}&=\Gamma e^{-ieV(\bar t-t_k)} \left\lvert \frac{\Lambda(t_k-\bar t)}\pi\right\rvert^{-2\bar\vartheta_+/\pi-1} \nonumber\\
&\times\left\lvert \frac{\Lambda (\bar t+\tau-t_k)}\pi\right\rvert^{2\bar\vartheta_+/\pi} \left\lvert \frac {\Lambda \tau}\pi\right\rvert^{-2\bar\vartheta_+/\pi}.
\end{align}
Except for the dimensionless unknown factor $\Gamma$ and the prefactor $\Lambda/\pi=\left(\Delta t\right)^{-1}$ 
which arises due to time discretization, the above asymptotics agrees in all power-laws with the 
exact result (\ref{eqn:DStarStarInverseExact}).
	
We now turn to the general situation of arbitrary $T_1$. The distribution function in this case is given by 
$f_{++}$ instead of $f_+$ which adds a discontinuity at $\mu_1=0$ (the one at $eV$ is now denoted by $\mu_2$). 
The asymptotics of the determinant $\Delta[4\vartheta_+(\bullet;t_k)]$ is determined by Eq.~(\ref{eq:non_T_series})
with
\begin{equation}
	c_{10} =\alpha_1+1, \quad c_{20}=\alpha_1, \quad c_{k1}=\beta_1+n_k\quad (k=1,2),
\end{equation}
where we have abbreviated
\begin{equation}
\alpha_1 \equiv 2\bar\vartheta_+/\pi, \quad \beta_1 \equiv \frac 1{2\pi i} \ln\left[R_1 e^{4i\bar\vartheta_+}+T_1\right],
\end{equation}
and the exponents
\begin{widetext}
\begin{align}
p_{12}&= (1+\alpha_1-\beta_1-n_1)(n_1-n_2-1)+(\beta_1+n_1)(n_2-n_1), \nonumber \\
p_{23}&= n_1-n_2-1)(\beta_1+n_2-\alpha_1)-(n_2-n_1)(\beta_1+n_2), \nonumber \\
p_{13}&= (1+\alpha_1-\beta_1-n_1)(\beta_1+n_2-\alpha_1)-(\beta_1+n_1)(\beta_1+n_2), \nonumber \\
q_{12}&= (1+\alpha_1-\beta_1-n_1)(\beta+n_1) + (n_1-n_2-1)(n_2-n_1)- (\beta_1-\alpha_1+n_2)(\beta_1+n_2).
\label{eqn:TripleStepExp}
\end{align}
One then has
\begin{multline} 
\label{eqn:AsTripleStep}
	(-1)^{1+k} \Delta[4\vartheta_+(\bullet;t_k),f_{++}] =\sum_{n_1,n_2} \Gamma_{(n_1,n_2)} e^{i\alpha_1\Lambda \tau+i\beta_1eV\tau} e^{ieV\left[n_1(t_k-\bar t)+n_2(\bar t+\tau-t_k)\right]}\\
	\times \left\lvert \frac{\Lambda (t_k-\bar t)}\pi\right\rvert^{p_{12}} \left\lvert \frac{\Lambda(\bar t+\tau-t_k)}\pi\right\rvert^{p_{23}} \left\lvert \frac{\Lambda\tau}\pi\right\rvert^{p_{13}} \left\lvert\frac{\pi eV}\Lambda\right\rvert^{q_{12}}.
\end{multline}
\end{widetext}
When deriving this asymptotics, we took into account that the phase factor $e^{i\Lambda(t_k-\bar t )} = (-1)^{1+k}$,
since $t_k=\bar t + (k-1)\Delta t$ with the infinitesimal time increment $\Delta t = \pi/\Lambda$, 
cf. Eq.~(\ref{eq:phase_lk}). The above relation is one of the main results of the section.
It yields the asymptotic value for $(g^{-1})_{1k}$, expressed through Eq.(\ref{eq:inv_g_1k}).
The determinant $\Delta[4\bar \vartheta_+, f_{++}]$ appearing in the latter relation can be found
exactly via Eq.~(\ref{eqn:GutmanAsDouble}), where one has to set $\delta = 4\bar \vartheta_+$
and $a=R_1$.

The following remark must be made concerning the above calculations. The result (\ref{eqn:AsTripleStep})
has been derived with the use of the asymptotic formula (\ref{eq:non_T_series}). It is
valid under the assumption $|(\tau_j-\tau_l)(\mu_k-\mu_m)|\gg 1$, which defines the range of
applicability to Eq.~(\ref{eqn:AsTripleStep}), namely $t_k-\bar t \gtrsim 1/eV$ and 
$\bar t +\tau -t_k \gtrsim 1/eV$. Below we examine another limit, when the time $t_k$
is close to either of two boundaries, $\bar t$ or $\bar t +\tau$.

To this end we represent the (normalized) determinant 
$\bar\Delta[\delta(t),f(\epsilon)] = \Delta[\delta(t),f(\epsilon)]/\Delta[\delta(t),T=0]$
in the equivalent form~\cite{Protopopov:2012}
\begin{widetext}
\begin{equation}
\label{eq:non_T_norm_series}
	\bar\Delta[\delta(t),f(\epsilon)] =\sum_{\lbrace n_{jk}\rbrace} \bar\Gamma_{\lbrace n_{jk}\rbrace}\ 
\exp\Biggl[i\sum_{1 \leq j \leq N_t} \sum_{1\leq k \leq N_\mu} \tau_j \beta_{jk} \mu_k  \Biggr]
		\prod_{1\le j<l\leq N_t}\prod_{1\le k<m\leq N_\mu}\ \Bigl[(\tau_l-\tau_j)(\mu_k-\mu_m)\Bigr]^{\gamma_{jl,km}},
\end{equation}
\end{widetext}
where
\begin{equation}
\gamma_{jl,km}= -c_{jk} c_{lm} - c_{jm} c_{lk}.
\end{equation}
The normalized determinant is cut-off ($\Lambda$) independent. All dependence on $\Lambda$ comes from
the zero temperature determinant, which up to a constant prefactor reads
\begin{align}
\Delta[\delta,T=0]&=\exp\Biggl[-i\sum_{1 \leq j \leq N_t} \Lambda\tau_j\frac{(\delta_j-\delta_{j-1})}{2\pi}\Biggr] 
\label{eq:Det_T0}\\
&\times \prod_{1\le j<l\leq N_t}\left\lvert \frac{\Lambda\left(\tau_j-\tau_l\right)}{\pi}\right\rvert^{
{(\delta_j-\delta_{j-1})(\delta_l-\delta_{l-1})}/{4\pi^2}}, \nonumber
\end{align}
where we have defined the phases $\delta_j \equiv \delta(t_j+0)$. Eq.~(\ref{eq:Det_T0}) is the particular
case of the asymptotics~(\ref{eq:non_T_series}) when the generalized Toeplitz problem has $N_t$
jumps of the phase	$\delta(t)$ in the time domain and the single (Fermi) edge at $\epsilon=0$.
The summation over branches of logarithms is not required here. The equivalence between 
two forms of the asymptotic expansion, Eqs.~(\ref{eq:Det_T0}) and (\ref{eq:non_T_series}),
follows from the sum rules,
\begin{equation}
\sum_{k=1}^{N_\mu} \beta_{jk} = \frac{\delta_j-\delta_{j-1}}{2\pi}, \qquad \sum_{j=1}^{N_t} \beta_{jk}=0.
\end{equation}
As before, representation~(\ref{eq:non_T_series}) holds provided $(\tau_l - \tau_j)(\mu_m-\mu_k) \gg 1$
for all $j>l$ and $m>k$. However, it enables a natural generalization to the situation, when this
condition is not satisfied. Namely, if for some set $(l,j,m,k)$ the opposite inequality is fulfilled,
the corresponding factor has to be omitted from the product in Eq.~(\ref{eq:non_T_series}). This
is the advantage of the normalized representation in comparison to Eq.~(\ref{eq:non_T_norm_series}).
In this way one can find the asymptotic form of $(D_{**}^{-1})(\bar t,t)$
if $t$ is close to $\bar t$ or $\bar t +\tau$.	

\begin{figure}[b]
\centering{\includegraphics[scale=0.6]{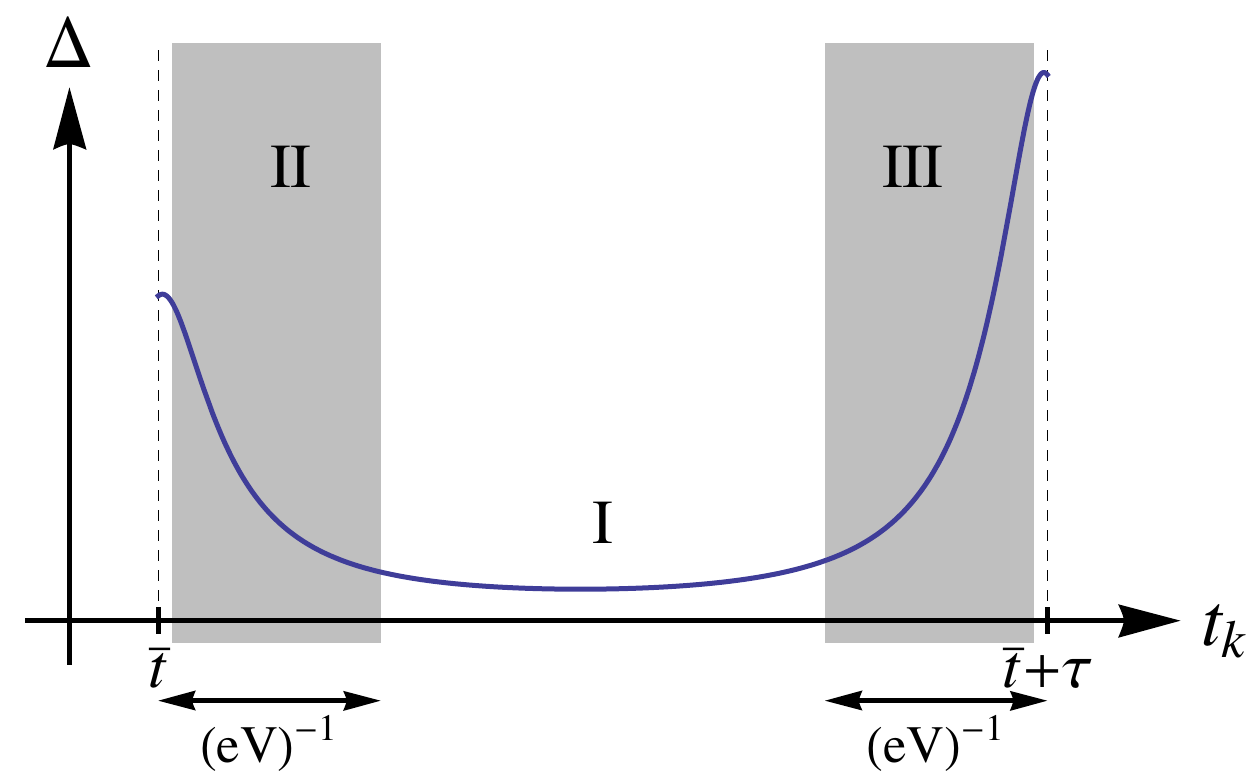}}
\caption{Sketch of the singular Fredholm determinant
$\Delta[4\vartheta_+(\bullet; t_k),f_{++}]$  as function of time $t_k$. 
In region I, the asymptotic 
expansion~(\ref{eqn:AsTripleStep}) is valid.   
In regions II and III, power-law exponents are different, and  $\Delta$
is given by 
Eqs.~(\ref{eqn:AsTripleStep1}) and (\ref{eqn:AsTripleStep2}), respectively.} 
\label{fig:Scetch_Delta}
\end{figure}

If $t_k-\bar t \lesssim 1/eV$, we obtain
\begin{widetext}
\begin{equation}
\label{eqn:AsTripleStep1}
	(-1)^{1+k} \Delta[4\vartheta_+(\bullet;t_k),f_{++}] =\sum_{n_2} \Gamma'_{n_2} e^{i\alpha_1\Lambda \tau+i(\beta_1 + n_2) eV\tau}
 \left\lvert \frac{\Lambda (t_k-\bar t)}\pi\right\rvert^{p'_{12}} 
\left\lvert \frac{\Lambda\tau}\pi\right\rvert^{p'_{13}} \left\lvert\frac{\pi eV}\Lambda\right\rvert^{q'_{12}},
\end{equation}
where we have introduced the exponents
\begin{equation}
p'_{12} = -(1+\alpha_1), \quad p'_{13} = -(\alpha_1-\beta_1-n_2)^2- (\beta_1+n_2)^2, \quad 
q'_{12} = 2(\alpha_1-\beta_1-n_2)(\beta_1+n_2).
\label{eq:exp_region_II}
\end{equation}
In the other limit, $\bar t +\tau-t_k \lesssim 1/eV$, the asymptotic expansion yields
\begin{equation}
\label{eqn:AsTripleStep2}
	(-1)^{1+k} \Delta[4\vartheta_+(\bullet;t_k),f_{++}] =\sum_{n_1} \Gamma''_{n_1} e^{i\alpha_1\Lambda \tau+i(\beta_1 + n_1) eV\tau}
 \left\lvert \frac{\Lambda (\bar t + \tau - t_k)}\pi\right\rvert^{p''_{23}} 
\left\lvert \frac{\Lambda\tau}\pi\right\rvert^{p''_{13}} \left\lvert\frac{\pi eV}\Lambda\right\rvert^{q''_{12}},
\end{equation}
with the exponents
\begin{equation}
p''_{23} = \alpha_1, \quad p''_{13} = -(1+\alpha_1-\beta_1-n_1)^2- (\beta_1+n_1)^2, \quad 
q''_{12} = 2(1+\alpha_1-\beta_1-n_1)(\beta_1+n_1).
\label{eq:exp_region_III}
\end{equation}
\end{widetext}
We notice, that Eqs.~(\ref{eqn:AsTripleStep1}), (\ref{eqn:AsTripleStep2}) and  (\ref{eqn:AsTripleStep})
represent the asymptotic expansion of the generalized Toeplitz determinant $\Delta[4\vartheta_+(\bullet;t_k),f_{++}]$
in the different domains of the variable $t_k$. It is straightforward to check, that 
asymptotic formulae~(\ref{eqn:AsTripleStep1}) and (\ref{eqn:AsTripleStep}) match each other at the scale 
$t_k-\bar t\simeq 1/eV$ because of the mutual relations
\begin{equation}
p'_{13}=p_{13} + p_{23}, \qquad q'_{12}-p'_{12}=q_{12}-p_{12} 
\label{eq:1_prime}
\end{equation} 
between the power-law exponents. Similarly, the expansion~(\ref{eqn:AsTripleStep2}) matches
Eq.~(\ref{eqn:AsTripleStep}) at the time scale $(\bar t~+~\tau~-~t_k) \simeq 1/eV$ due to 
analogous relations
\begin{equation}
p''_{13}=p_{13} + p_{12}, \qquad q''_{12}-p''_{23}=q_{12}-p_{23}. 
\end{equation} 
The sketch of the determinant $\Delta[4\vartheta_+(\bullet; t_k),f_{++}]$ as the function of the
time $t_k$ is shown in Fig.~\ref{fig:Scetch_Delta}.

To summarize this section, we have found the discretized representation $(g^{-1})_{1k}$ for
the inverse of the single-channel counting operator ${\cal D}_{**}(\bar t, t_k)$, see
Eq.~(\ref{eq:inv_g_1k}). The generalized Toeplitz determinant $\Delta[4\vartheta_+(\bullet; t_k),f_{++}]$,
appearing in this formula, is given by the asymptotic Eqs.~(\ref{eqn:AsTripleStep}) and 
(\ref{eqn:AsTripleStep1},\ref{eqn:AsTripleStep2}). Accordingly, the denominator $\Delta[4\bar\vartheta_+,f_{++}]$ 
can be found with the use of Eq.~(\ref{eqn:GutmanAsDouble}).

\subsection{Interference current in the strong coupling limit}
\label{s4.3}

In Sec.~\ref{s4.2} we have discussed the asymptotic properties of
singular Fredholm determinants and have found the inverse kernel ${\cal
D}_{**}^{-1}(\bar t,t)$
of the single-channel counting operator. We are now going to use these results
to evaluate the AB conductance in the strong-coupling limit.

We start by considering the particle number $N_{+-}$, which is given by Eq.~(\ref{eqn:CurrentExpPreExp})
of Sec.~\ref{s3.2}. Making use of the relation ${\cal A}_{\rm ferm} = -i\ln {\rm Det} {\cal D}$
together with Eq.~(\ref{eqn:preExp}), we can represent $N_{+-}=N^*_{-+}$ in the form
\begin{align} 
\label{eqn:MZICoherentNumber}
N_{+-}\!\!=\!\!-i(R_2 T_2)^{1/2}\, e^{i\Phi}\!\!\! \int\!\! d \bar t\, e^{i{\cal A}^{(0)}}
\frac{{\rm Det} \tilde{\cal D}}{{\rm Det} \tilde{\cal D}^{(0)}}
\langle+| \tilde f \tilde{\cal D}^{-1}(\bar t,\bar t)|-\rangle,
\end{align}
where we have defined the action 
${\cal A}^{(0)}={\cal A}_{\rm ferm}\Bigl|_{T_1=0} + \delta {\cal A}$ and the
operator $\tilde{\cal D}^{(0)}=\tilde{\cal D}\Bigl|_{T_1=0}$ in the absence of
edge-to-edge tunneling. Eq.~(\ref{eqn:MZICoherentNumber})
is evaluated at the optimal phases $\vartheta^q_{*\eta}(t)=\bar\vartheta_\eta w_{\bar t,\tau}(t)$ with
$\bar\vartheta_\eta = -\eta\pi/\nu$, which have different sign for upper and lower arms of the MZI.
It turns out that the integrand is in fact independent of time $\bar t$ and thus
the integral is formally divergent. This amounts to an infinite number of
electrons counted during an infinite measuring time in a stationary situation.
The stationary current its obtained by dropping the $\bar t$-integral and
putting, say, $\bar t=0$. 
Using Eqs.~(\ref{eq:mu_fD_kappa}), (\ref{eq:plus_minus}) one arrives at
\begin{align} 
e^{-1}I_{+-} = &-i(R_2 T_2)^{1/2}\, e^{i\Phi + 2i\bar\vartheta_+}
\frac{{\rm Det} \tilde{\cal D}}{{\rm Det} \tilde{\cal D}^{(0)}}\,e^{i{\cal A}^{(0)}} \nonumber \\
&\times \int_{\bar t}^{\bar t+\tau} dt' \,{\cal D}_{**}^{-1}(\bar t, t')f_{+-}(t'-\bar t) 
\label{eqn:MZICoherentCurrent}
\end{align}
Let us now use the result of Sec.~\ref{s4.1} where the evaluation of the matrix determinant 
${\rm Det}\,{\cal D}$ has been reduced to the product of single-channel determinants.
For the scalar operator ${\cal D}_{**}$ we introduce ${\cal D}_{**}^{(0)}={\cal D}_{**}\Bigl|_{T_1=0}$.
Taking into account the factorization formula~(\ref{eq:Det_D}) and the block structure of
the matrix ${\cal D}_*$, given by Eq.~(\ref{eq:D_star3}), one can write
\begin{equation}
\frac{{\rm Det} \tilde{\cal D}}{{\rm Det} \tilde{\cal D}^{(0)}} = 
\frac{{\rm Det} \tilde{\cal D}_{**}}{{\rm Det} \tilde{\cal D}^{(0)}_{**}}=
\frac{\Delta[4\bar\vartheta_+, f_{++}]}{\Delta[4\bar\vartheta_+, f_{+}]}
\end{equation}
(to derive this relation we have made use of the fact that the operator $\tilde {\cal D}_0$ is $T_1$--independent). 
The dimensionful operator kernel ${\cal D}_{\ast\ast}^{-1}(t_1,t_2)$ and its discretized dimensionless counterpart $g^{-1}_{ij}$ are related by the energy factor
\begin{align*}
	W= {\cal D}^{(0) -1}_{\ast\ast}(\bar t,t_k)/\left(g^{-1}\right)^{(0)}_{1k} \propto \Lambda
\end{align*}
where the label ``$(0)$'' denotes the $T_1\to 0$ limit. 
 With the help of Eq.~(\ref{eq:inv_g_1k}) for the matrix element $g^{-1}_{1k}$ the expression for
the current $I_{+-}$ is then reduced to the form 
\begin{align} 
	e^{-1} I_{+-}& = (R_1 T_1 R_2 T_2)^{1/2}\ e^{i\Phi+2i\bar\vartheta_+} e^{i{\cal A}^{(0)}}
	\int_{0}^{\tau}\!\! d t_k\, 
\times W\nonumber \\
&\times \frac{(-1)^{1+k}\Delta[4\vartheta_+(\bullet;t_k),f_{++}]}{\Delta[4\vartheta_+,f_+]}  
	(f_+(t_k)-f_-(t_k)).
\label{eq:I_pm_int}
\end{align}
Here we have substituted the off-diagonal density matrix element $f_{+-} = i(R_1 T_1)^{1/2}(f_+-f_-)$. 
In the formula above, one needs to perform further the integration over time $t_k$ and
to evaluate the action ${\cal A}^{(0)}$ in the absence of tunneling between two edges of the MZI.
These steps of calculations are discussed below.

\subsubsection{The time integral over $t_k$}
\label{s4.3.1}

To perform the time integration let us consider the $t_k$--dependent part in Eq.~(\ref{eq:I_pm_int}),
\newcommand{\Jo}{{\mathcal J}}
\begin{equation}
		\Jo(t_k) \equiv (-1)^{1+k}\Delta[4\vartheta_+(\bullet;t_k),f_{++}] (f_+(t_k)-f_-(t_k)). 
\end{equation}
According to the asymptotic analysis of the previous Sec.~\ref{s4.2}, this integrand
has the power-law singularities which give the dominant contribution to the integral~(\ref{eq:I_pm_int})
around $t_k\sim 0$ and $t_k \sim \tau$, provided the real part of the corresponding power-law
exponents is negative. In general, for any time  $t_k \in (0,\tau)$ the integrand is a superposition of 
powers-law terms,
\begin{align}
\Jo(t_k) \sim \sum_{\{n\}}\,t_k^{\tilde p_{12}} (\tau-t_k)^{\tilde p_{23}} \tau^{\tilde p_{13}} (eV)^{\tilde q_{12}} 
\Lambda^{\tilde \gamma+1},
\end{align}
where the exponent $\tilde \gamma\equiv \tilde p_{12}+\tilde p_{23} +\tilde p_{13}-\tilde q_{12}$ 
ensures the correct dimensionality (which is inverse time). As discussed previously, the sum here runs over 
integers $n_1$, $n_2$ or both depending on whether the time $t_k$ lies in the region II, III or I, 
respectively (see Fig.~\ref{fig:Scetch_Delta}). For instance, in the region I the function ${\cal J}$,
in accordance with Eq.~(\ref{eqn:AsTripleStep}), has the above scaling behavior with 
$\tilde p_{12} = p_{12}-1$, $\tilde p_{23} = p_{23}$, $\tilde p_{13} = p_{13}$ and $\tilde q_{12}=q_{12}$.
Equation (\ref{eqn:TripleStepExp}) shows that by choosing $\lvert n_1\rvert$ and
$\lvert n_2\rvert$ sufficiently 
large, $\Re p_{12}$ and $\Re p_{23}$, respectively, can be easily made negative.  Therefore the integral
over $t_k$ will be determined by a vicinity of the end points of the time
interval $(0,\tau)$.
 
First, let us examine the limit of short times $t_k \ll \tau$. One has
to consider two asymptotic
regions. For $\Lambda^{-1}\lesssim t_k \lesssim (eV)^{-1}$  (region II) we can use the short time
expansion
\begin{align}
		f_+(t_k)-f_-(t_k) = \frac i{2\pi} \frac{e^{-ieVt_k}-1}{t_k} \simeq \frac{eV}{2\pi},
\end{align}
which gives the powers $\tilde p_{12}=p'_{12}$, $\tilde p_{23}=0$, $\tilde p_{13}=p'_{13}$, and $\tilde q_{12}=q'_{12}+1$. Evaluating the $t_k$--integral over the region II for some given 
integer $n_2$, we find
\begin{eqnarray}
\int^{1/eV}_{1/\Lambda}\!\!\! d t_k\, \Jo(t_k) & \sim & (eV\tau)^{p'_{13}}
(eV/\Lambda)^{q'_{12}+1-p'_{13}} \nonumber \\
&\times & 
(\Lambda t_k)^{p'_{12}+1}\Biggl\rvert_{1/eV}^{1/\Lambda}\nonumber\\
		& \sim & (eV\tau)^{p'_{13}}
(eV/\Lambda)^{1+\alpha_1^2+\alpha_1},
		\label{eq:Int_Region_II}
\end{eqnarray}
where we kept only the dominant contribution. Here $\alpha_1 = -2/\nu$ and, cf. Eq.~(\ref{eq:exp_region_II}),
\begin{align}
		p'_{13} = -2\left(n_2-\frac{\alpha_1-2\beta_1}2\right)^2 -\frac{\alpha_1^2}2
		\label{eq:p_13_prime}
\end{align}
The term which gives the leading contribution to the current from the region II is then found
by maximizing ${\rm Re}\, p'_{13}$ with respect to $n_2$.  
The fact that the exponent $\Lambda$ is independent on $n_2$ is not a coincidence. It encodes renormalization effects due to high-energy virtual excitations. In contrast, the arbitrary integers which encode different branches of $\ln \tilde g$ are relevant for intermediate energies $0<\epsilon<eV$ only, and thus do not affect the high energy scale $\Lambda$.

Let us further look onto longer times, $(eV)^{-1} \lesssim t_k \ll \tau$ from region I,  
where we can approximate $f_+(t_k)-f_-(t_k)\sim 1/t_k$. Hence the powers are 
$\tilde p_{12}=p_{12}-1$, $\tilde p_{23}=p_{23}$, $\tilde p_{13}=p_{13}$, and $\tilde q_{12}=q_{12}$. 
By choosing some intermediate time scale $(eV)^{-1} \lesssim t \ll \tau$ as the upper cut-off,
the integral reads
\begin{eqnarray*}
		\int_{1/eV}^t\!\!\! d t_k\,\Jo(t_k) & \sim &
(eV\tau)^{p_{13}+p_{23}} \nonumber \\
& \times &(eV/\Lambda)^{q_{12}-p_{13}-p_{23}}
(\Lambda t_k)^{p_{12}} \Biggl\rvert_{1/eV}^t.
\end{eqnarray*}
Under the assumption $\Re p_{12}<0$, the upper boundary $t$ is irrelevant. Using
relations~(\ref{eq:exp_region_II})
and (\ref{eq:1_prime}), we obtain exactly the same asymptotics as in Eq.~(\ref{eq:Int_Region_II}), 
\begin{align}
		\int_{1/eV}^t\! d t_k\, \Jo(t_k) \sim (eV\tau)^{p'_{13}} (eV/\Lambda)^{1+\alpha_1^2+\alpha_1}.
\end{align}

We turn now to the analysis of the integral~(\ref{eq:I_pm_int}) around the
second singularity 
$t_k\sim\tau$. Close to this end point one has $f_+(t_k)-f_-(t_k)\sim \tau^{-1}$. Following 
the same line of reasoning as above, we consider two asymptotic regions (I and III). 
The time integral over the region III for a given integer $n_1$ yields 
(we recall that we consider the case $\nu\geq 2$)
 \begin{align}
\int^{\tau-\Lambda^{-1}}_{\tau-(eV)^{-1}} &d t_k\, \Jo(t_k) 
\sim (eV\tau)^{p''_{13}-1} (eV/\Lambda)^{2+\alpha_1^2+2\alpha_1} \nonumber \\
&\times \Bigl(\Lambda(\tau-t_k)\Bigr)^{1+\alpha_1}\Bigg\rvert_{\tau-(eV)^{-1}}^{\tau-\Lambda^{-1}}\nonumber\\
		& \sim (eV\tau)^{p''_{13}-1} (eV/\Lambda)^{1+\alpha_1^2+\alpha_1},
\label{eqn:TauSingularity1}
\end{align}
 The exponent $p''_{13}-1$ can be read from the definition~(\ref{eq:exp_region_III}). 
 It is convenient to rewrite it in the form analogous to Eq.~(\ref{eq:p_13_prime}),
\begin{align}
		p''_{13}-1=-2\left(n_1-\frac{\alpha_1+1-2\beta_1}2\right)^2-\frac{(\alpha_1+1)^2}2-1.
\end{align} 
which explicitly shows that ${\rm Re}\, p''_{13}-1$ can be maximized with respect to $n_1$.
 
It remains to estimate the time integral when $t_k$ lies in the region I. 
Introducing as above some intermediate time scale $t$ satisfying $(eV)^{-1}\ll t<\tau-(eV)^{-1}$ 
(which in case of $\Re p_{23}<-1$ will be irrelevant as an integral boundary) 
one obtains
	\begin{align}
		\int_{t}^{\tau-(eV)^{-1}}\! d t_k\, \Jo(t_k) \sim (eV\tau)^{p''_{13}-1} (eV/\Lambda)^{1+\alpha_1^2+\alpha_1}.
	\end{align} 

In the following we are interested in the integers $n_2$ and $n_1$ which maximize the exponents,
$\Re p'_{13}$ and $\Re p''_{13}-1$. To this end we write $\alpha_1=M+m$ with $M\in\mathbb Z$ and $\lvert m\rvert \le 1/2$. Then for even $M$ leading contributions come from $n_2=M/2$ and $n_1=(M+1)/2\pm1/2$, and for odd $M$ they come from $n_2=M/2\pm1/2$ and $n_1=(M+1)/2$. Straightforward analysis shows that for integer filling
fractions $\nu\ge 2$ in all optimal contributions we have $\Re p'_{13}\ge \Re p''_{13}-1$.

These observations lead us to the conclusion that, with all oscillatory terms $e^{ieV\tau n_2} e^{i\alpha_1\Lambda \tau}$ and $t_k$-independent contributions,
\begin{align}
\Delta[4\bar\vartheta_+,f_+]^{-1} &\propto e^{-i \alpha_1(\Lambda+eV)\tau} 
\left(\frac{\Lambda\tau}\pi\right)^{ \alpha_1^2},
\end{align}
taken into account, the leading terms of the $t_k$-integral for $\nu\ge 2$ are
\begin{widetext}
\begin{align}
\label{eq:Int_tk}
\int_{1/\Lambda}^{\tau-1/\Lambda} d t_k\,{\cal J}(t_k)/\Delta[4\bar\vartheta_+,f_+]  &= 
\frac{\Lambda}\pi\ e^{-i\alpha_1 eV\tau+i\beta_1 eV\tau} (eV/\Lambda)^{1+\alpha_1} \\
&\times\left(\sum_{n_2} \Gamma'_{n_2}  e^{ieV\tau n_2}(eV\tau)^{p'_{13}+\alpha_1^2}
			+\sum_{n_1}\Gamma''_{n_1} \left(e^{-ieV\tau}-1\right)  
           e^{ieV\tau n_1}(eV\tau)^{p''_{13}-1+\alpha_1^2}\right) \nonumber
\end{align}
\end{widetext}	 
where $\Gamma'_{n_2}$ and $\Gamma''_{n_1}$ are some unknown dimensionless constants. This expansion contains all terms in leading order of $(eV/\Lambda)$ (also those subleading in $(eV\tau)$). 
 
\subsubsection{Action ${\cal A}^{(0)}$ in the absence of tunneling}
\label{s4.3.2}
Let us now evaluate the action of the system when inter-edge tunneling is absent, 
\begin{align}
i{\cal A}^{(0)}={\rm Tr}\Ln\left[\mathds{1}-\hat f +e^{2i\hat \vartheta^q}\hat f\right]-2i{\rm Tr}\hat \vartheta^q f_0.
\end{align}
In this subsection the traces extend over all $\nu$ upper and $\nu$ lower inner channels. We combined all $2\nu$ distribution functions $f_\lambda$ and phases $\vartheta^q_\lambda$ into $2\nu\times 2\nu$-matrices $\hat f$ and $\hat \vartheta^q$. Due to the Dzyaloshinskii-Larkin theorem we anticipate that only first-and-second-order-in-$\vartheta$ terms are non-vanishing for the action above (it is worth reminding here that throughout this section 
the distribution functions were assumed to be the Fermi-like). Hence we expand,
\begin{align}
i{\cal A}^{(0)} &= {\rm Tr}\left[\Ln \left[\mathds{1}+\left(2i\hat \vartheta^q-2\hat \vartheta^{q2}\right)\hat f\right]-2i\hat \vartheta^qf_0\right]\nonumber \\
&=2i  {\rm Tr}\,\hat \vartheta^q\left(\hat f-f_0\right)-2\Tr\,\hat \vartheta^q (\mathds{1}-\hat f)\hat \vartheta^q\hat f.
\end{align}
Consider now three local operators $A$, $B$, $C$, where by definition $A(t_1,t_2)=A(t_1) \delta(t_1-t_2)$ etc. 
Evaluating the following trace (one should carefully take into account here the non-local in time 
structure of the Fermi-distribution function), one obtains
\begin{align} 
\label{eqn:commuteTrace}
\Tr_t A [B,f_0]C&=\int\!\! d t_2\, \lim_{t_1\to t_2} A(t_1) 
\frac i{2\pi} \frac {B(t_1)-B(t_2)}{t_1-t_2+i 0}C(t_2)\nonumber \\
&=\frac i{2\pi}\int\!\! d t\, A(t) \dot B(t) C(t).
\end{align}
For Fermi-like non-equilibrium distribution functions, $f_1(t)=e^{-ieV_1t} f_0(t)$, the above relation 
implies
\begin{align} 
\label{eqn:voltageTrace}
{\rm Tr}_t A(f_1-f_0)C=\frac{eV_1}{2\pi} \int\!\! d t\, A(t)C(t).
\end{align}
Let us assume that among the channels belonging to the upper edge of the MZI, the outer channel is biased by $V$ 
and all the rest ones by $V_0$. At the same time, all channels on the lower edge are grounded. 
This gives the first, ``zero mode'' contribution to the action,
\begin{align}
		2i {\rm Tr}\hat \vartheta^q\left(\hat f-f_0\right)
		= i \frac{e\left[V+(\nu-1)V_0\right]\tau}{2\pi}\ 2\bar \vartheta_+.
		\label{eq:theta_expansion1}
\end{align}
The quadratic contribution to the action $i{\cal A}^{(0)}$ is UV divergent and needs to be cutoff by 
at the scale $\Lambda\sim\omega_c$, yielding
	\begin{align}
	\label{eq:theta_expansion2}
		-2{\rm Tr}\vartheta^q \left(\mathds{1}-\hat f\right)\vartheta^q \hat f= 
-\frac {2\nu\bar \vartheta^{2}_+}{\pi^2}\ln\omega_c\tau=-\frac 2\nu\ln\omega_c\tau.
	\end{align}
In passing we note, that since ${\cal A}^{(0)}$ is a purely Gaussian contribution,
it can be equivalently evaluated by averaging the phase 
$e^{i{\cal A}^{(0)}}=\left\langle e^{i\vartheta^f_+(0)-i\vartheta^b_-(0)}\right\rangle_0$
over the Gaussian action of the MZI in the limit $T_1\to 0$.

\subsubsection{Current, Conductivity and Visibility}
\label{s4.3.3}
The results of previous subsections enable us to evaluate the Aharonov-Bohm current in the MZI.
Setting the voltage of all outer channels to zero, $V=0$, and defining the exponents
\begin{align}
	p'(n_2)&\equiv -2\left(n_2-\frac{\alpha_1-2\beta_1}2\right)^2+1+\frac{\alpha_1^2}{2}+\alpha_1,\\
	p''(n_1)&\equiv -2\left(n_1-\frac{\alpha_1+1-2\beta_1}2\right)^2-\frac 12+\frac{\alpha_1^2}{2},
\end{align}
one can represent the coherent current contribution in the form
\newcommand{\jo}{{\mathcal I}_0}
\begin{equation}
I_{\rm coh}=2 \Re I_{+-} =\frac e{2\pi\tau} (R_1 T_1 R_2 T_2)^{1/2} 2\,\Re e^{i\Phi}\ \jo,
\end{equation}
where $\jo$ is the normalized amplitude of the Aharonov-Bohm current. 
Collecting Eqs.~(\ref{eqn:MZICoherentCurrent}), (\ref{eq:Int_tk}), (\ref{eq:theta_expansion1}) 
and (\ref{eq:theta_expansion2}) together, we finally arrive at
\begin{widetext}
\begin{multline}
		\jo = 2e^{-2\pi i/\nu} e^{i(\beta_1 + 1/\nu)eV\tau}
		 \times \left(\sum_{n_2}\Gamma'_{n_2} e^{in_2eV\tau} \left(eV\tau\right)^{p'(n_2)}+\sum_{n_1}\Gamma''_{n_1} e^{in_1eV\tau} \left(e^{-ieV\tau}-1\right)\left(eV\tau\right)^{p''(n_1)}\right).
		 \label{eq:I0_series}
\end{multline}
\end{widetext}
The following table shows the power-law exponents corresponding to the terms which give
the dominant contribution to the series~(\ref{eq:I0_series}) at each filling
factor $\nu$:
	\begin{center}
	\begin{tabular}{c|l}
		$\nu$ & leading powers\\
		\hline
		2 & $p'(0)=p'(-1)=0=p''(0)$\\
		3 & $p'(-1$) and $p'(0)$\\
		$\ge 4$ & $p'(0)$ and $p''(0)$ if $T_1<1/2$ or $p''(1)$ if $T_1>1/2$
	\end{tabular}
	\end{center}	
Taking these leading terms into account we obtain the results presented in Sect.~\ref{s2.2}, 
namely Eq.~(\ref{eq:I_0}) with the exponents (\ref{eqn:MZIExpLowNu}) and (\ref{eqn:MZIExpHighNu}).	
We have also checked the validity of analytical asymptotics~(\ref{eq:I0_series}) by straightforward
numerical evaluation of Eq.~(\ref{eq:I_pm_int}) for the AB current. The perfect agreement between
the analytical and numerical approaches, demonstrated in Fig.~\ref{fig:MZI_grid}, provides
additional support towards the conjecture of Ref.~\onlinecite{Protopopov:2012}.

We close this section by providing the reader with a qualitative physical picture, underlying 
the result (\ref{eq:I0_series}).  We begin with inspection of the phase pulse $\delta(t)=4\vartheta_+(t;t_k)$
which according to our discussion in Sec.~\ref{s4.2} determines the resolvent (${\cal D}^{-1}$) of
the counting operator (see Fig.~\ref{fig:PhaseWPulse}). First, we note that in our model of the maximally long-ranged
Coulomb interaction the counting phase $4\vartheta_+(t) = (4\pi/\nu) q(t)$, with $q(t)$ being the
``background'' charge on the upper arm of the MZI. The phase $\delta(t)$ is different from $4\vartheta_+(t)$
by the $2\pi$-pulse of the duration $t_k$, which describes injection of the interfering ``transport''
electron at time $\bar t = 0$ into the interferometer via the QPC1 and its annihilation at time $t_k$ 
(the electrons leaves the MZI through the QPC2). As we see it from the calculations of the Sec.~\ref{s4.3.1},
in the high-voltage limit $eV \tau \gg 1$, the critical exponents $p'$ are associated with many-particle scattering processes with short $t_k$, $t_k \simeq \hbar/eV$. We can interpret it as the event when one electron enters 
the MZI through the QPC1 and another electron leaves the MZI shortly afterward on a time scale $\sim \hbar/eV$ via 
the QPC2.  
The possibility for an electron to propagate through the system in a time $t_k\ll \tau$ is due to the long-range nature of Coulomb interaction in our model 
(a similar situation will occur in a model with a strong short-ranged interaction, where such a process can be mediated by the exchange of plasmons, which have a velocity exceeding by far the bare velocity of electrons).
On the contrary, the second critical exponent $p''$ is due to many-body scattering events with time $t_k\simeq \tau$. 
In this case the MZI is excited into a state with an extra electron for a long time $\tau \gg \hbar/eV$
(it will be, e.g., the only possibility in the limit of weak short-ranged interaction when the behavior of 
the MZI is very close to the one with non-interacting electrons). 

Next, we associate the integers $n_1$ and $n_2$ with the number of inelastically excited electron-hole
pairs on the MZI in the corresponding time intervals (at $0<t<t_k$ and $t_k < t < \tau$, respectively).
More precisely, one can interpret each term in the asymptotic expansion for the current ${\cal J}(t_k)$
as the product of forward (electron-like) and backward (hole-like) many-particle scattering 
amplitudes, $A^{b}(\{n^-\}; t_k) \times A^f(\{n^{+}\}; t_k)$. An electron, propagating through the upper arm of
the MZI leaves a trace in the bath of particle-hole excitations, which is encoded 
in the numbers $n^+_1$ and $n^+_2$ of the excited electron-hole pairs after the electron's injection and annihilation. 
The dominant e-h pairs correspond to excitations from one Fermi edge to the other, so that the typical energy of an electron-hole pair is $\hbar \omega \simeq eV$. The corresponding many-particle 
amplitude should behave as
\begin{equation*}
A^f(\{n^{+}\}; t_k) \sim e^{-i n^+_1 eV t_k - i n_2^+ eV (\tau -t_k)}.
\end{equation*}
Similarly, the backward amplitude is characterized by numbers $n_1^-$ and $n^-_2$.   
In the case when the relative numbers $n_{1} = n^+_1-n^-_1$ and $n_{2} = n^+_2-n^-_2$
between the forward and backward scattering amplitudes
are non-zero one obtains an interference term in the current with a phase
which is linear in voltage. The two dominant terms
with inequivalent phases in the series for the current $I_{+-}$ produce the lobe pattern in the
visibility of the Aharonov-Bohm oscillations. For example, at $\nu=3$ such two terms
are those corresponding to scattering processes characterized by the short time $t_k \sim \hbar/eV$ but 
having different numbers $n_2=-1$ and $n_2=0$, resp.

\newcommand{\eq}{{({\rm eq})}} 
\newcommand{\norm}{{\rm norm}} 
\newcommand{\act}{{\cal A}}
\newcommand{\ferm}{{\rm ferm}}
\newcommand{\dd}{{\rm d}}

\section{Numerical approach}
\label{s5}

In this section we present a numerical evaluation scheme for the interference current. 
Results, obtained within this scheme, corroborate and complement the analytical 
study of Sec.~\ref{s4} based on the particular form of the Fredholm
counting operator ${\cal D}$.
After making a few introductory remarks regarding the numerical formula for the Aharonov-Bohm current,
which is suitable for a practical implementation, we consider two cases when we
did not succeed to obtain the analytical asymptotics and had to evaluate the
Fredholm determinants numerically. This is, first, the case of non-equilibrium
incoming distribution (Sec.~\ref{s5.1}) and, second, the regime of intermediate
interaction strength $E_c\tau \sim 1$ (Sec.~\ref{s5.2}).

As argued in Sec.~\ref{s4.3}, the interference current is obtained from
Eq.~(\ref{eqn:CurrentExpPreExp}) 
by dropping the divergent $\bar t$-integral and putting $\bar t=0$. For the purpose of the present section
we rewrite this formula in the equivalent form
\begin{align} 
\label{eqn:currentNormEq}
	I_{+-} = I_{+-}^\eq \left[\Det \tilde{\cal D}\right]_\norm \left[\bra + \tilde f\tilde {\cal D}^{-1}(0,0) 
\ket -\right]_\norm.
\end{align}
Here,
\begin{align*}
	I_{+-}^\eq = -i \left(R_2 T_2\right)^{1/2} e^{i\Phi} e^{i\act_{\ferm}+i\delta\act} \bra+\tilde f\tilde {\cal D}^{-1}(0,0)\ket-\Big\rvert_{eV=\epsilon_0}
\end{align*}
denotes the near-to-equilibrium current, evaluated at very small voltages $\epsilon_0\tau\ll 1$, while the label ``norm'' refers to quantities which are normalized with respect to near-to-equilibrium values, i.e.
\begin{align*}
	\left[\Det \tilde {\cal D}\right]_\norm = {\Det \tilde {\cal D}}/{\Det \tilde {\cal D}\Big\rvert_{eV=\epsilon_0}}.
\end{align*}
The density matrix $\tilde f$ can be removed from the matrix element appearing in Eq.~(\ref{eqn:currentNormEq}) 
by using a relation
\begin{align*}
	\xUnit-\tilde{\cal D}^{-1} = \left(e^{2i\hat \vartheta^q}-\xUnit\right)\tilde f \tilde {\cal D}^{-1},
\end{align*}
which leads to
\begin{align*}
	\left[\bra + \tilde f\tilde {\cal D}^{-1}(0,0) \ket -\right]_\norm = \frac{\bra+ \tilde {\cal D}^{-1}(0,0) \ket -}{\bra+ \tilde {\cal D}^{-1}(0,0) \ket -\Big\rvert_{eV=\epsilon_0}}.
\end{align*}
By solving an appropriate Riemann-Hilbert problem we were able to show that close to equilibrium all effects of non-equilibrium and interaction, like dephasing and renormalization, are absent for the Aharonov-Bohm current. 
In agreement with the Landauer-B\"uttiker result it depends linearly on voltage, thus
\begin{align*}
	I^\eq_{+-} = \left(R_1T_1R_2T_2\right)^{1/2} e^{i\Phi}\ \frac{e\epsilon_0}{2\pi}.
\end{align*}
Corrections to this equilibrium expression are encoded in the normalized terms in Eq.~(\ref{eqn:currentNormEq}). By discretizing the integral operator ${\cal D}$ the latter become amenable to numerical evaluation.
Below we discuss two different discretization schemes. In practice, the choice between two
depends mainly on the strength of Coulomb interaction.

\subsection{Non-equilibrium incoming distribution}
\label{s5.1}

We consider the situation when the incoming electron beam is diluted and
is driven out of equilibrium even before scattering at the first QPC. More
specifically, we focus on the setup (that has been realized experimentally)
with an additional QPCO with a transparency $0<R<1$ placed outside of the
interferometer and diluting the incoming beam. The incoming then
distribution function acquires the  
double-step form, $f_+(\epsilon)=T_0 \theta(-\epsilon)+R_0\theta(eV-\epsilon)$.

As in Sec.~\ref{s4}, we focus here on the limit of strong interaction
$E_c\tau\gg 1$, where the counting  
phase~(\ref{eq:w_func}) is a window function and the counting operator $\cal D$
is of block Toeplitz form with a two-channel structure. Note that its
decomposition into single-channel operators that was performed in  
Sec.~\ref{s4.1} requires trivial incoming distribution functions $f_+$,
$f_-$, i.e.\ zero-temperature Fermi distributions, possibly with different
chemical potentials. Thus the non-equilibrium form of $f_+$ constitutes an
obstacle for a further analytical evaluation. At this stage, we do not know
whether there is an analytical way to overcome this problem. We thus resort to
a numerical evaluation of the determinants.

To proceed numerically we use the same discretization procedure
as in Sec.~\ref{s4.2}, where the energy cutoff $\Lambda$ is introduced and
the symbol $\hat g(\epsilon)$ of $\tilde{\cal D}$ is required to satisfy
periodic boundary conditions in energy domain, 
\begin{align*}
	 \hat g(\epsilon) = e^{i\hat {\bar \vartheta} \epsilon/\Lambda}
\left[\xUnit+(e^{2i\hat{\bar \vartheta}}-\xUnit) \tilde f(\epsilon)\right]. 
\end{align*}
Now the symbol $\hat g(\epsilon)\in {\mathbb C}^{2\times 2}$ has an additional
$2\times 2$-channel structure, the same holds for the diagonal  matrix $\hat
{\bar \vartheta}$ with diagonal entries $\bar\vartheta_+$, $\bar\vartheta_-$. 
In the discretized representation the counting operator $\tilde {\cal D}$ then reads
\begin{widetext}
\begin{align} \label{eqn:DDiscreteToeplitz}
	\tilde{\mathcal D}_{j-k} &= \int_{-\Lambda}^\Lambda\!\frac{\dd \epsilon}{2\Lambda}\, e^{-i \epsilon\frac\pi\Lambda [j-k]}\ \hat g(\epsilon)\nonumber\\
	&= \frac i{2\pi} \frac{e^{2i\hat{\bar\vartheta}}-\xUnit}{j-k-\hat {\bar\vartheta}/\pi} \begin{pmatrix}
		R_1R_0 e^{-i \pi \frac{eV}\Lambda \left[j-k-\bar\vartheta_+/\pi\right]}+T_0+T_1R_0 & i\left(R_1 T_1\right)^{1/2} R_0 \left(e^{-i \pi \frac{eV}\Lambda \left[j-k-\bar\vartheta_+/\pi\right]}-1\right)\\
		-i\left(R_1 T_1\right)^{1/2} R_0 \left(e^{-i \pi \frac{eV}\Lambda \left[j-k+\bar\vartheta_+/\pi\right]}-1\right) & T_1 R_0 e^{-i \pi \frac{eV}\Lambda \left[j-k+\bar\vartheta_+/\pi\right]}+R_1+T_1T_0
	\end{pmatrix}.
\end{align}
\end{widetext}

The resulting visibility which follows from the above approach is shown and discussed in Sec.~\ref{s2.2}, 
see Fig.~\ref{fig:Inj_noneq}. A good convergence was already achieved for moderate matrix sizes 
with $N=\Lambda \tau/\pi\sim 100$. While we do not have a complete analytical
form of the Aharonov-Bohm current in this case, the dephasing rate can be
deduced from the leading large-$\tau$ 
asymptotic behavior of $\Det \tilde{\cal D} \sim e^{-\tau/\tau_\phi}$. Making
use of the results for the block Toeplitz determinants~\cite{Widom:1974}, we
arrive at
\begin{align}
	\tau_\phi^{-1} &= -\int\!\frac{\dd \epsilon}{2\pi}\, \Re\ln\det \left[\xUnit-\hat f+s_1^\dagger e^{2i\hat{\bar \vartheta}} s_1 \hat f\right]\\
	&=-\frac{eV}{2\pi} \Re\ln \left[T_0+R_0\left(R_1 e^{-i2\pi/\nu} +T_1 e^{i2\pi/\nu}\right)\right] \nonumber
\end{align}
In the case $\nu=2$ the dephasing rate is simplified to $\tau_\phi^{-1} = -({eV}/{2\pi}) \ln |2 R_0-1|$,
as has been already mentioned in Sec.~\ref{s2.2}.

\subsection{Intermediate interaction strength}
\label{s5.2}

We now consider the discretization scheme applicable in the case of a moderate
charging energy, $E_c\tau\sim 1$. 
In this general case the correlation function (\ref{eq:J-greater-t}) has to be evaluated numerically
(see, e.g., Fig.~\ref{fig:JNumAna} in Sec.~\ref{s3.2} which shows the result
for $\omega_c\tau=25$).  
Owing to finite $E_c$, the counting phase $\vartheta_+(t)$ is not a
piecewise constant function of time anymore, but rather acquires oscillations in
time. As an illustration, the phase $\theta_+(t)$ is 
plotted in Fig.~\ref{fig:Phase_plus} for two different charging energies. Thus,
the situation is 
different from the previous subsection, since the counting operator is no
longer of Toeplitz form. 
Despite this complication, a numerical treatment based on
Eq.~(\ref{eqn:currentNormEq}) is nevertheless possible, 
but a time discretization of the kernel ${\cal D}(t,t')$  
has to be performed directly in the time domain 
without any reference to the conjugate energy representation. To this end, we
rely on the approach similar to the one used in Ref.~\onlinecite{Snyman:2007}
(see Supplemental Material of that work)  and interpret 
the zero-temperature Fermi distribution function $f_0$ in terms of the Cauchy principal value and 
the Dirac delta distribution,
\begin{align*}
	f_0(t) = \frac i{2\pi} \frac 1{t+i0}=\frac i{2\pi}\mathcal P \frac 1 t +\frac 12 \delta(t).
\end{align*}
Consequently, the discretization of the non-equilibrium single-particle density matrix,
$\tilde f_{jk} = \hat s_1 \hat f^1(t_i, t_j) \hat s_1^\dagger \Delta t$,  with $t_j=(j-1)\Delta t$
and $\Delta t = \pi/\Lambda$ yields
\begin{widetext}
\begin{align*}
	\tilde f_{jk} &= \left(1-\delta_{jk}\right) \frac i{2\pi} \frac 1{j-k} +\frac 12 \delta_{jk} +
\begin{pmatrix}
R_1 & \left(R_1 T_1\right)^{1/2} \\ -i\left(R_1 T_1\right)^{1/2}& T_1 
\end{pmatrix} 
\left[\left(1-\delta_{jk}\right) \frac i{2\pi} \frac{e^{-i\pi \frac{eV}\Lambda(j-k)}-1}{j-k}+\delta_{jk} \frac{eV}{2\Lambda}\right],\\
\end{align*}
\end{widetext}
with $\Lambda\gg E_c$ being a high-energy cut-off.
Making use of above expression one can further construct the discretized matrix of the counting operator
\begin{align}
\label{eq:D_disr}
	\tilde {\cal D}_{jk} = \delta_{jk} \xUnit_2- \left(e^{2i\hat\vartheta^q(t_j)}-\xUnit_2\right)\circ\tilde f_{jk}
\end{align}
where ``$\circ$'' denotes matrix multiplication with respect to channel indices. 

The discretized form outlined above is very general, since it allows for
arbitrary time-dependent phases. Note, however, that for the case of
window-function time dependence of the ``counting'' phase (i.e., Toeplitz case)
this regularization yields a result which is manifestly $2\pi$ periodic in
$2\vartheta^q$, with nonanalyticity at points $(2n+1)\pi$. This 
should be contrasted to the analytic and non-periodic
behavior of the Toeplitz determinant within the proper regularization discussed
above. The corresponding difference between the present problem and that of FCS
has already been mentioned in the end of Sec.~\ref{s4.1.1}. We have checked that
for $|2\vartheta^q| < \pi$ both regularization schemes, Eqs.
(\ref{eqn:DDiscreteToeplitz}) and (\ref{eq:D_disr}) produce identical results,
although the convergence of the second scheme is generally worse. 

In the case of a finite charging energy the phase changes
continuously with time, and there is no more problem with the
regularization (\ref{eq:D_disr}), independently of how large the values
acquired by the phase are. We used the matrix size $N = \Lambda \tau/\pi \sim 500$,
which was sufficient to obtain numerical results for the visibility
at $E_c \sim 1/\tau$ with good precision. The results were presented
and discussed in Sec.~\ref{s2.3}. 

Finally, we have calculated analytically the long-$\tau$ asymptotics of the
action ${\cal A}_{\rm ferm} = -i \ln {\rm Det}{\cal D}$,
which gives the out-of-equilibrium dephasing rate~(\ref{eq:Deph_rate_Charging}) of the 
Aharonov-Bohm oscillations.

\section{Summary}
\label{concl}

In this paper, we have discussed an exactly solvable model of a quantum
Hall electronic Mach-Zehnder 
interferometer for arbitrary integer filling factor $\nu$. The model is specified by a form of
{\it e-e} interaction restricted to the inner part of the interferometer and two single-particle
scattering matrices of quantum point contacts (QPCs). Our main results can be summarized as follows:

\begin{enumerate}[(i)]

\item Making use of the non-equilibrium functional bosonization approach, we have established the 
exact solution of the above model in terms of the resolvent of the Fredholm integral operator
--- single-particle ``counting operator'' ${\cal D}$, which is related to the
problem of electron full counting statistics (FCS). The time-dependent
scattering phase
$\vartheta_+(t)$ of the operator ${\cal D}$ encodes all information about the interaction
in the system. The link between the initial many-body problem with Coulomb interaction and
single-particle quantities is established by virtue of the real-time instanton technique
which becomes exact for the specific type of the Keldysh action describing the MZI.  

\item The focus of our study was on the model with ``maximally long-range''
Coulomb interaction
characterized by the electrostatic charging energy $E_c$. In the limit of strong interaction
$E_c \gg 1/\tau$ (here $\tau$ is the electron flight time through the MZI) the scattering phase
$\vartheta_+(t)$ becomes a piecewise constant ``window'' function and the
operator $\cal D$ simplifies to the
block Toeplitz form. In the absence of external dephasing, we were able to
get rid of the matrix structure of $\cal D$ and have expressed the result
in terms of singular Fredholm determinants that may be viewed as a
generalization of Toeplitz determinants with Fisher-Hartwig singularities.  
This has allowed us to evaluate the Aharonov-Bohm (AB) conductance in a closed
analytical form. At a moderate charging energy 
$E_c \sim 1/\tau$ and/or in the situation when the incoming distribution is
made non-equilibrium by an additional QPC placed outside of MZI, we have
obtained the results for the visibility by evaluating the determinants
numerically. 

\item Results of our theory at $E_c \sim 1/\tau$ match in all principal aspects the experimental 
observation in many designs of Mach-Zehnder interferometers at filling factor $\nu$=2. If
the transmission coefficient $T_1$ of the QPC1 is close to $1/2$ the visibility dependence 
on external bias shows a number of ``lobes'', their amplitude is being
suppressed with the 
increase of voltage. The AB-phase dependence is close to a piecewise constant function with jumps 
equal to $\pi$ at minima of the visibility.  The visibility is further suppressed when
the MZI is subjected to an out-of-equilibrium shot noise, generated by the QPC0 placed outside
the interferometer. We have quantified the dephasing
rate $1/\tau_\phi$ which governs the decay
of AB oscillations with bias in terms the transmission of QPCs, filling factor $\nu$ and
the strength of {\it e-e} interaction $E_c\tau$.

\item Our analytical results in the limit of strong interaction $E_c \tau \gg 1$ 
show an intimate connection between the observed ``lobe'' structure in the
visibility, on one hand,  
and multiple branches in the asymptotics of singular integral determinants, on
the other hand. 
In more physical terms, this is the many-body interference effect  
resulting from the quantum superposition of many-particle scattering amplitudes 
with the mutual phase differences which are linear in external bias.
We derived the non-equilibrium quantum critical exponents, which 
depend both on the transmission $T_1$ and the filling factor $\nu$.
They are attributed to the Anderson orthogonality catastrophe under 
out-of-equilibrium conditions and describe the power-law dependence 
of the above many-particle amplitudes on voltage.

\end{enumerate}

Before closing, we mention some future research directions. An
extension of the presented approach to the fractional quantum Hall edge states devices,
comprising two (or more) QPCs which couple the co-propagating edge modes, would
be of great interest. Another important research direction is the
analysis of the asymptotic behavior of {\it block} Toeplitz determinants 
with Fisher-Hartwig singularities (and, more generally, block determinants,
with symbols that have multiple energy and time singularities). This would
not only permit to obtain closed analytical results in the model with moderate
interaction strength but would likely have multiple further applications.

\section{Acknowledgments}
\label{sec:acknowledgments}

This work was supported by the collaborative research grant  SFB/TR12 of the Deutsche Forschungsgemeinschaft
and by German-Israeli Foundation. D.B. is grateful to the TKM Institute at KIT
for the hospitality.

\bibliography{my_biblio}
\bibliographystyle{apsrev4-1}

\end{document}